%% file: MH_review.tex
\documentclass[twoside,12pt]{article}
\usepackage{epsfig,color,ulem,hyperref,float,amsmath}

\newcommand{\be}{\begin{equation}}
\newcommand{\ee}{\end{equation}}
\newcommand{\bea}{\begin{eqnarray}}
\newcommand{\eea}{\end{eqnarray}}

\begin{document}

\title{ \vspace{1cm} Neutrino Mass Hierarchy}
\author{X. Qian$^1$\footnote{Email: xqian@bnl.gov} and P. Vogel$^2$\footnote{Email:pvogel@caltech.edu} \\
\\
$^1$ Physics Department, Brookhaven National Laboratory, Upton, NY, USA \\
$^2$ Kellogg Radiation Laboratory, California Institute of Technology, Pasadena, CA, USA \\
}
\maketitle
\begin{abstract} 
The neutrino mass hierarchy, i.e., whether the $\nu_3$ neutrino mass eigenstate 
is heavier or lighter than the $\nu_1$ and $\nu_2$ mass eigenstates, is one of the remaining 
undetermined fundamental features of the neutrino Standard Model. Its determination would 
represent an important step in the formulation of the generalized model, and would have a 
profound impact on the quest of the nature of neutrinos (Dirac or Majorana) and the search for a 
theory of flavor. In this review, we summarize the status of experimental and theoretical work 
in this field and explore the future opportunities that emerge in light of the recently 
discovered non-zero and relatively large third neutrino mixing angle $\theta_{13}$.
\end{abstract}
\tableofcontents

\input{intro.tex}

\input{methods.tex}


\input{experiments.tex}

\input{summary.tex}

\section{Acknowledgments}
We thank Brett Viren, Chao Zhang, Steve Kettell, Wei Tang for 
reading the manuscript. This material is support in part by 
the U.S. Department of Energy, Office of Science, 
Office of High Energy Physics, Early Career Research program 
under contract number DE-SC0012704.

\bibliographystyle{unsrt}
\bibliography{MH_review}{}
\end{document}

%% file: intro.tex
\section{Introduction}~\label{sec:introduction}
Intense experimental and global analysis effort over the last several decades led 
to only a very small number of observations signifying that the standard 
model of elementary particles is incomplete and that ultimately it requires 
a generalization. Among them the existence of non-vanishing neutrino masses
has a prominent position. Despite the obvious difficulty of experiments 
involving neutrinos, it is now clear beyond any doubt that neutrinos have 
masses, albeit tiny, and that they mix, unlike the assumptions of the standard 
model. Thanks to the discoveries of the last twenty years, we now have a simple 
three-flavor description of these new phenomena that explains most of the data. 
Despite the success, this is not the end of the story. Further experimental and 
theoretical effort is needed so that the next step, formulation of a new theory 
that encompasses the new insight into the neutrino physics, can be accomplished.
 
Experiments with solar, atmospheric, reactor, and accelerator neutrinos have determined, 
with remarkable accuracy, most of the necessary parameters 
that describe the three-generation oscillations. However, several crucial pieces 
are still missing: the neutrino mass hierarchy (or the ordering of the neutrino 
masses), the magnitude of the CP (charge and parity) phase $\delta_{CP}$, the absolute scale 
of the neutrino mass, and whether neutrinos are Dirac or Majorana fermions.
This review is devoted to the problem of the neutrino mass hierarchy. 
 
In the following, we assume that the neutrinos $\nu_e$, $\nu_{\mu}$ and $\nu_{\tau}$ 
(so-called flavor eigenstates) that couple with the gauge bosons $W^{\pm}$ through weak 
interactions are coherent superpositions of three mass eigenstates 
$\nu_i, i=1,2,3$, i.e.
\begin{equation}
\left( \begin{array}{c} \nu_{e} \\ \nu_{\mu} \\ \nu_{\tau} \end{array} \right) = 
\left ( \begin{array}{ccc} U_{e1} & U_{e2} & U_{e3} \\
	U_{\mu1} & U_{\mu2} & U_{\mu3} \\
	U_{\tau1} & U_{\tau2} & U_{\tau3} 
\end{array} \right) \cdot \left( \begin{array}{c} \nu_{1} \\ 
\nu_{2} \\ \nu_{3} \end{array} \right) ,
\end{equation}
 where $U$ is a unitary 3$\times$3 matrix. In other words, we assume that 
there are three flavor and mass neutrino states and dismiss, 
for the purpose of this work, the possible existence of additional 
sterile neutrinos. This matrix implicitly contains possible charged lepton 
mixing; it is assumed that the corresponding charged lepton matrix matrix is diagonal 
in the flavor basis.

The lepton mixing matrix $U$~\cite{ponte1,ponte2,Maki} (usually denoted as the PMNS matrix 
for Pontecorvo, Maki, Nakagawa and Sakata who were the pioneers of the field) can be parametrized  
in terms of the three angles $\theta_{13}$,
$\theta_{23}$, and $\theta_{12}$ and the $CP$ phase $\delta_{CP}$.
\begin{eqnarray}
U &=& \left( \begin{array}{ccc} 
	1 & 0 & 0 \\
	0 & c_{23} & s_{23} \\
	0 & -s_{23} & c_{23} 
	\end{array}\right) \cdot 
	\left( \begin{array}{ccc}
	 c_{13} & 0 & s_{13} e^{-i\delta_{CP}} \\
	0 & 1 & 0 \\
	-s_{13} e^{i\delta_{CP}} & 0 & c_{13} 
	\end{array} \right) \cdot
	\left( \begin{array}{ccc}
		c_{12} &  s_{12} & 0 \\	
	-s_{12} & c_{12} & 0\\
	0   & 0 & 1 
	\end{array} \right) \cdot
        \left(\begin{array}{ccc}
          e^{i \alpha}& 0& 0\\
          0& e^{i \beta}& 0\\
          0& 0& 1\end{array}
          \right)
\nonumber \\
&=& \left( \begin{array}{ccc}
 c_{13}c_{12} & c_{13}s_{12} & s_{13} e^{-i \delta_{CP}}\\
  -c_{23}s_{12}-s_{13}s_{23}c_{12}e^{i\delta_{CP}} & 
c_{23}c_{12}-s_{13}s_{23}s_{12}e^{i\delta_{CP}} &
c_{13}s_{23} \\
 s_{23}s_{12}-s_{13}c_{23}c_{12}e^{i\delta_{CP}} & 
-s_{23}c_{12}-s_{13}c_{23}s_{12}e^{i\delta_{CP}} &
c_{13}c_{23}
\end{array} \right) \cdot
        \left(\begin{array}{ccc}
          e^{i \alpha}& 0& 0\\
          0& e^{i \beta}& 0\\
          0& 0& 1\end{array}
          \right)
\end{eqnarray}
where e.g. $c_{ij} = \cos\theta_{ij}$ and $s_{ij} = \sin\theta_{ij}$, etc. 
Here, $\alpha$ and $\beta$ (both unknown at the present) are so-called Majorana phases 
that are decoupled from the phenomenon of neutrino oscillation. 

\begin{table}[h] 
\label{tab:PMNS}
\caption{Neutrino oscillation parameters taken from Ref.~\protect\cite{GonzalezGarcia:2012sz}. 
For the atmospheric mass-squared splitting ($\Delta m^2_{3x}$), the best fit results for both 
the normal and the inverted mass hierarchy are shown. These values are used in all the following 
plots except where noted.}  
\medskip
\renewcommand{\arraystretch}{1.1} \centering 
\begin{tabular}{|c|cc|}
\hline 
parameter & best fit value $\pm~1\sigma$ & 3$\sigma$ range\\
\hline \hline
$\sin^2\theta_{12}$ & $0.304^{+0.012}_{-0.012}$ & (0.270, 0.344) \\
$\theta_{12}$ (degrees) & $33.48^{+0.77}_{-0.74}$ & (31.30, 35.90) \\
\hline 
$\sin^2\theta_{23}$ & [$0.451^{+0.001}_{-0.001}$] or $0.577^{+0.027}_{-0.035}$  & (0.385, 0.644)\\
$\theta_{23}$ (degrees) & [$42.2^{+0.1}_{-0.1}$] or $49.4^{+1.6}_{-2.0}$ & (38.4, 53.3) \\\hline 
$\sin^2\theta_{13}$ & $0.0219^{+0.0010}_{-0.0011}$ & (0.0188, 0.0251) \\
$\theta_{13}$ (degrees) & $8.52^{+0.20}_{-0.21}$ & (7.87, 9.11) \\\hline 
$\delta_{CP}$ (degrees) & $251^{+67}_{-59}$ & (0, 360) \\\hline \hline
$\Delta m^2_{21}$ $\times 10^{-5}$ eV$^2$ & $7.50^{+0.19}_{-0.17}$ & (7.03, 8.09) \\
\hline 
(normal) $\Delta m^2_{31}$ $\times 10^{-3}$ eV$^2$ & $+2.458^{+0.046}_{-0.047}$ & (+2.325, +2.599) \\
(inverted) $\Delta m^2_{32}$ $\times 10^{-3}$ eV$^2$ & $-2.448^{+0.047}_{-0.047}$ & (-2.590, -2.307) \\
\hline
\end{tabular}
\end{table}

\begin{figure}[ht]
\begin{centering}
\includegraphics[width=1.0\textwidth]{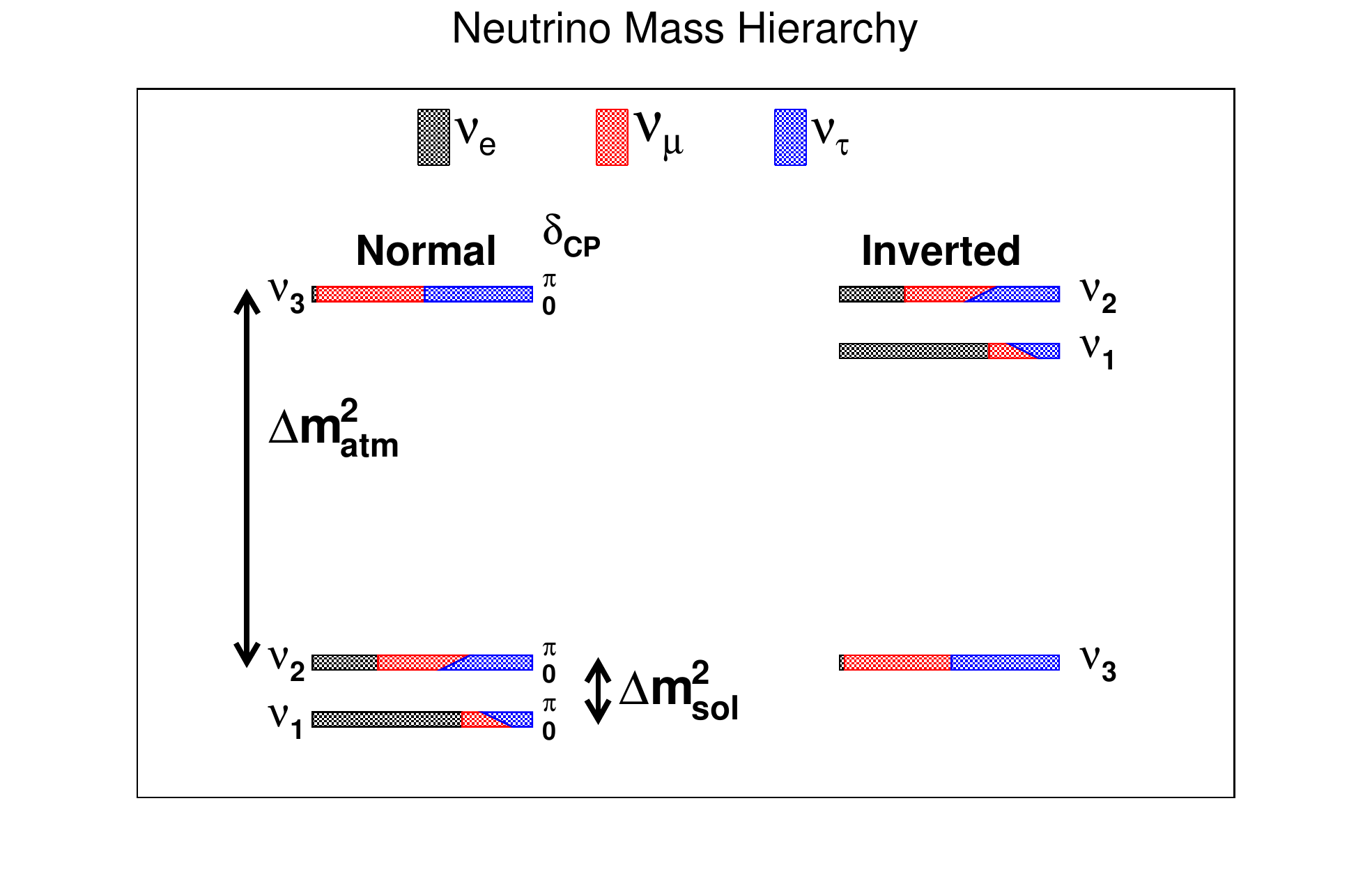}
\par\end{centering}
\caption{\label{fig:pattern}  
Pattern of neutrino masses for the normal and inverted hierarchies is shown as mass 
squared. Flavor composition of the mass eigenstates as the function of the unknown CP 
phase $\delta_{CP}$ is indicated. $\Delta m^2_{atm} \sim |\Delta m^2_{31}| \sim |\Delta m^2_{32}|$
and $\Delta m^2_{sol} \sim \Delta m^2_{21}$ stands for the atmospheric and the solar mass-squared 
splitting, respectively. }
\end{figure}

Table~\ref{tab:PMNS} summarizes the present knowledge of neutrino 
masses and mixings including neutrino mixing angles, the CP phase $\delta_{CP}$,
and neutrino mass-squared differences $\Delta m^2_{ij} = m^2_i - m^2_j$,
based on the recent fit~\cite{GonzalezGarcia:2012sz} after the Neutrino 2014 
conference~\cite{nu2014}.
A comparable result has also been obtained in Ref.~\cite{Forero:2014bxa}. 
The pattern of neutrino masses and mixings is schematically 
shown in Fig.~\ref{fig:pattern}. At the present time, we cannot decide whether 
the $\nu_3$ neutrino mass eigenstate is heavier or lighter than 
the $\nu_1$ and $\nu_2$ neutrino mass eigenstates in nature. 
The scenario, in which the $\nu_3$ 
is heavier, is referred 
to as the normal mass hierarchy (NH). The other scenario, in which the $\nu_3$ is 
lighter, is referred to as the inverted mass hierarchy (IH). 

Superficially, it appears that the difference between the two mass orderings depicted 
in Fig.~\ref{fig:pattern} is striking, and hence it should be relatively easy to decide which one of them
is realized in nature. As we show in detail further this is not the case and, in fact, the 
determination of the true mass hierarchy (MH) is very challenging. However, despite the 
difficulties, it is vitally important to solve this problem, and correspondingly a great 
effort is devoted to its solution.

   Lets first explain the nomenclature. Since the neutrinos $\nu_1$ has the largest 
component of the electron neutrino $\nu_e$ while $\nu_3$ has the smallest 
component of $\nu_e$, the normal hierarchy in some crude sense resembles the 
mass ordering  of the charged leptons, hence it is denoted as `normal' . Obviously, 
the inverted hierarchy represents the opposite situation. Note that the word hierarchy 
is used throughout this work even if the lightest neutrino mass $m_{L}$ ($\nu_1$ for the NH 
and $\nu_3$ in the IH) could be comparable to the other two masses, i.e. when all 
three masses are much larger than the $\sqrt{\Delta m^2_{ij}}$ (so called degenerate 
scenario).

   Why is it important to determine the MH? The ultimate goal of all experiments and 
theories centered on neutrinos is to formulate the particle physics model that explains 
the observed neutrino masses and mixing patterns, and relates them to the well known 
charged lepton masses (and possibly to the quark masses and mixings). Clearly, 
explaining MH is a crucial part of that, and its determination would 
strengthen, or eliminate, roughly half of the proposed models. 

   Neutrino masses can be also constrained or ultimately determined by cosmology 
and astrophysics, where the sum of the masses, $m_{cosmo}=\Sigma m_i$ is the relevant 
quantity. So far upper limits on that sum, more and more restrictive, were reported~\cite{Agashe:2014kda}. 
Depending on the MH, a lower limit of $m_{cosmo}$ is 
$\sim 0.05$ eV for NH and $\sim 0.1$ eV for IH. Thus, knowing the MH changes 
the lower limit by a factor of two, and correspondingly restrict the range of the sum.

   Most theoretical models of neutrino mass assume that neutrinos are massive 
Majorana fermions. The best way to test such a hypothesis is to search for the 
neutrinoless double beta decay $0\nu\beta\beta$; its rate is proportional to 
the square of the effective neutrino mass $m_{ee} := ||U_{e1}|^2m_1 + |U_{e2}|^2m_2e^{2\alpha i} 
+ |U_{e3}|^2m_3e^{2\beta i}|$. That quantity is 
restricted from below,  $m_{ee} \ge  14$ meV (taking into account 
the 3$\sigma$ error bars of the oscillation parameters) for IH 
while $m_{ee} = 0$ is possible for NH.
Thus, if IH is realized in nature, the next generation of the $0\nu\beta\beta$ 
experiments can decide whether neutrino are Majorana fermions or not.

Similarly to most of the parameters describing neutrino mass and mixing, the neutrino 
MH can be accessed through the neutrino flavor oscillation. 
As shown in Table~\ref{tab:PMNS}, there are two small parameters in the neutrino oscillation 
description; the mixing angle $\theta_{13}$ ($\sin^2 \theta_{13}  \sim 0.022$) and the ratio 
$\Delta m^2_{21}/\Delta m^2_{31}$  ($\sim 3\%$). Due to this feature, most oscillation results 
are reasonably well described in the framework of mixing only two neutrinos, instead of three. 
In this case, the probability of flavor change in the vacuum and the oscillation length are 
given by
\begin{equation}
P(\nu_l \rightarrow \nu_{l'}) = \sin^2 2\theta \cdot \sin^2 \left( 1.27 \cdot \frac{\Delta m^2 ({\rm eV^2}) 
\cdot L ({\rm m})}{E ({\rm MeV})}\right), 
~L_{vacuum} {\rm (m)} = \frac{ 2.48 \cdot E_{\nu} {\rm (MeV)}}{\Delta m^2 {\rm (eV^2)}}
\end{equation}
and, obviously, the sign of $\Delta m^2$ (the mass hierarchy) cannot be 
determined in such case. 

Therefore, in order to determine MH, i.e. to find effects that 
are sensitive to the sign of $\Delta m^2_{31}$ or $\Delta m^2_{32}$, one has to either go beyond 
the vacuum oscillation or go beyond the simple framework of two-neutrino mixing.
Correspondingly, there are two direct ways to determine MH.  In the first 
method, instead of the neutrino propagation in vacuum, one uses the propagation in matter 
where an additional phase from the interaction between neutrinos and the matter constituents 
makes the hierarchy determination possible. This effect is commonly referred to as the 
the Mikheyev-Smirnov-Wolfenstein (MSW) effect~\cite{Wolfenstein:1978ue,Mikheev:1986wj,Mikheev:1986gs} 
or the matter effect. In the second method, one explores the aforementioned small 
difference ($\Delta m^2_{21}$) between $\Delta m^2_{31}$ and $\Delta m^2_{32}$. In either case, 
carefully designed experiments are required. In the following, without going into details, 
we show the basic principles of both methods. 

The oscillation phenomenon has its origin in the phase difference between the coherent 
components of the  neutrino flavor eigenstates caused by the finite masses $m_i$  
of the mass eigenstates $\nu_i$ ($i=$ 1,2,3). When neutrinos propagate in matter, an additional contribution 
to the phase appears. That phase affects the electron neutrinos and antineutrinos only, since only electron 
neutrinos/antineutrino interact with electrons in matter through the charged weak current 
($W^{\pm}$-boson exchange). All three neutrino flavors interact with matter constituents through 
neutral current (Z-boson exchange) with same strength. Thus, these phases do not have effects on the 
neutrino oscillation. The corresponding effective potential and the matter oscillation length are
\begin{equation}\label{eq:matter}
V_C = \sqrt{2}G_FN_e ,  ~~ L_{matter} = \frac{2 \pi}{ \sqrt{2}G_FN_e} = \frac{1.7 \times 10^7 {\rm (m)}}{\rho {\rm (g/cm^3)} Y_e} ~,
\end{equation}
where $G_F$ is the Fermi constant, $N_e$ is the electron density in matter, 
$\rho$ is the matter density, and $Y_e \sim 0.5$ is the electron fraction.

For the illustrative case of two flavors of neutrinos propagating in matter,
the equation of motion for neutrinos in the flavor basis then has the form
\begin{equation}
i \frac{d}{dx} \left( \begin{array}{c}  \nu_e \\ \nu_{\alpha} \end{array} \right) =
2 \pi \left( \begin{array}{cc} -\kappa \frac{\cos2\theta}{L_{vacuum}} + \frac{1}{L_{matter}} &
\kappa \frac{\sin2\theta}{2 L_{vacuum}}  \\
\kappa \frac{\sin2\theta}{2 L_{vacuum}} & 0 \end{array} \right)
\left( \begin{array}{c}  \nu_e \\ \nu_{\alpha}  \end{array} \right)  ~,
\label{eq:sch_fl}
\end{equation}
where $\kappa = {\rm sign} (m_2^2 - m_1^2)$. For antineutrinos the sign in front of 
$1/L_{matter}$ is reversed.
Here one can see clearly that in matter, unlike in vacuum, the oscillation pattern depends on 
whether $m_2$ is larger or smaller than $m_1$. Note that the sign of $\Delta m^2_{21}$ can 
be compensated by the sign of the $\cos2\theta$, i.e. by the octant of the mixing angle 
$\theta$. The ordinary matter contains electrons, thus what really matters is whether 
$\nu_e$ is dominantly the lighter or the heavier one of the matter eigenstates ($\nu_1$ or $\nu_2$).

Therefore, from the study of solar neutrinos, where the matter effects are crucial, we know 
that $\Delta m^2_{21} > 0$, provided that we define, as it is customary, that $\nu_1$ is the 
mass eigenstate that is the dominant component of the electron neutrinos $\nu_e$. However, 
given the size of the matter oscillation length, it is clear that in order to use the matter 
oscillations in Earth to determine the sign of $\Delta m^2_{32}$ and 
$\Delta m^2_{31}$, very long baselines are needed.

\begin{figure}[ht]
\begin{centering}
\includegraphics[width=1.0\textwidth]{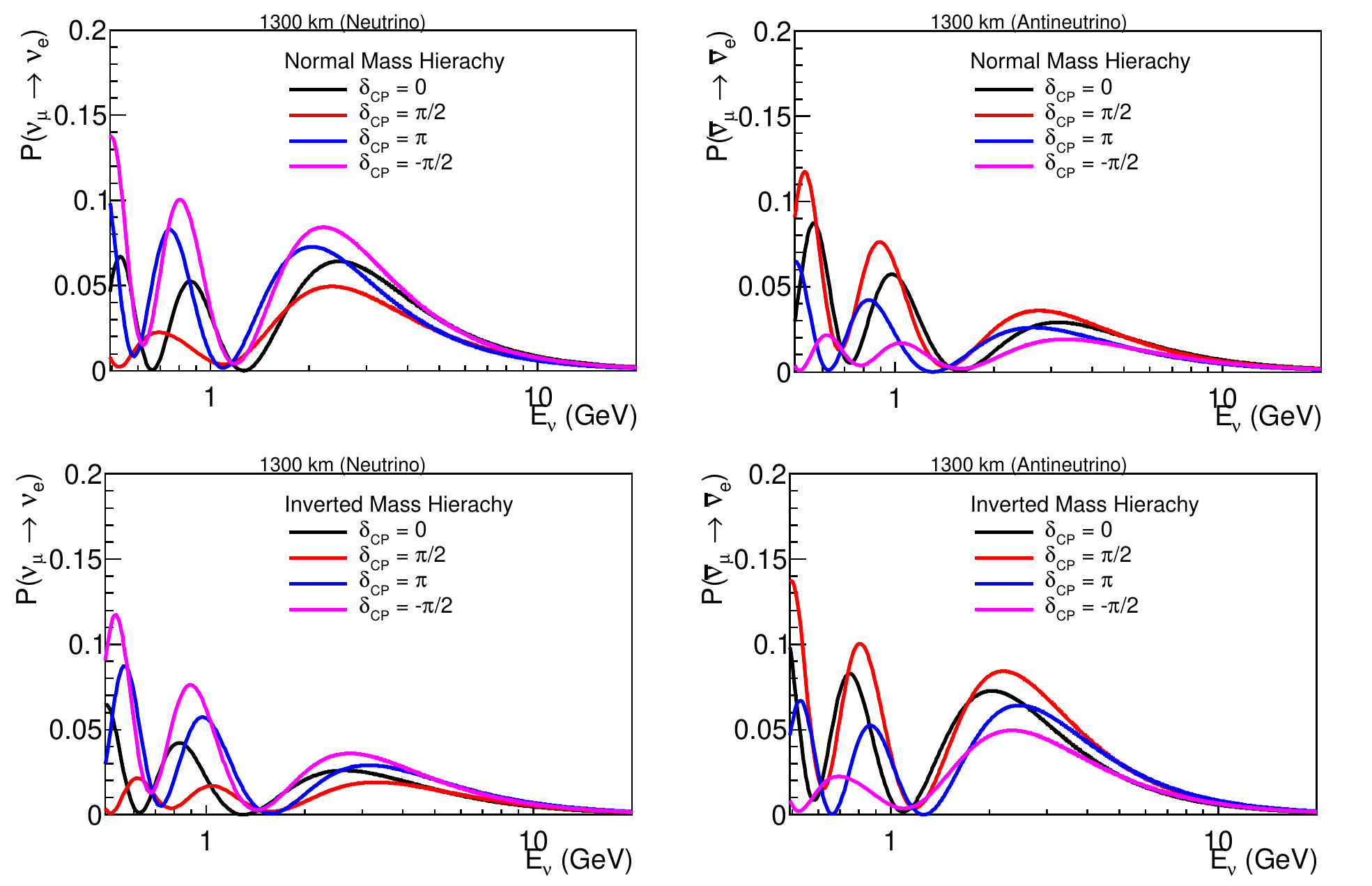}
\par\end{centering}
\caption{\label{fig:LBNE_osc} 
Electron neutrino and antineutrino appearance oscillation probabilities as a function of the neutrino 
energy $E_{\nu}$ at $L=1300$ km for the indicated four values of the CP phase $\delta_{CP}$.
The matter density $\rho$ and the electron fraction $Y_e$ are assumed to be 
2.7 g/cm$^3$ and 0.5, respectively. 
$\sin^2\theta_{13}$, $\sin^2\theta_{12}$, $\sin^2\theta_{23}$, $\Delta m^2_{31}$, and 
$\Delta m^2_{21}$ are assumed to be 0.0219, 0.304, 0.5, 2.4$\times10^{-3}$ eV$^{2}$, and 
7.65$\times10^{-5}$ eV$^{2}$, respectively.
Top (bottom) two panels correspond to the normal 
(inverted) mass hierarchy. Left (right) panels correspond to the neutrino 
(antineutrino) oscillation. In the normal (inverted) mass hierarchy, the neutrino (antineutrino) 
appearance are enhanced, and the antineutrino (neutrino) appearance are suppressed. }
\end{figure}

Matter effect in earth may be used for the determination of MH in two ways.
The first method utilizes the appearance of new neutrino flavors with accelerator neutrino 
sources, in practice the appearance of the electron neutrinos $\nu_{e}$ in a beam that begins 
as the muon neutrinos $\nu_{\mu}$. The probability that a beam of $\nu_{\mu}$ 
propagating in the matter of a constant electron density $N_e$ will be transformed into $\nu_e$ 
as a function of the neutrino energy $E_{\nu}$ and the neutrino travel distance $L$ is
\begin{equation}
P(\nu_\mu \rightarrow \nu_e) \approx P_0 + P_{\sin\delta} + P_{\cos \delta} + P_3,
\end{equation}
where 
\begin{eqnarray}
P_0 &=& \sin^2\theta_{23} \frac{\sin^22\theta_{13}}{(A-1)^2} \sin^2[(A-1)\Delta], \\
P_3 &=& \alpha^2\cos^2\theta_{23}\frac{\sin^22\theta_{12}}{A^2}\sin^2(A\Delta), \\
P_{\sin\delta} &=& -\alpha \frac{\sin2\theta_{12}\sin2\theta_{13}\sin2\theta_{23}\cos\theta_{13}\sin\delta_{CP}}{A(1-A)}\sin\Delta \sin(A\Delta) \sin[(1-A)\Delta], \\
P_{\cos\delta} &=& \alpha \frac{\sin2\theta_{12}\sin2\theta_{13}\sin2\theta_{23}\cos\theta_{13}\cos\delta_{CP}}{A(1-A)}\cos\Delta\sin(A\Delta)\sin[(1-A)\Delta],
\end{eqnarray}
and where 
\begin{equation}
\Delta = \frac{\Delta m^2_{31}L}{4E_{\nu}},~ \alpha = \frac{\Delta m^2_{21}}{\Delta m^2_{31}}, ~ A=\sqrt{2}G_{F}N_{e}\frac{2E_{\nu}}{\Delta m^2_{31}}. \nonumber
\end{equation}
When the neutrinos are replaced by antineutrinos the signs of the CP 
phase $\delta_{CP}$ and of the matter effect signifying the parameter $A$ are reversed. 
Given known values of mixing angles, when measurements at optimally selected $L/E_{\nu}$ 
with both neutrinos and antineutrinos are performed, the matter effect and 
the $CP$ violation can be disentangled and the sign of MH (i.e. 
the sign of $\Delta$) can be determined. Fig.~\ref{fig:LBNE_osc} shows the 
oscillation probabilities vs. $E_{\nu}$ at 1300 km baseline with distinguishing features of 
MH. A constant matter density is assumed. At 1300 km, 
the difference in oscillation probabilities between a constant matter density and realistic 
earth density profile is negligible.

In addition, MH could be determined using the atmospheric neutrinos 
exposed to the matter effect by their propagation through Earth. Resonant oscillation~\cite{Chizhov:1999az,Akhmedov:1999va,Chizhov:2000tn} occurs 
either for neutrinos in the case of NH, or antineutrinos for IH. A detector that could resolve $\nu$ from $\bar{\nu}$ needs only to demonstrate 
for which state the resonance occurs. Detectors that can only determine neutrino flavor 
can take advantage of
the difference in the atmospheric neutrino flux between $\nu$ and $\bar{\nu}$ and the 
differences in interaction cross sections at the detection to determine the hierarchy. 
Details of the needed strategy and 
the corresponding challenges with these two approaches will be described in the next section.

Beside the matter effect, MH can be determined by exploring the small 
difference ($\Delta m^2_{21}$) between $\Delta m^2_{31}$ and $\Delta m^2_{32}$ in the three-flavor
neutrino framework with neutrino and antineutrino disappearance. 
The survival probability for a neutrino of flavor $l$ in vacuum is given by
\begin{equation}
P_{ll} = 1 - \left[ A^l_{21} \sin^2 \Delta_{21} + A^l_{31} \sin^2 \Delta_{31} + A^l_{32} 
\sin^2 \Delta_{32} \right] ~,
\end{equation}
where $A^l_{ij} = 4 |U^{li}|^2 |U^{lj}|^2$ and $\Delta_{ij} = \Delta m^2_{ij} L/4E_{\nu}$. 
This general formula, however,  does not show explicitly the difference in the survival 
probability for the two hierarchies. In the case
of practical importance, namely for the electron antineutrinos, the formula 
can be transformed into
\begin{equation}\label{eq:mphi}
P_{ee} = 1 - 2 s^2_{13}c^2_{13} -4 c^4{13}s^2_{12}c^2_{12}\sin^2 \Delta_{21} 
+ 2s^2_{13}c^2_{13} \sqrt{1 - 4s^2_{12}c^2_{12} \sin^2 \Delta_{21}} \cos(2\Delta_{32} 
\pm \phi_{ee}) ~,
\end{equation}
where the angle $\phi$ that characterizes the difference of the two hierarchies is
\begin{equation}
\sin \phi_{ee} = \frac{c^2_{12} \sin 2 \Delta_{21}}{\sqrt{1 - 4 s^2_{12} c^2_{12} 
\sin^2 \Delta_{21}}}
~,~ \cos \phi_{ee} = \frac{ c^2_{12} \cos 2 \Delta_{21} + s^2_{12}}{\sqrt{1 - 4 s^2_{12} 
c^2_{12} \sin^2 \Delta_{21}}}.
\end{equation}
Here the $\pm$ sign in the last term in Eq.~\eqref{eq:mphi}
contains the information regarding the mass hierarchy. 
The plus (minus) sign corresponds to the normal (inverted) mass hierarchy. 
The analogous formula for the survival probability of the muon neutrinos is
\begin{eqnarray}
P_{\mu\mu} &=& 1 - 2 |U_{\mu 3}|^2 \cdot (|U_{\mu 1}|^2 + |U_{\mu 2}|^2) - 4 |U_{\mu 1}|^2 
|U_{\mu 2}|^2 \sin^2 \Delta_{21} \nonumber \\
&-& 4 |U_{\mu 3}|^2 \cdot \left(\sqrt{|U_{\mu1}|^4+|U_{\mu2}|^4+2|U_{\mu1}|^2|U_{\mu2}|^2
\cos2\Delta_{21}}\right) 
\cdot \cos (2\Delta_{32} \pm \phi_{\mu\mu}) ~,
\end{eqnarray}
where $U_{\mu i}$ are the elements of the second row of the PMNS matrix and now the angle $\phi$ is
\begin{equation}
\sin \phi_{\mu\mu} = \frac{|U_{\mu 1}|^2 \sin 2 \Delta_{21}}{\sqrt{|U_{\mu1}|^4+|U_{\mu2}|^4+2|U_{\mu1}|^2|U_{\mu2}|^2
\cos2\Delta_{21}}}
~,~ \cos \phi_{\mu\mu} = \frac {|U_{\mu 1}|^2 \cos 2 \Delta_{21} + |U_{\mu 2}|^2} {\sqrt{|U_{\mu1}|^4+|U_{\mu2}|^4+2|U_{\mu1}|^2|U_{\mu2}|^2
\cos2\Delta_{21}}}
\end{equation} 

By an accurate measurement of the survival probability and its $L/E_{\nu}$ dependence, 
combining with the accurate knowledge of at least some of the parameters involved, 
MH (the sign in front of $\phi$) can be determined. 
Detailed description of the issues involved will be given again in the next section.

Furthermore, there is a possibility of determining MH based on 
the observation of core-collapse supernovae. There are at best only a handful of the 
galactic core-collapse supernovae per century 
and their neutrino emission and detection is a complicated matter. On the other hand, 
neutrino sky coverage by the sufficiently large detectors has been on for about three 
decades already, and will continue in the foreseeable future. There is, therefore, a good 
chance that a high statistics SN neutrino sample will eventually become a reality and 
that the neutrino mass hierarchy can be deduced from its analysis. It is possible, 
perhaps even likely, that the SN neutrino spectra are not universal; they might depend 
on the progenitor mass, accretion or rotation. Nevertheless, some general features exist.
Only recently it has been recognized that in the deep core region where the neutrino 
densities are very  large the neutrino self-interaction could result in a hierarchy 
dependent collective flavor conversion. (For a review see Ref.~\cite{Duan:2010bg}).  
Generally, this flavor conversion is not expected to affect the $\nu$ and $\bar{\nu}$ 
spectra in the case of NH, but will alter significantly and differently 
the $\nu$ and $\bar{\nu}$ spectra in the case of IH.  In addition, the neutrino 
spectra will be affected by the resonant matter effect in the outer lower mass density 
layers of the supernova. Resonance occurs for neutrinos in the case of NH 
and antineutrinos in the case of IH. Also, the detected signal could 
be possibly altered by the matter effect during the transitions through earth (for general 
references see the review in Ref.\cite{Raffelt:2010zza}). Given the fast development of the supernova 
explosion and neutrino production and propagation theory, we will not describe the possibilities 
of MH determination based on the future SN neutrino 
observation in further detail.

\begin{figure}[H]
\begin{centering}
\includegraphics[width=0.8\textwidth]{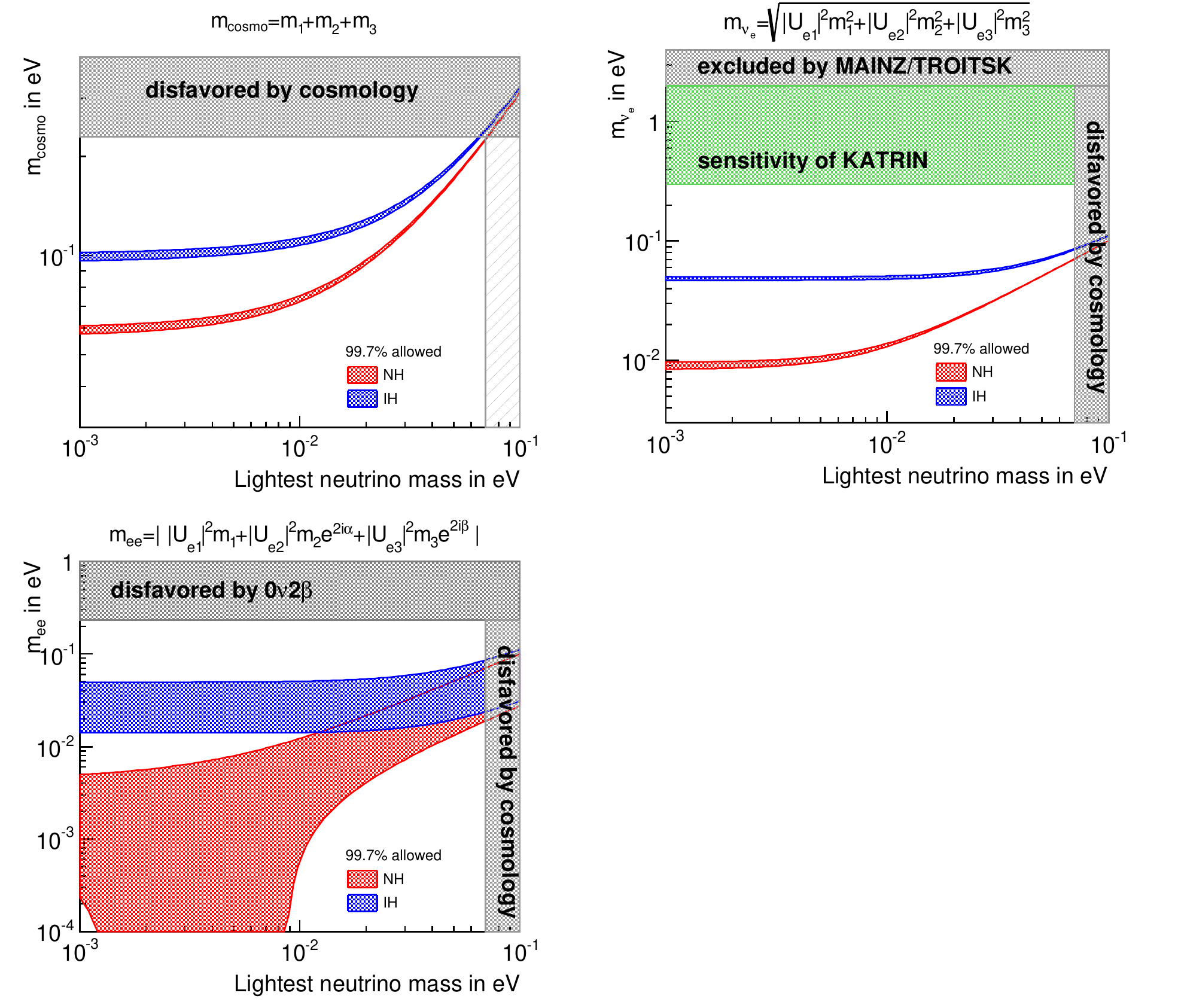}
\par\end{centering}
\caption{\label{fig:mass} 
The top left, top right, and bottom left panels show the $m_{cosmos}$ (the summation of 
neutrino masses), $m_{\nu_e}$ (the effective neutrino mass probed in the beta decay experiments),
and $m_{ee}$ (the effective neutrino mass probed in experiments searching for the neutrinoless double
beta decay) vs. the lightest neutrino mass $m_{L}$ for both the normal and inverted mass hierarchy 
scenarios. The bands cover the allowed region given 3$\sigma$ neutrino mixing range in 
Table \protect\ref{tab:PMNS} and all possible values of Majorana phases.}
\end{figure}

Finally, some limited information on MH can come from 
cosmology, the study of the beta decay endpoint, and the study of the 
neutrinoless $\beta\beta$ decay. In cosmology, the sum of the neutrino masses 
$m_{cosmos} = \Sigma_i m_i$ (top left panel of Fig.~\ref{fig:mass})
\begin{eqnarray}
m_{cosmos} = \left\{ \begin{array}{lr}
m_{L} + \sqrt{m^2_L + \Delta m^2_{21}} + \sqrt{m^2_L + |\Delta m^2_{31}|} & ({\rm Normal~Hierarchy}) \\
m_{L} + \sqrt{m^2_L + |\Delta m^2_{32}|} + \sqrt{m^2_L + |\Delta m^2_{31}|}& ({\rm Inverted~Hierarchy})
\end{array}\right.
\end{eqnarray}
 could be determined. If that sum is below the minimum mass of $\sim 0.1$ eV corresponding 
to the inverted hierarchy, the existence of the normal hierarchy is indicated. 
Similarly, in the study of the beta decay spectrum near the endpoint, 
if the effective neutrino mass $m_{\nu_e} = \sqrt{|U_{e1}|^2m_1^2+|U_{e2}|^2m_2^2+|U_{e3}|^2m_3^2}$ 
(top right panel of Fig.~\ref{fig:mass})
\begin{eqnarray}
m_{\nu_e} = \left\{ \begin{array}{lr}
\sqrt{m_L^2 + c_{13}^2s_{12}^2\Delta m_{21}^2+s_{13}^2|\Delta m_{31}^2|} & ({\rm Normal~Hierarchy}) \\
\sqrt{m_L^2 + c_{13}^2c_{12}^2|\Delta m_{31}^2|+c_{13}^2s_{12}^2|\Delta m_{32}^2|} & ({\rm Inverted~Hierarchy})
\end{array}\right.
\end{eqnarray}
 is measured to be smaller than $\sim0.05$ eV
corresponding to the minimum value of the inverted hierarchy,  the existence of the normal hierarchy is 
suggested. In analogy, if the neutrinoless $\beta\beta$ decay is reliably observed 
and the deduced neutrino Majorana mass $m_{ee} = ||U_{e1}|^2m_1 + |U_{e2}|^2m_2e^{2\alpha i} 
+ |U_{e3}|^2m_3e^{2\beta i}|$ 
\begin{eqnarray}
m_{ee}^2= \left\{ \begin{array}{lr}
(c_{12}^2c_{13}^2 m_L + c_{13}^2s_{12}^2\sqrt{m_L^2+\Delta m^2_{21}}\cos{2\alpha} + s_{13}^2\sqrt{m_L^2 + |\Delta m^2_{31}|}\cos{2\beta})^2 +  & \\
(c_{13}^2s_{12}^2\sqrt{m_L^2+\Delta m^2_{21}}\sin{2\alpha} + s_{13}^2\sqrt{m_L^2 + |\Delta m^2_{31}|}\sin{2\beta})^2 & ({\rm Normal~Hierarchy}) \\
(c_{12}^2c_{13}^2 \sqrt{m_L^2 + |\Delta m^2_{31}|} + c_{13}^2s_{12}^2\sqrt{m_L^2+|\Delta m^2_{31}|}\cos{2\alpha} + s_{13}^2m_L\cos{2\beta})^2 +& \\
(c_{13}^2s_{12}^2\sqrt{m_L^2+|\Delta m^2_{31}|}\sin{2\alpha} + s_{13}^2m_L\sin{2\beta})^2 & ({\rm Inverted~Hierarchy})
\end{array}\right.
\end{eqnarray}
with $\alpha$ and $\beta$ being the Majorana phases 
(bottom left panel of Fig.~\ref{fig:mass}) is again below its minimal value corresponding 
to the inverted hierarchy, the normal hierarchy is implied. Note, however, that the proportionality 
between the double beta decay rate and $m^2_{ee}$ is only one, albeit the simplest, of the 
possible ways that the decay could occur. 

The rest of this review is organized as follows. In the next section, the various 
experimental methods with sensitivity to the neutrino mass hierarchy are 
described in more detail.  In Sec.~\ref{sec:experiments}, the proposed experiments 
are described and discussed. Finally, in the last section the entire subject is briefly 
summarized.

%% file: methods.tex
\section{Various Methods to Determine the Mass Hierarchy}\label{sec:methods}

Two known ways to determine the neutrino mass hierarchy (MH) were introduced 
in the previous section. The first one depends on the charge-current forward coherent scattering 
between $\nu_e$ (or $\bar{\nu}_{e}$) and the electrons in Earth (matter effect). The second one 
explores the small difference between $\Delta m^2_{31}$ and $\Delta m^2_{32}$. 
Both methods are now of practical significance due to the recently discovered non-zero,
and relatively large, third neutrino mixing angle $\theta_{13}$. In this section,
we describe the practical approaches to resolve MH that use available 
neutrino sources, namely i) the accelerator neutrinos, ii) the atmospheric neutrinos, 
and iii) the reactor antineutrinos. We also review the statistical interpretation 
of the mass hierarchy sensitivity, which is crucial for understanding the physics 
reach of various experiments to be discussed in the next section. 

\input{theta13.tex}

\input{accelerator.tex}

\input{atm.tex}
\input{reactor.tex}

\input{stat.tex}

%% file: theta13.tex
\subsection{The ``Large'' Third Neutrino Mixing Angle $\theta_{13}$}

In this section we briefly review the experimental discovery of the non-zero value of
$\theta_{13}\approx 8.4^{\circ}$, which opens doors to MH determination. 

Historically, the first attempt to determine the value of $\theta_{13}$ 
was done by the CHOOZ~\cite{Chooz1,Chooz2} and Palo Verde~\cite{PaloVerde} experiments 
in late 1990s and early 2000s. Both experiments were reactor antineutrino experiments 
searching for oscillations of $\bar{\nu}_e$ at baselines of $\sim$1 km with a 
single-detector configuration. No oscillations were observed in either experiment 
and an upper limit of $\sin^22\theta_{13}<0.12$ was set at 90\% confidence level 
(C.L.) by CHOOZ. Almost 10 years later, in 2011, several hints suggested a non-zero $\theta_{13}$. 
The first one was based on a tension~\cite{Fogli:2011qn} between the KamLAND 
$\bar{\nu}_{e}$ disappearance measurement (a reactor antineutrino experiment 
at an average baseline of $\sim$180 km) and the solar neutrino measurements 
(e.g. ratio of $\nu_e$ charge-current to $\nu_{e,\mu,\tau}$ neutral-current interactions from SNO). 
Subsequently, accelerator neutrino experiments MINOS~\cite{minos} and 
T2K~\cite{t2k} reported on their searches of $\nu_{\mu}$ to $\nu_e$ appearance 
oscillation that is sensitive to $\theta_{13}$. In particular, T2K disfavored 
the $\theta_{13}=0$ hypothesis at 2.5$\sigma$~\cite{t2k}. In early 2012, 
the Double CHOOZ reactor antineutrino experiment reported that the $\theta_{13}=0$ 
hypothesis was disfavored at 1.6$\sigma$ with one detector~\cite{dc}.

\begin{figure}[H]
\begin{center}
\includegraphics[width=0.5\textwidth]{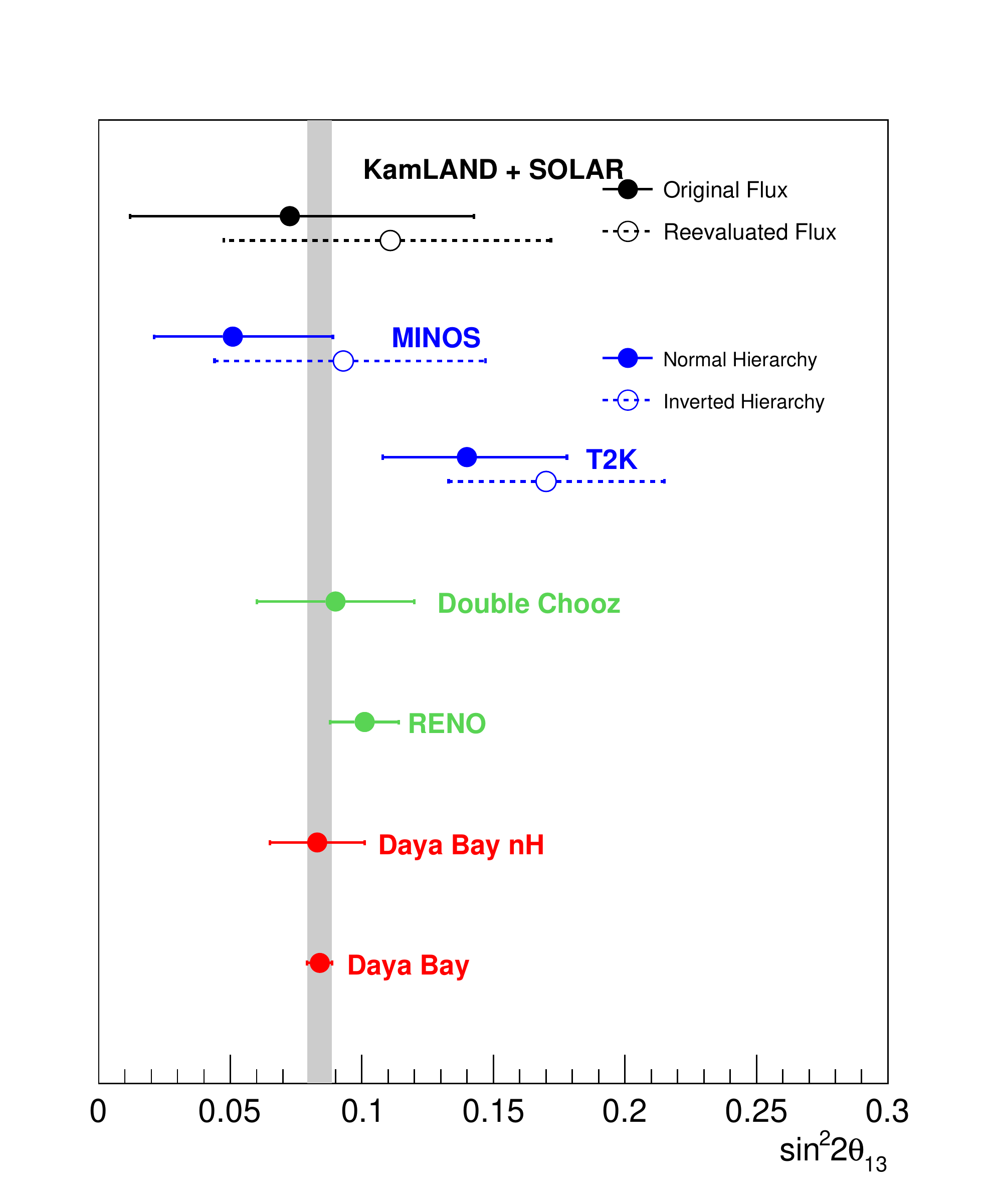}
\includegraphics[width=0.49\textwidth]{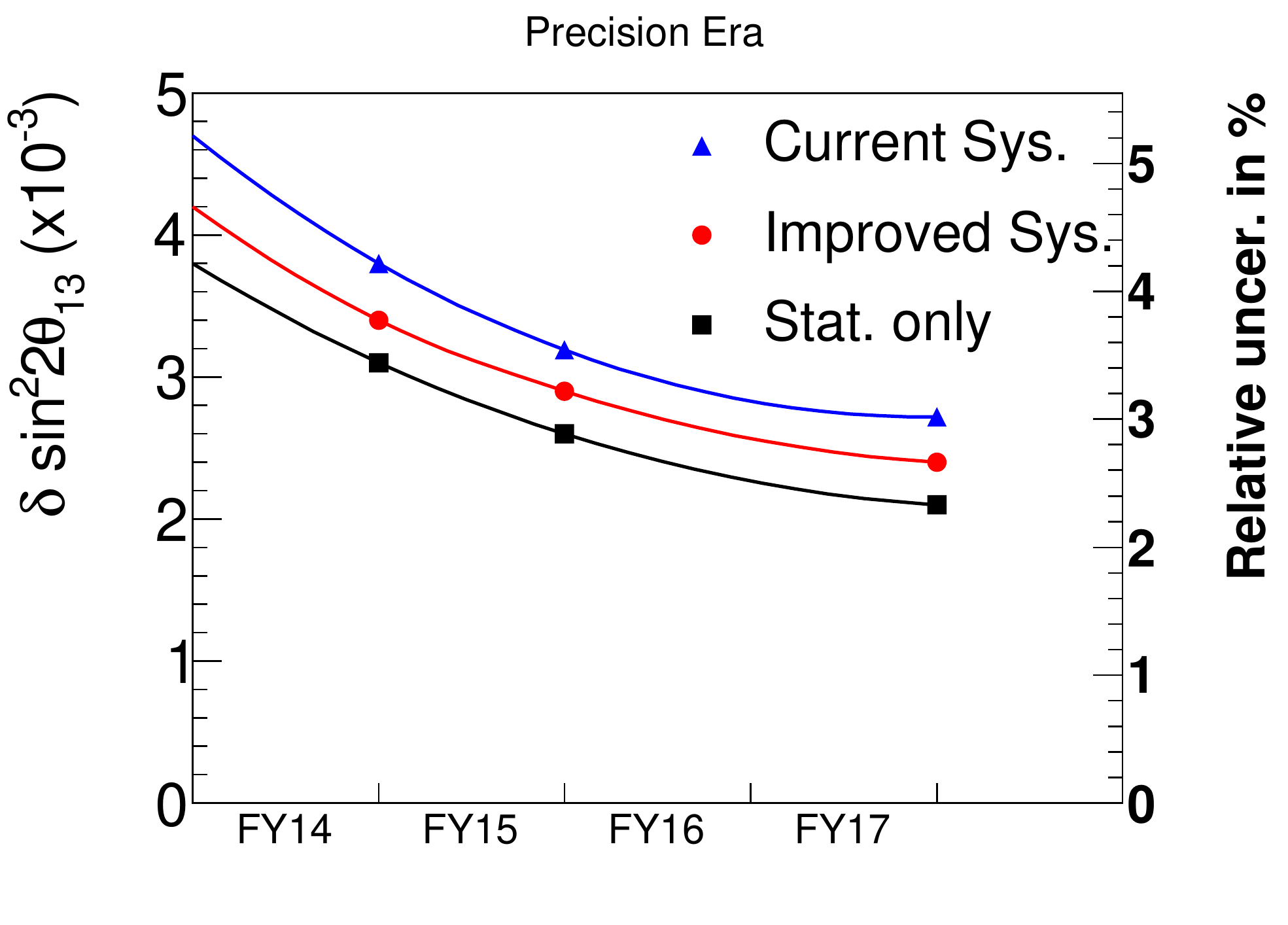}
\end{center}
\caption{\label{fig:theta13_global} (Left) 
The current global status of the $\theta_{13}$ measurements. Results are taken from 
KamLAND+SOLAR~\protect\cite{Fogli:2011qn}, MINOS~\protect\cite{Evans:2013pka}, 
T2K~\protect\cite{Abe:2013hdq}, Double CHOOZ~\protect\cite{Abe:2013sxa}, 
RENO~\protect\cite{2013arXiv1312.4111S}, Daya Bay (nH)~\protect\cite{An:2014ehw}, 
and Daya Bay~\protect\cite{zhang_neutrino2014}. Among them, T2K and MINOS 
have assumed $\delta_{CP}=0$.
(Right) Uncertainty on the Daya Bay 
measurement of $\sin^22\theta_{13}$ in the precision era (FY14-FY17, or 2015-2018) under different 
assumptions from Ref.~\protect\cite{dyb_run}.}
\end{figure}

In March 2012, the Daya Bay reactor antineutrino experiment reported the 
discovery of non-zero $\theta_{13}$ with a $>$5$\sigma$ significance~\cite{dayabay}. 
About one month later, RENO confirmed the Daya Bay discovery with a 
$4.9\sigma$ significance~\cite{reno}. And later in that year, Daya Bay 
increased the significance to 7.7$\sigma$ with a 
larger data set~\cite{An:2013uza}. By then, non-zero $\theta_{13}$ was firmly 
established. The left panel of Fig.~\ref{fig:theta13_global} shows the 
current global status of $\sin^22\theta_{13}$ measurements compiled with 
the latest results from each experiment. The precision of this important 
parameter is still being improved by the current-generation experiments. In particular, 
the right panel of Fig.~\ref{fig:theta13_global} shows the expected uncertainty of 
$\sin^22\theta_{13}$ from the Daya Bay experiment; by the end of 
the experiment it is expected to reach a better than 3\% measurement precision
of $\sin^22\theta_{13}$.

%% file: accelerator.tex
\subsection{Mass Hierarchy from Accelerator Neutrinos Appearance}\label{sec:acc}

The accelerator neutrino experiments provided and continue to provide crucial inputs 
to the development of the 3-flavor PMNS model. Lederman, Schwartz, and Steinberger
already in 1960's discovered the muon flavor neutrinos~\cite{Danby:1962nd} at the Brookhaven 
National Laboratory (BNL) and showed that the lepton flavor appears to be conserved in the weak 
interaction. The tau flavor neutrinos $\nu_{\tau}$, the second to last elementary particle in the 
Standard Model experimentally confirmed, was discovered by the DONUT collaboration 
at the Fermi National Accelerator Laboratory (Fermilab)~\cite{Kodama:2000mp} in 2001. 
The K2K long-baseline experiment~\cite{Ahn:2006zza} confirmed the observation of 
the neutrino oscillation at energies comparable to those produced in the atmosphere.
The OPERA experiment provided the evidence for 
$\nu_{\mu}$ to $\nu_{\tau}$ appearance~\cite{Agafonova:2014bcr}. The MINOS long-baseline 
experiment~\cite{Adamson:2014vgd} provided the most precise measurement of the atmospheric 
mass-squared splitting $\Delta m^2_{32}$. 

Very recently, the T2K experiment~\cite{Abe:2013hdq} discovered 
for the first time ever the neutrino appearance oscillation ($\nu_e$ appearance from $\nu_{\mu}$), 
at more than 5$\sigma$ 
statistical significance, and confirmed 
the ``large'' non-zero $\theta_{13}$ discovered in the Daya Bay experiment~\cite{An:2012eh}. 
The T2K experiment~\cite{Abe:2013hdq} together with the reactor neutrino experiments 
in Refs.~\cite{An:2013zwz,Ahn:2012nd,Abe:2012tg} provide an initial interesting hint of the 
leptonic CP violation. The most precise measurement of the atmospheric mixing angle $\theta_{23}$ 
also comes from the T2K experiment. Besides the possible determination of MH, the accelerator neutrino experiments are currently the only known reliable 
way to measure CP phase $\delta_{CP}$, as well as determining the $\theta_{23}$ octant: whether the 
$\nu_3$ contains more muon neutrino $\nu_{\mu}$ or tau neutrino $\nu_{\tau}$. Furthermore, 
the measurement of the third neutrino mixing angle $\theta_{13}$ from accelerator 
experiments together with the measurement of $\theta_{13}$ from reactor antineutrino 
experiments provides a nice unitarity test of the 
PMNS matrix~\cite{Qian:2013ora}.

\begin{figure}[H]
\begin{center}
\includegraphics[width=0.95\textwidth]{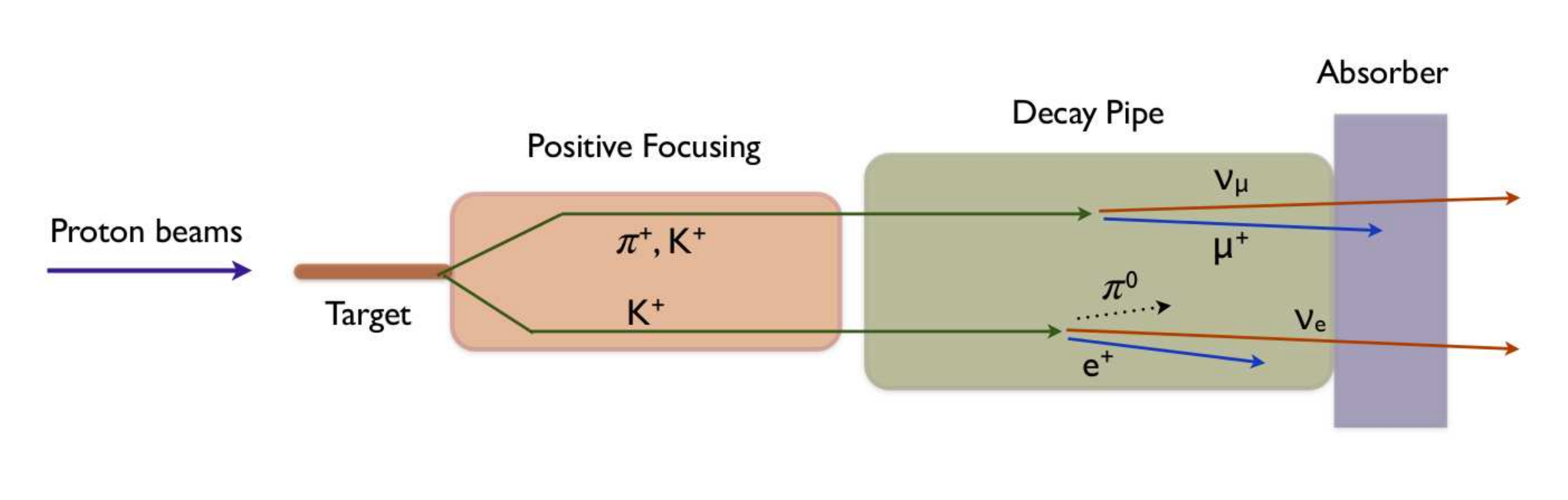}
\end{center}
\caption{\label{fig:accelerator} A schematic view of the conventional neutrino beamline. 
The magnetic field is toroidal and assumed to focus positive charged particle. The muon 
neutrinos and antineutrinos are mainly generated through the two-body decay of pions and kaons. The 
electron neutrinos and antineutrinos, one of the main backgrounds for the electron neutrino/antineutrino 
appearance measurement, are mainly generated through the three-body decay of kaons and 
muons. The dashed line illustrates the direction of very short-lived $\pi^{o}$.}
\end{figure}

Accelerator neutrinos are generated by a high-energy proton beam striking 
a nuclear target to produce pions and kaons, which in turn decay into neutrinos. 
A schematic view of a conventional neutrino beamline is shown in Fig.~\ref{fig:accelerator}. 
First, energetic secondary pions and kaons are produced when high-energy proton beams 
interact with the nuclear target. Second, some of the charged pions and kaons within 
certain momentum range are focused by magnets (also known as horns), so that they 
are approximately traveling in parallel with the incident proton beam direction.
The polarity of the magnet can be selected by changing the current 
in order to focus either the positive or negative charged particles. The charged pions
and kaons then travel through a long decay pipe to provide 
enough time for them to decay. Finally, a thick absorber is placed at the end of 
decay pipe to absorb the muons (decay products of pions and kaons) and other remaining 
charged particles. The "on-axis" neutrino beam refers to the case where the neutrino 
beam direction to detector is in parallel with the incident
proton beam direction. In comparison, the "off-axis" neutrino beam refers to the case 
when there is a angle difference between these two directions.  The ``off-axis" beams 
pioneered by the E889 proposal at BNL~\cite{Beavis-BNL-52459} result in
a narrower neutrino energy distribution.

\begin{figure}
\begin{center}
\includegraphics[width=1.0\textwidth]{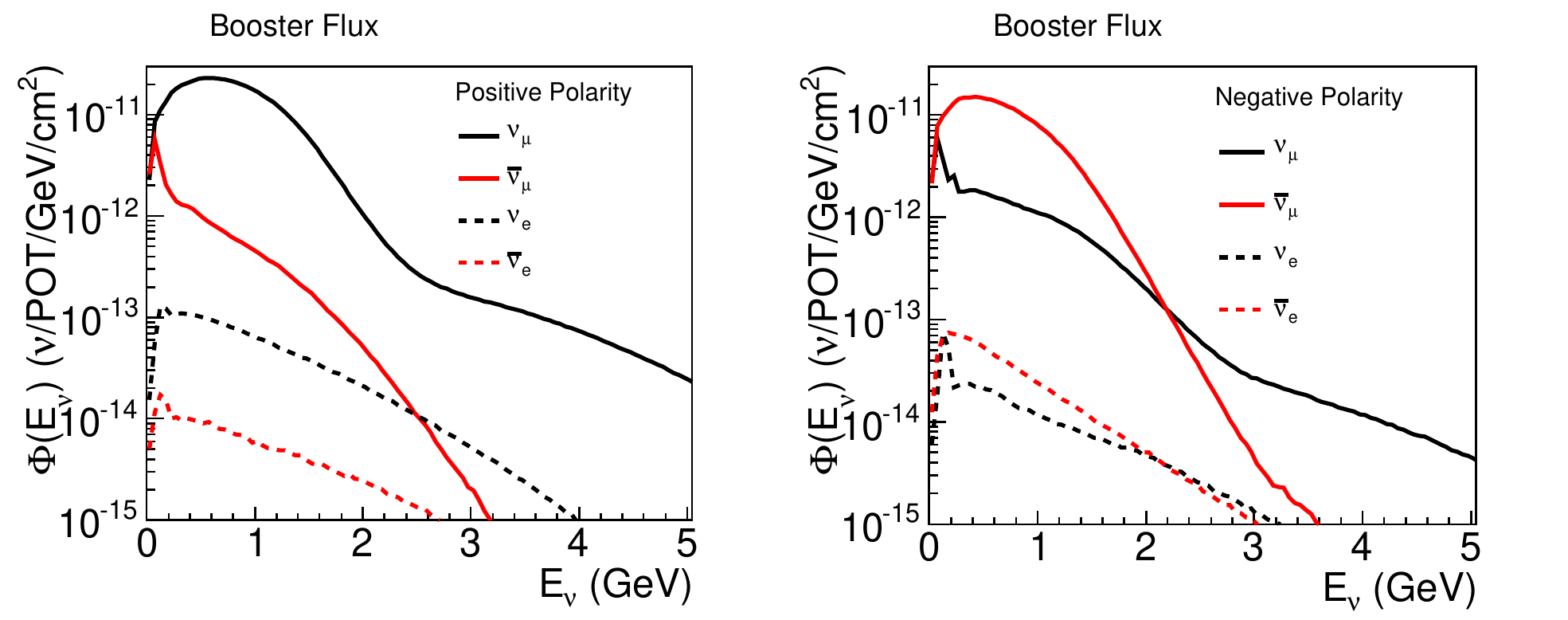}
\end{center}
\caption{\label{fig:booster} Neutrino fluxes for different species are shown 
for the Fermilab Booster neutrino beamline~\cite{AguilarArevalo:2008yp}. 
Left (right) panel represents the positive (negative) polarity which focuses 
positive (negative) charged particles in the magnetic horn. Besides the 
intrinsic $\nu_e$ and $\bar{\nu}_e$ fluxes, there are also sizable wrong-sign neutrinos 
($\bar{\nu}_{\mu}$ in the positive polarity mode and $\nu_{\mu}$ in the negative 
polarity mode), which are produced by the wrong-sign charged pions and kaons. 
Reproduced with the permission from Ref.~\cite{AguilarArevalo:2008yp}. }
\end{figure}

The primary neutrino species are $\nu_{\mu}$ (or $\bar{\nu}_{\mu}$) through positive
(negative) charged pion and kaon 2-body decay: 
\begin{eqnarray}
\pi^{+} \rightarrow \mu^{+} + \nu_{\mu} ~~~~& (\pi^{-} \rightarrow \mu^{-} + \bar{\nu}_{\mu}) & ~~~~~{\rm B.R.} \sim 100\% ~\label{eq:piondecay}\\
K^{+} \rightarrow \mu^{+} + \nu_{\mu}~~~~ &(K^{-} \rightarrow \mu^{-} + \bar{\nu}_{\mu})& ~~~~~{\rm B.R.} \sim63\%, 
\end{eqnarray}
where B.R. stands for the branching ratio. Values are taken from Ref.~\cite{Agashe:2014kda}. There are also 
small amounts of irreducible $\nu_e$ and $\bar{\nu}_e$, mainly produced in the 3-body 
muon and kaon decays:
\begin{eqnarray}
\mu^{+} \rightarrow e^+ + \nu_{e} + \bar{\nu}_{\mu} ~~~~~ &\mu^- \rightarrow e^- + \bar{\nu}_e + \nu_{\mu}& ~~~~~ {\rm B.R.} \sim 100\% ~\label{eq:muondecay}\\
K^+ \rightarrow \pi^{o} + e^+ + \nu_{e} ~~~~~ &K^- \rightarrow \pi^{o} + e^- + \bar{\nu}_{e}& ~~~~~ {\rm B.R.} \sim 5\% \\
K^{o}_{L} \rightarrow \pi^{+} + e^{-} + \bar{\nu}_{e} ~~~~~ &K^{o}_{L} \rightarrow \pi^{-} + e^{+} + \nu_{e}& ~~~~~ {\rm B.R.} \sim 41\%. 
\end{eqnarray}
These intrinsic $\nu_{e}$ and $\bar{\nu}_e$ in the beam are one of the main backgrounds for the 
$\nu_{\mu} \rightarrow \nu_{e}$ and $\bar{\nu}_{\mu} \rightarrow \bar{\nu}_{e}$ appearance measurements. 
As an example, 
Fig.~\ref{fig:booster} shows the neutrino fluxes at the Fermilab Booster neutrino beamline,  
where the incident beam proton kinetic energy is about 8 GeV. 

The accelerator-produced neutrinos are generally detected through the charged-current 
interaction with nucleon/nuclei in the detector. In an oscillation 
experiment, there is typically a detector placed near the neutrino production
region which is intended to measure the essentially unoscillated neutrino
flux and a second detector placed far from the source to be sensitive to oscillations.
The commonly used detector technologies for 
far detectors include water Cherenkov, liquid scintillator, magnetized iron plus scintillator, 
and liquid argon time projection chamber. The involved target nucleon/nuclei are 
typically $^{1}$H, $^{12}$C, $^{16}$O, $^{40}$Ar, and $^{56}$Fe. 

\begin{figure}[H]
\centering
\includegraphics[width=0.49\textwidth]{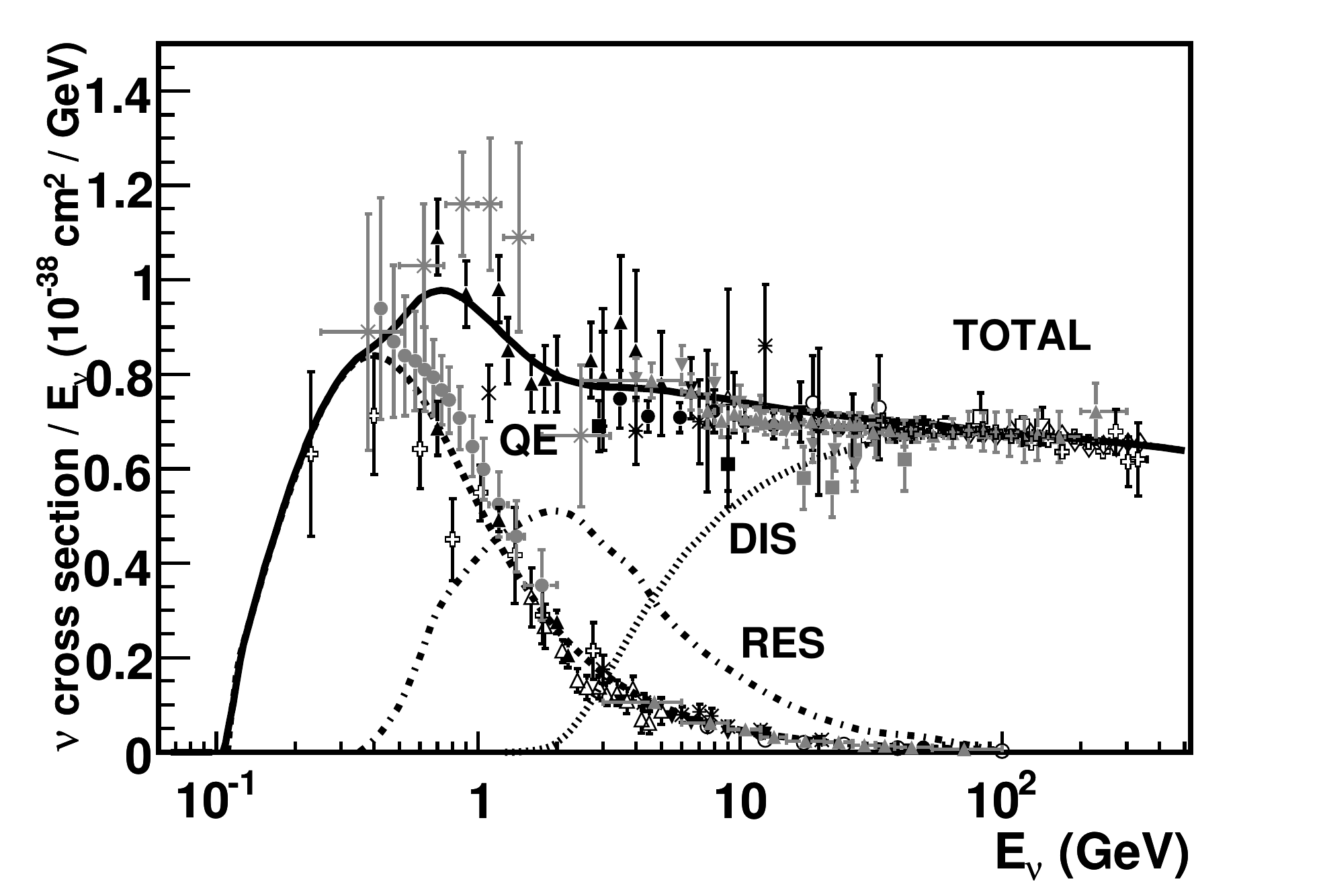}
\includegraphics[width=0.49\textwidth]{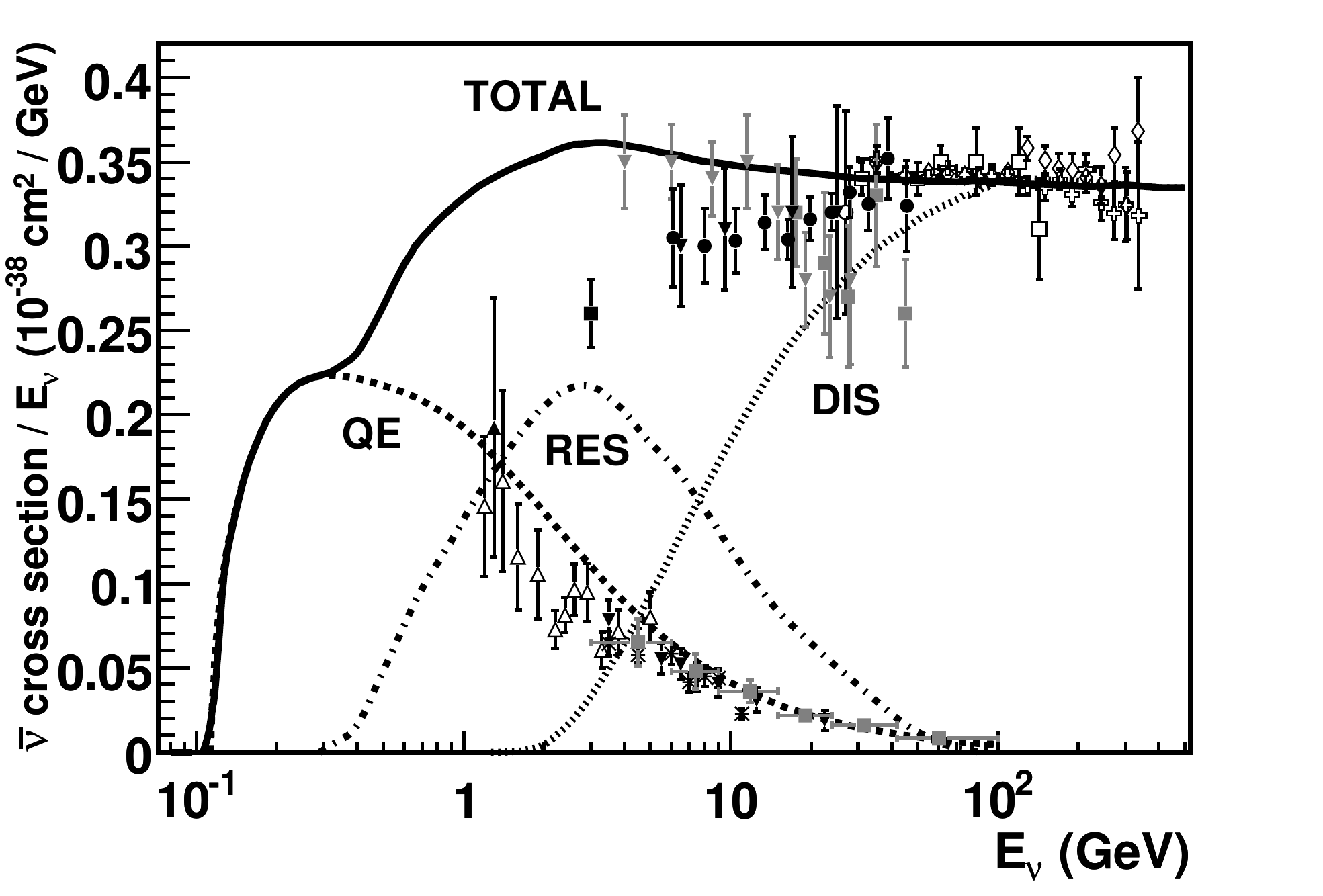}
\caption{\label{fig:xs_section} Summary of neutrino-nucleon interaction 
cross sections is reproduced with the permission from from Ref.~\cite{Formaggio:2013kya}. 
Left (right) panel shows the ratio of neutrino (anti-neutrino) cross section 
with respect to the neutrino energy.  Three main reaction 
mechanisms based on the assumption that the neutrino 
interacts with one of the bound nucleon (``N'') in the target: 
quasi-elastic (QE), resonance (RES), and deep-inelastic 
scattering (DIS) are shown in addition to the total interaction
cross section.}
\end{figure}

Fig.~\ref{fig:xs_section} shows the neutrino-nucleon and antineutrino-nucleon interaction cross sections. 
There are three main components based on the assumption that the neutrino 
interacts with one of the bound nucleons (``N'') in the target.  

In the quasi-elastic scattering (QE)
\begin{equation}
\nu + ``N" \rightarrow l^- + N' ~~~~~~ \bar{\nu} + ``N" \rightarrow l^+ + N', 
\end{equation}
the nucleon is converted into its isospin partner.
In the resonance scattering (RES)
\begin{equation}
\nu + ``N" \rightarrow l^- + N^* ~~~~~~ \bar{\nu} + ``N" \rightarrow l^+ + N^*, 
\end{equation}
the nucleon is excited to a resonance state. 
And in the deep inelastic scattering (DIS)
\begin{equation}
\nu + ``q" \rightarrow l^- + X ~~~~~~ \bar{\nu} + ``q" \rightarrow l^+ + X, 
\end{equation}
the neutrino is assumed to interact with valence or sea quarks ($``q"$) 
in the nucleon.

Besides these reaction mechanisms, neutrinos can also interact with more than 
one of the bound-state nucleons (i.e. meson-exchange currents and short-range correlations).
In addition, neutrinos can 
also interact with the entire nucleus by the elastic scattering
\begin{equation}
\nu + A \rightarrow l^- + A' ~~~~~~ \bar{\nu} + A \rightarrow l^+ + A',
\end{equation}
where the A' is the daughter nucleus or by the coherent meson production
\begin{equation}
\nu + A \rightarrow l^- + A + m^+ ~~~~~~ \bar{\nu} + A \rightarrow l^+ + A + m^-,
\end{equation}
where $m$ represents a meson (such as $\pi$ or $\rho$). The cross sections
of the elastic and the coherent scattering in general are much smaller than the 
QE/RES/DIS cross sections. 
 
For neutrino oscillation measurements, the knowledge of the neutrino energy $E_{\nu}$ 
is crucial, as the neutrino oscillation probability is expressed as a function of $E_{\nu}$ and 
the neutrino traveling distance $L$. Depending on the reaction mechanism, there 
are different strategies to reconstruct the neutrino energy (see e.g. 
Ref.~\cite{Formaggio:2013kya} and the references therein). For example, the 
quasi-elastic scattering is assumed to be a two body to two body scattering. 
For nucleons at rest in the lab frame, the neutrino energy can be reconstructed 
from the charged lepton kinematics (momentum and direction) together with the incident
neutrino momentum direction. For the target nucleons bound in the nucleus,
the neutrino energy reconstruction at QE kinematics $\nu_{l} + ``n'' \rightarrow l + p$ can be written in this crude
but simple and often used approximation as:
\begin{equation}
E^{QE}_{\nu} = \frac{2(M_n - E_B)E_{l}-\left[(M_n-E_B)^2+m_l^2-M_p^2\right]}{2\left[M_n-E_B-E_l+p_l\cos(\theta_l)\right]},
\end{equation}
with $M_n$, $M_p$, and $M_l$ are the mass of neutron, proton, and charged lepton, respectively. 
$E_l$, $p_l$, and $\theta_{l}$ are the energy, momentum, and polar angle of the scattered charged lepton. 
$E_B$ is the separation energy.
The above method has a systematic bias, as the reaction mechanism should include the Fermi motion,  
involvement of more than one nucleon 
in the initial state (such as the short-range correlation), and final state interactions. The correct modeling of the 
reaction mechanism in the neutrino event generator is thus crucial. 
Another example is the deep inelastic scattering. The neutrino energy can be 
reconstructed by the calorimetry method, which sums the charged lepton energy with 
the energy of the hadronic system. In this method, the correct modeling of the 
hadronic energy response in the simulation and the correct estimation of the
missing energy due to neutral particles (neutron and neutrino)  becomes essential. 

\begin{figure}[H]
\centering
\includegraphics[width=1.0\textwidth]{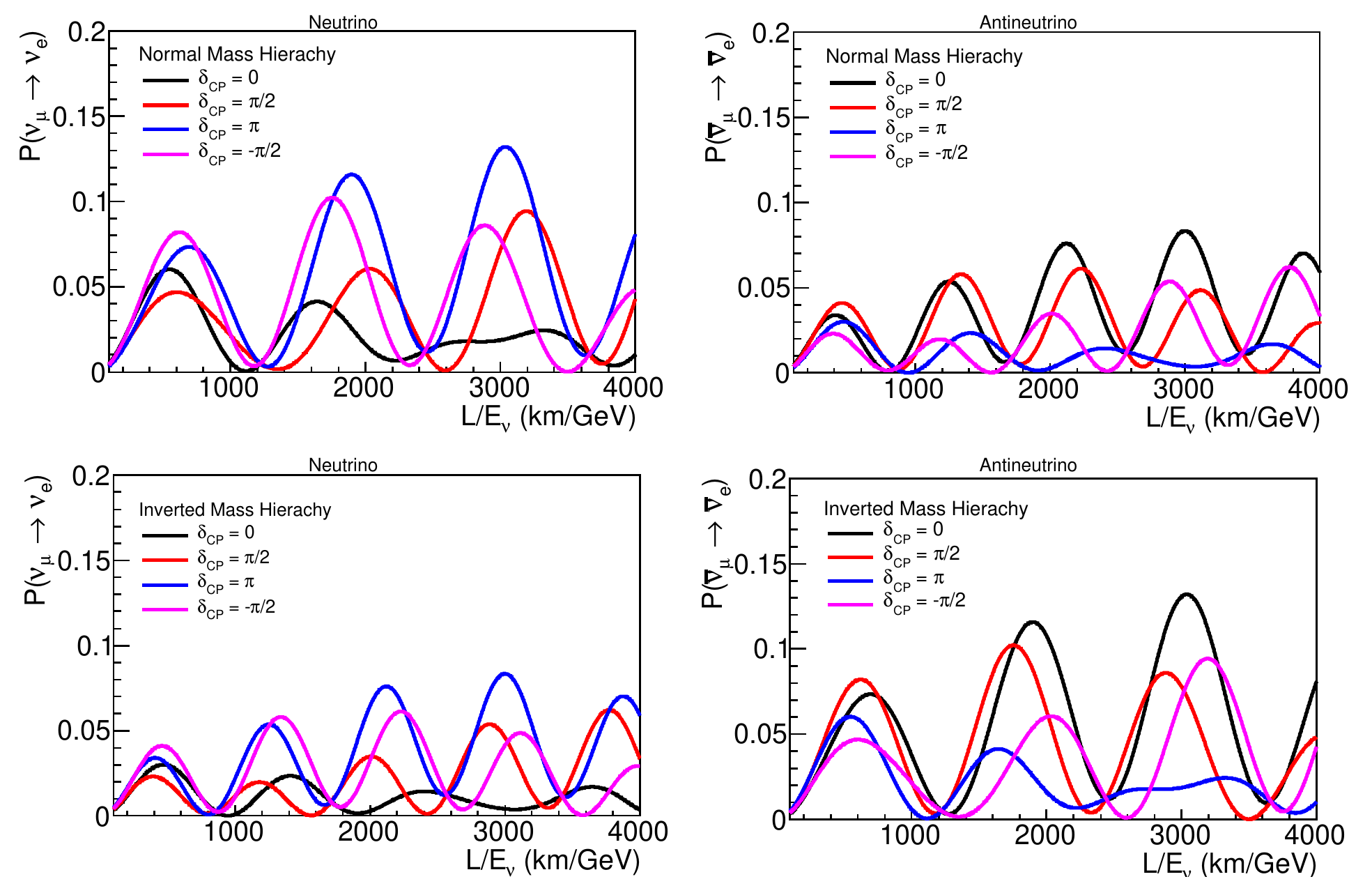}
\caption{\label{fig:acc_MH} The $\nu_{\mu}$ to $\nu_e$ appearance
oscillation probabilities are shown vs. the ratio $L/E_{\nu}$. 
The matter density $\rho$ and the electron fraction $Y_e$ are assumed to be 
2.7 g/cm$^3$ and 0.5, respectively.
$\sin^2\theta_{13}$, $\sin^2\theta_{12}$, $\sin^2\theta_{23}$, $\Delta m^2_{31}$, and 
$\Delta m^2_{21}$ are assumed to be 0.0219, 0.304, 0.5, 2.4$\times10^{-3}$ eV$^{2}$, and 
7.65$\times10^{-5}$ eV$^{2}$, respectively.
Top (bottom) panels correspond to the normal (inverted) mass hierarchy. 
Left (right) panels correspond to the neutrino (antineutrino) oscillation. 
The plot is made by scanning distance $L$ at a fixed neutrino energy $E_{\nu}$= 2 GeV. 
Different curves represent different values of the CP phase $\delta_{CP}$.}
\end{figure}

Fig.~\ref{fig:acc_MH} shows the calculated $\nu_{\mu}$ to $\nu_e$ appearance 
oscillation probability vs. the ratio $L/E_{\nu}$. Due to the matter effect, 
the different MH leads to different behavior for the neutrino 
and antineutrino oscillations. In particular, at the first oscillation maximum 
$L/E_{\nu} \sim 510$ km/GeV, the normal (inverted) MH 
would enhance the neutrino (antineutrino) oscillations, but
suppress the antineutrino (neutrino) oscillations. Therefore, 
by precision measurements of the $\nu_{\mu}$
to $\nu_e$ and the $\bar{\nu}_{\mu}$
to $\bar{\nu}_e$ oscillations, MH can be 
determined. Comparing to Fig.~\ref{fig:LBNE_osc}, which is 
evaluated at fixed baseline 
$L=1300$ km, Fig.~\ref{fig:acc_MH} is made by scanning $L$
at fixed neutrino energy $E_{\nu}=2$ GeV. Due to this choice, the matter 
effect is significant in the second and higher oscillation maxima
in Fig.~\ref{fig:acc_MH}. On the other hand, it is not as important for the second 
and higher oscillation maxima in Fig.~\ref{fig:LBNE_osc} as the neutrino 
energy is much lower.

For neutrino oscillation experiments, the neutrino flavor at detection can 
be tagged by the charged lepton flavor. For the appearance experiments, the
background determination is obviously essential. Besides the intrinsic $\nu_e$ and 
$\bar{\nu}_e$ beam backgrounds, there are two additional backgrounds. 
The first one is the neutral-current $\pi^o$ production
\begin{equation}
\nu + ``N" \rightarrow \nu + \pi^o + X.
\end{equation} 
Due to its very short lifetime ($\sim$8.5$\times$10$^{-17}$s in its at-rest frame), 
$\pi^o$ would immediately
decay after production. About 99\% of $\pi^o$ would decay into two gammas. 
If one of the gammas is missed in the reconstruction (if it has low energy or 
because it is too close to the other gamma), the $\pi^o$ could be misidentified 
as an $e^\pm$ event. The second one is the charged-current $\nu_{\tau}$ 
scattering, when the neutrino energy is high enough to produce a $\tau$ lepton, 
\begin{equation}
\nu_{\tau} + ``N" \rightarrow \tau^- + X ~~~~~~ \bar{\nu}_{\tau} + ``N" \rightarrow \tau^+ + X. 
\end{equation}
About 18\% of $\tau^{\pm}$ would decay through 
\begin{equation}
\tau^- \rightarrow e^- + \bar{\nu}_e + \nu_{\tau} ~~~~~~ \tau^+ \rightarrow e^+ + \nu_e + \bar{\nu}_{\tau},
\end{equation}
in which the e$^\pm$ could mimic the $\nu_e$/$\bar{\nu}_e$ charged-current interaction.

%% file: atm.tex
\subsection{Mass Hierarchy from Atmospheric neutrinos}\label{sec:atm}

In 1998, the Super Kamiokande collaboration reported a zenith angle dependent deficit of 
atmospheric muon neutrinos, inconsistent with the expected atmospheric 
neutrino flux~\cite{Fukuda:1998mi}. This seminal result pioneered the entire field of 
the neutrino flavor oscillations. Now, with the next-generation gigantic detectors, 
atmospheric neutrinos provide a new opportunity to determine MH.

\begin{figure}[H]
\begin{center}
\includegraphics[width=0.51\textwidth]{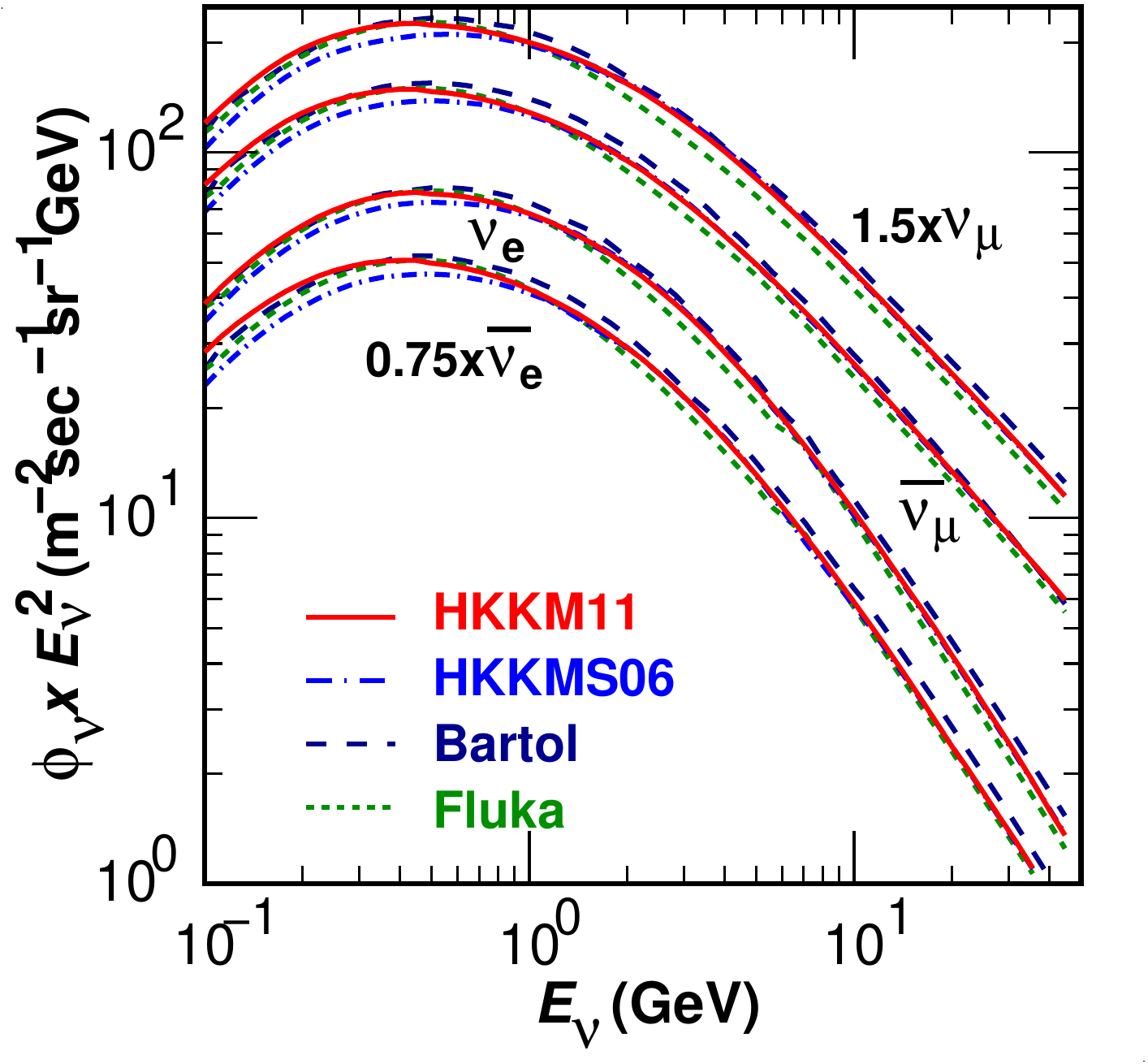}
\includegraphics[width=0.48\textwidth]{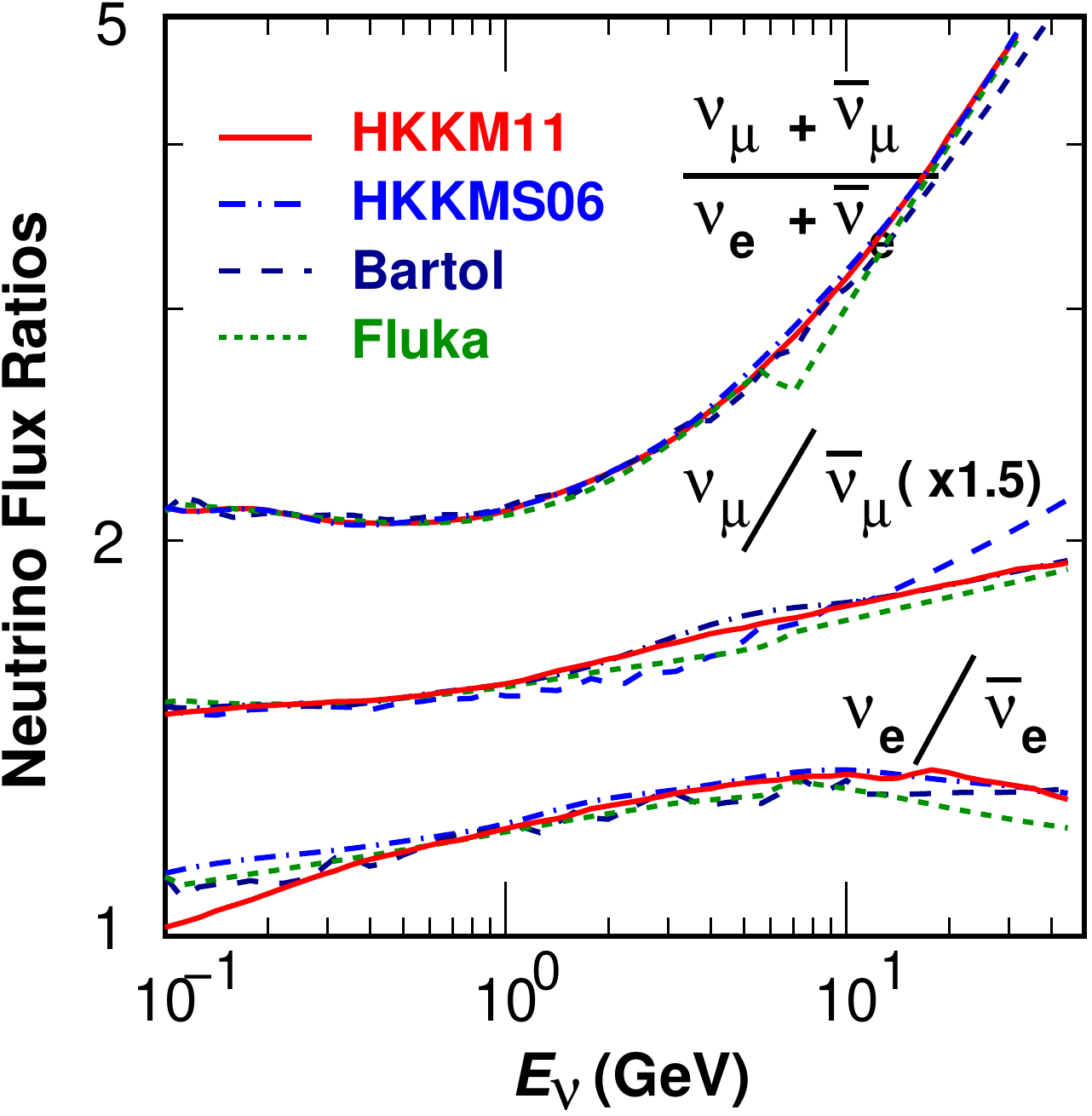}
\end{center}
\caption{\label{fig:atm_flux} Atmospheric neutrino flux prediction 
is reproduced with the permission from Ref.~\protect{\cite{Honda:2011nf}}. 
The left panel shows the atmospheric neutrino fluxes calculated for the 
Kamioka site averaged over all directions. The right panel shows the corresponding flux ratios. 
Four models, HKKM11~\protect\cite{Honda:2011nf}, 
HKKMS06~\protect\cite{Honda:2006qj},
Bartol~\protect\cite{Barr:2004br}, and Fluka~\protect\cite{Battistoni:2002ew} 
are compared. }
\end{figure}

Atmospheric neutrinos are generated by high-energy cosmic rays colliding 
with nuclei in the upper atmosphere. The collision produces a shower of hadrons
dominated by pions with a small amount of kaons. Similarly to the accelerator 
neutrinos, atmospheric neutrinos are generated through the decay of pions, 
muons, and kaons. The positively charged pions decay to a positively charged 
muon and a muon neutrino. The positively charged muons further decay to 
a positron, an electron neutrino, and a muon antineutrino. The decay chain 
of negatively charged pions is analogous. Therefore, one naively 
expects a two-to-one ratio of muon neutrinos to electron neutrinos, 
when the momentum distribution of decaying pions and muons is neglected.  
Fig.~\ref{fig:atm_flux} shows the comparison of the atmospheric neutrino 
flux calculations for the Kamioka site for different neutrino flavors, where above 2 GeV
neutrino energy the ratio of muon neutrino to the electron neutrino begins
to increase. 

In addition, the ratios of neutrinos to antineutrinos are
larger than unity. This is because the positively charged pions, 
containing  the $u$ and $\bar{d}$ quarks, are expected to have a higher production 
yield than the the negatively charged pions  containing  the $d$ and $\bar{u}$
quarks. This, in turn, is caused by the fact that the primary cosmic rays are
typically the high-energy protons, containing two $u$ and one $d$ valence quarks.

\begin{figure}[H]
\begin{center}
\includegraphics[width=1.0\textwidth]{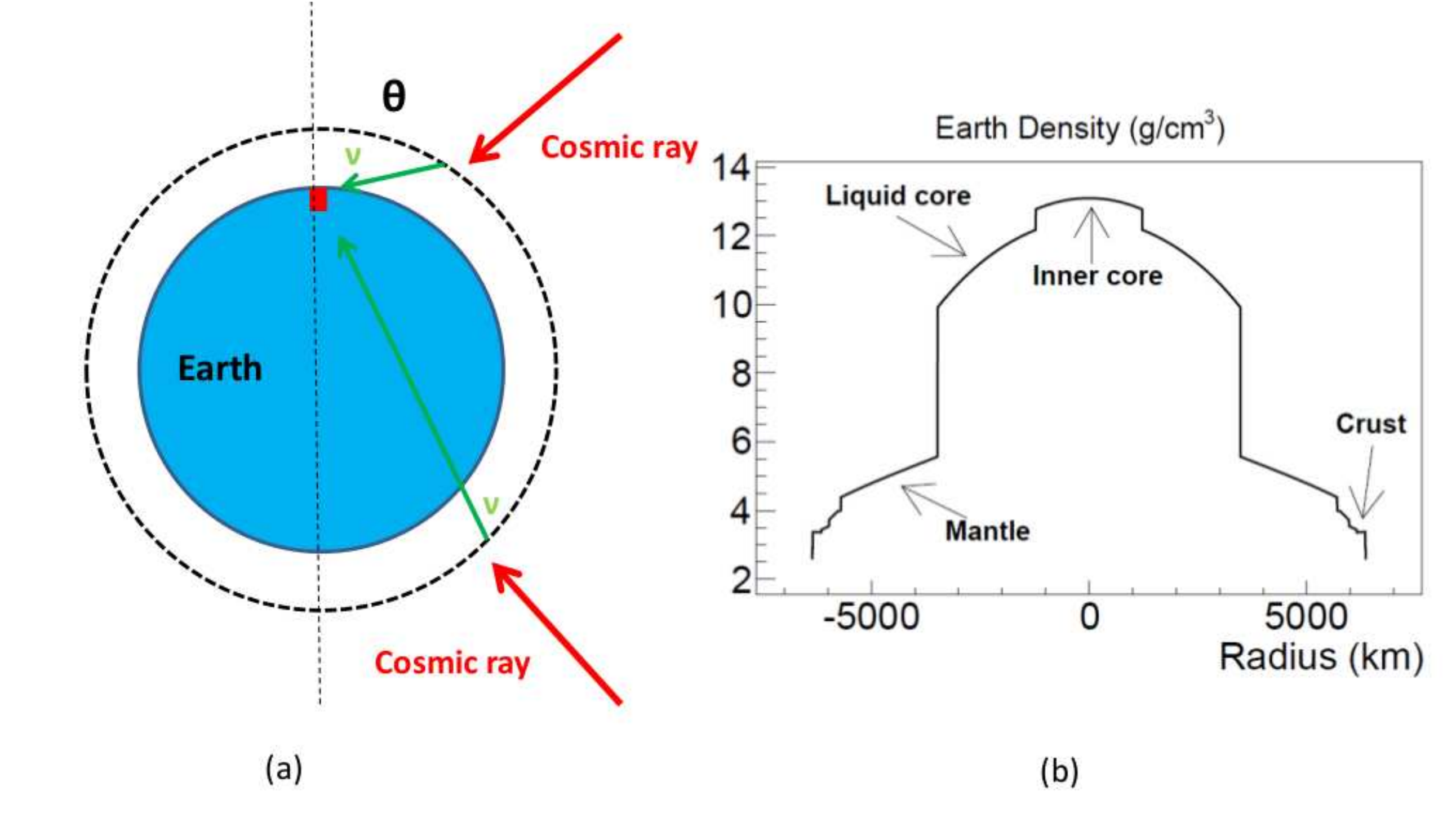}
\end{center}
\caption{\label{fig:earth} Left panel: a schematic view of the atmospheric 
neutrinos generation and detection. The zenith angle $\theta$ of the neutrinos 
are indicated. Right panel: The matter density of the Earth is shown as a function
of the radius. The data is based on the Preliminary Reference Earth Model 
(PREM)~\protect{\cite{Dziewonski1981}}.}
\end{figure}

 The zenith angle of the 
atmospheric neutrinos is directly correlated with the neutrino traveling 
distance for a detector near the earth surface as shown in the left panel of Fig.~\ref{fig:earth}. 
The neutrino detection 
relies on the charged-current neutrino-nuclei interaction as discussed in the 
previous section. In particular, as shown in Fig.~\ref{fig:xs_section}, 
the neutrino interaction cross section is much larger (about factor of 2 at high 
energies) than the antineutrino interaction cross section.

Similarly to the accelerator neutrino case, MH
can be determined through the matter effect as elaborated in Ref.~\cite{Franco:2013in,
Ribordy:2013xea,Winter:2013ema,Ge:2013ffa,Capozzi:2015bxa} among others. It is necessary, however, to take 
into account that the earth matter density is not constant, as shown in the right panel 
of Fig.~\ref{fig:earth}.

\begin{figure}[H]
\begin{center}
\includegraphics[width=1.0\textwidth]{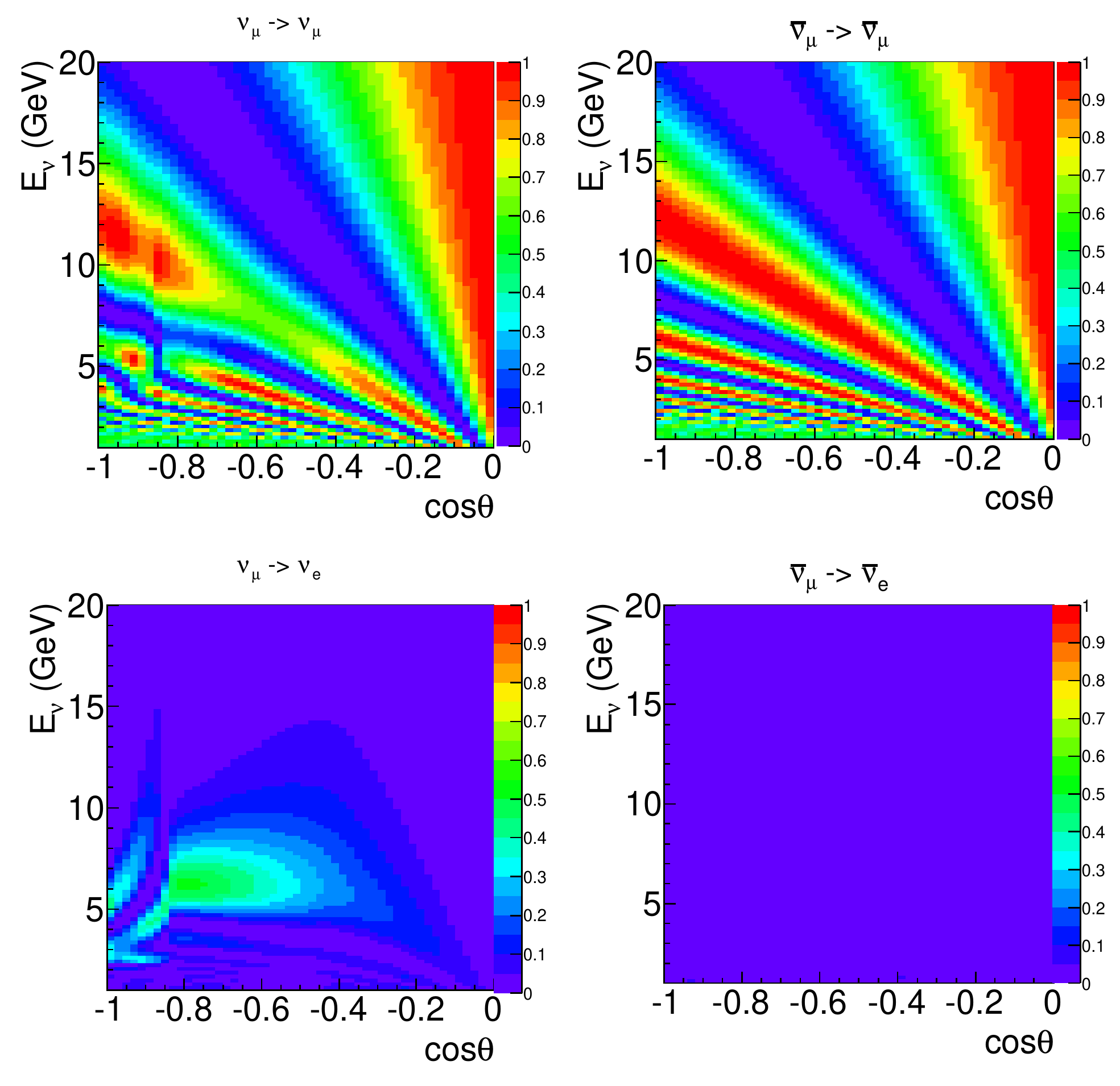}
\end{center}
\caption{\label{fig:atm_numu} Atmospheric muon neutrino oscillations calculated
with the {\it nuosc} package~\cite{nuosc}
are shown as a function of the zenith angle $\cos\theta$ and neutrino energy 
$E_{\nu}$. Top (bottom) panels show the muon neutrino disappearance (electron neutrino 
appearance). Left (right) panels show the neutrino (antineutrino) oscillations.
The normal mass hierarchy and zero CP phase $\delta_{CP} = 0$ is assumed. 
For the inverted mass hierarchy, the neutrino and antineutrino oscillation patterns 
are exchanged. }
\end{figure}

\begin{figure}[H]
\begin{center}
\includegraphics[width=1.0\textwidth]{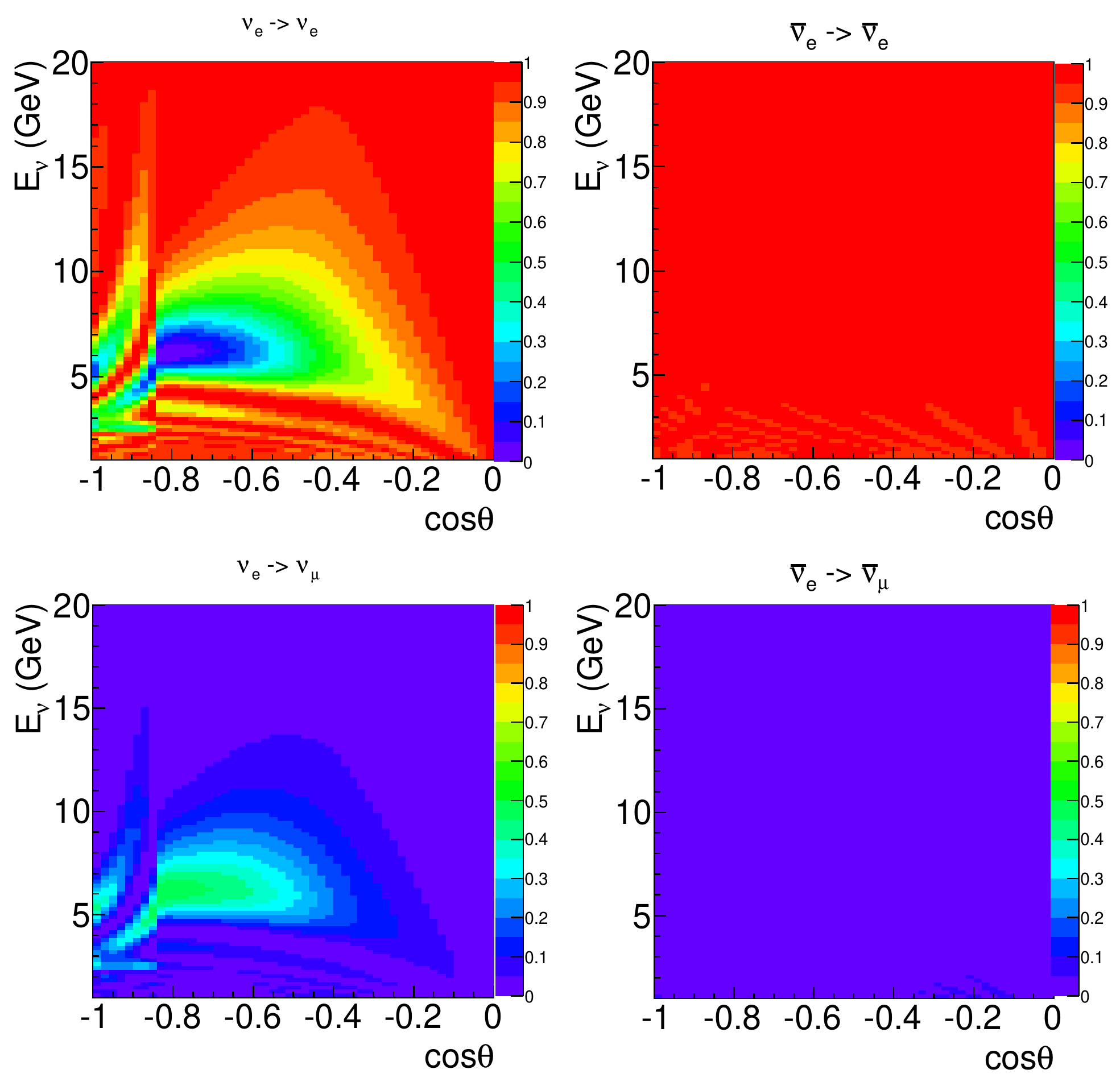}
\end{center}
\caption{\label{fig:atm_nue} Atmospheric electron neutrino oscillations calculated
with the {\it nuosc} package~\cite{nuosc}
are shown as a function of the zenith angle $\cos\theta$ and neutrino energy 
$E_{\nu}$. Top (bottom) panels show the electron neutrino disappearance (muon neutrino 
appearance). Left (right) panels show the neutrino (antineutrino) oscillations.
The normal mass hierarchy and zero CP phase $\delta_{CP} = 0$ is assumed. 
For the inverted mass hierarchy, the neutrino and antineutrino oscillation patterns 
are exchanged.}
\end{figure}

Figs.~\ref{fig:atm_numu} and ~\ref{fig:atm_nue} show the calculated oscillation probability 
for the atmospheric muon neutrinos and electron neutrinos, respectively. In NH 
scenario, we have the following features
\begin{itemize}
\item Muon neutrino disappearance $\nu_{\mu} \rightarrow \nu_{\mu}$ and 
muon antineutrino disappearance $\bar{\nu}_{\mu} \rightarrow \bar{\nu}_{\mu}$: 
there is a resonance for the neutrinos at energy $E_{\nu} \sim 5$ GeV and zenith
angle $\cos\theta \sim -0.95$ (neutrinos traveling from the other side of the 
earth) compared to the antineutrinos.
\item Muon neutrino to electron neutrino appearance $\nu_{\mu} \rightarrow \nu_e$ and muon antineutrino to electron antineutrino appearance $\bar{\nu}_{\mu} \rightarrow \bar{\nu}_e$: 
there is a much larger oscillation in the neutrinos than in the antineutrinos. 
\item Electron neutrino disappearance $\nu_e \rightarrow \nu_e$ and 
electron antineutrino disappearance $\bar{\nu}_e \rightarrow \bar{\nu}_e$:
there is a much larger oscillation in the neutrinos than in the antineutrinos. 
\item Electron neutrino to muon neutrino appearance $\nu_e \rightarrow \nu_{\mu}$ and electron antineutrno to muon antineutrino appearance $\bar{\nu}_e \rightarrow \bar{\nu}_{\mu}$: 
there is a much larger oscillation in the neutrinos than in the antineutrinos. 
\end{itemize}
In IH scenario, the behavior of neutrino oscillations
are exchanged with those of antineutrino oscillations.

The atmospheric neutrino flux contains four neutrino species 
($\nu_{\mu}$, $\bar{\nu}_{\mu}$, $\nu_e$, $\bar{\nu}_e$) and the detectors are usually
only sensitive to the neutrino flavor. Therefore, one cannot disentangle the
muon neutrino disappearance from the electron neutrino to muon neutrino appearance.
Similarly, one cannot disentangle  
the electron neutrino disappearance from the muon neutrino to electron neutrino appearance. 
Nevertheless, they can still be sensitive to the matter effect, 
as the modification of neutrino oscillations due to the matter effect is large. 
In addition, as the neutrino flux is higher than the antineutrino flux and the 
neutrino-nuclei cross section is larger than the antineutrino-nuclei  cross section, 
one could determine MH with a detector that is not sensitive 
to the lepton (in practice muon) charge. The sensitivity 
to MH can be further enhanced with a magnetized detector 
which is sensitive to the muon charge. 

%% file: reactor.tex
\subsection{Mass Hierarchy from Reactor Antineutrinos}\label{sec:reactor}

Reactor antineutrino experiments have a glorious history. In 1956, Cowan and Reines 
first experimentally discovered 
neutrinos at the Savannah River reactor power plant in the U.S.~\cite{nu_dis}. 
In 2005, the KamLAND experiment in Japan observed the neutrino oscillations in the solar 
sector~\cite{KamLAND_spec}. Recently, the Daya Bay experiment in China, the 
RENO experiment in Korea, and the Double-Chooz experiment in France 
reported the discovery of non-zero $\theta_{13}$, the third 
neutrino mixing angle in 2012~\cite{dayabay,reno,dc}. In the future, besides the determination of 
the mass hierarchy, the reactor neutrino experiments will also provide precision
measurements of neutrino mixing parameters including the mixing angles $\theta_{13}$ and 
$\theta_{12}$ and the mass-squared splittings $\Delta m^2_{32}$ and $\Delta m^2_{21}$.

\begin{figure}[H]
\begin{centering}
\includegraphics[width=0.6\textwidth]{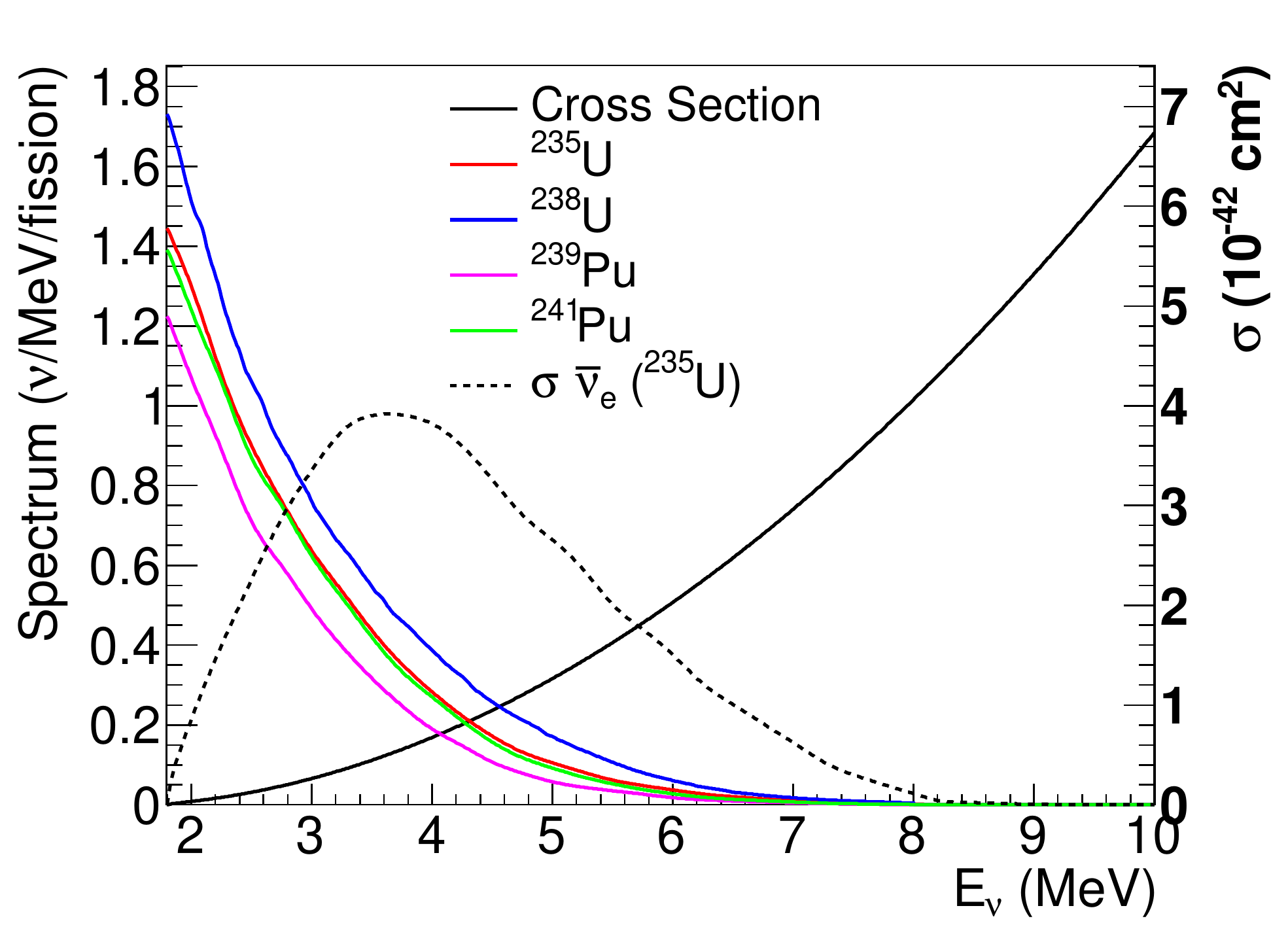}
\par\end{centering}
\caption{\label{fig:IBD} 
$\bar{\nu}_e$ energy spectra (four curves with negative slopes) for 
$^{235}$U~\cite{Schreckenbach:1985ep}, $^{238}$U~\cite{Vogel:1980bk}, 
$^{239}$Pu~\cite{Hahn:1989zr}, and $^{241}$Pu~\cite{Hahn:1989zr} are shown. 
The curve with the positive slope represents the cross section of the 
inverse beta decay (IBD) process. The convoluted IBD spectrum, measured 
in experiments,  is shown as the dotted line. }
\end{figure}

Nuclear power reactors produce electricity by the sustained nuclear chain reaction, and are 
essentially pure electron antineutrino $\bar{\nu}_e$ sources. In every nuclear fission 
$\sim$200 MeV energy is released together with on average six $\bar{\nu}_e$ 
emitted by the $\beta$-decay of the fission fragments.~\footnote{There is also 
a small component of the electron neutrinos $\nu_e$ with energy $\sim$0.1 MeV from the neutron 
activation of the shielding materials.} For each 1 gigawatt (GW) of the reactor thermal power, about 
2$\times$10$^{20}$ $\bar{\nu}_e$ are emitted isotropically every second, 
making nuclear reactors one of the most powerful man-made neutrino sources. 
$^{235}$U and $^{238}$U, together with the $^{239}$Pu and $^{242}$Pu produced by breeding 
in the reactor, are the main sources of the reactor $\bar{\nu}_{e}$. The 
corresponding $\bar{\nu}_e$ energy spectra are shown in Fig.~\ref{fig:IBD}.

The primary method to detect reactor $\bar{\nu}_e$ is using the inverse beta decay (IBD) 
reaction $\bar{\nu}_e + p \rightarrow e^{+} + n$ in which electron antineutrino 
interacts with a free proton. The final state particles are a positron 
and a neutron. An IBD event is represented by a coincident signal consisting
of i) a prompt signal induced by the positron ionization and annihilation 
inside the detector such as a liquid scintillator (LS) detector and ii) 
a delayed signal produced by the neutron capture on a proton or 
a nucleus (such as Gd) with a large neutron capture cross section. 
In particular, the neutron capture on Gd releases 
multiple gammas with a total energy $\sim$8 MeV. With 0.1\% Gd-doped LS, 
the average time between the prompt and the delayed signal is about 30 $\mu$s.
In comparison, for neutron capture on hydrogen a single gamma with 
energy of $\sim$2.2 MeV is emitted. For liquid scintillator $C_{n}H_{\sim2n}$, the 
average time between the prompt and the delayed signal is about 200 $\mu$s.

\begin{figure}[H]
\begin{centering}
\includegraphics[width=0.5\textwidth]{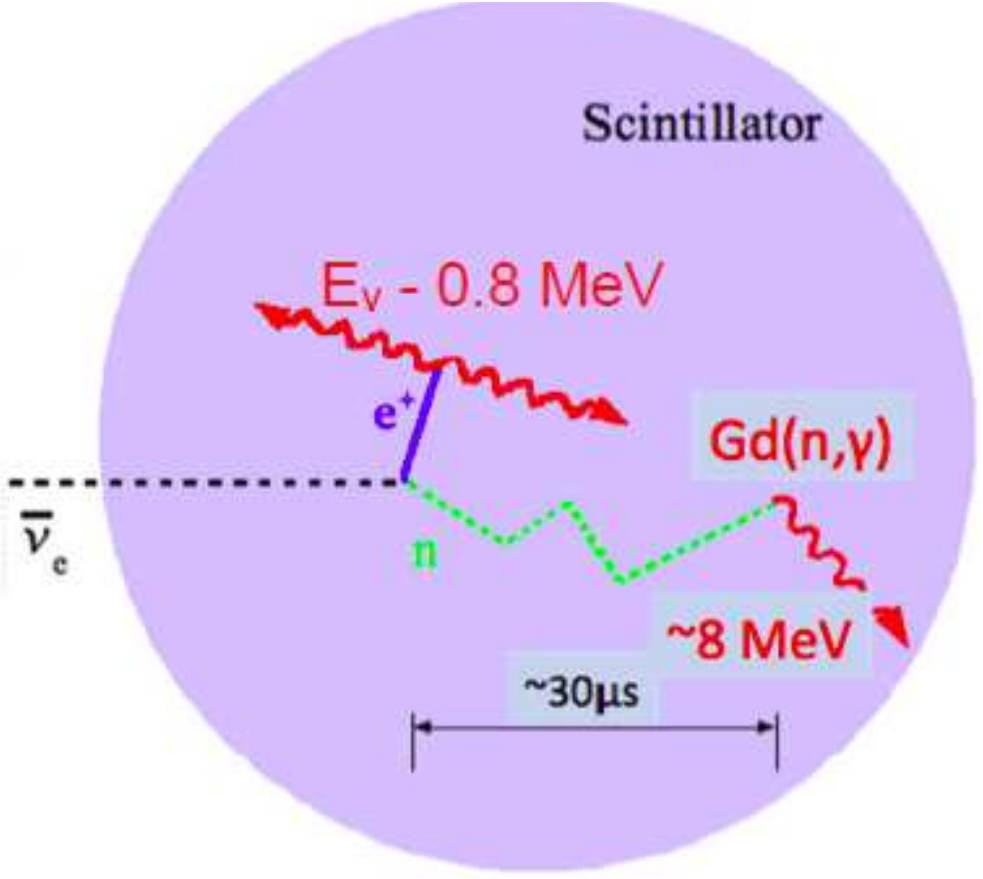}
\par\end{centering}
\caption{\label{fig:IBD1} 
The detecting principle of IBD is shown.
This figure is reproduced with the permission from Ref.~\protect\cite{Qian:2014xha}. 
}
\end{figure}

Fig.~\ref{fig:IBD1} illustrates the reaction and detection principle of the 
IBD process. The cross section of the IBD process is shown as the solid 
black line in Fig.~\ref{fig:IBD} with units showing on the right axis. 
The convoluted energy spectrum, i.e. the product of the reactor antineutrino flux and 
the IBD cross section, is plotted as the dashed line. 
It begins at the threshold of $\sim$1.8 MeV neutrino energy $E_{\nu}$, 
and peaks around 4 MeV. The tail of the spectrum extends beyond 8 MeV. 
Due to the time and space correlations, IBD can be easily distinguished from 
radioactive backgrounds which mostly consist of only a single signal. 
Furthermore, the energy of the prompt signal (also known as the visible 
energy $E_{vis}$) is directly related to the neutrino energy: 
\begin{equation}
E_{\nu} = E_{e^+} + m_n +T_n - m_p\approx E_{prompt} + 0.78~MeV,
\end{equation} 
where $m_e$, $m_n$, and $m_p$ are the mass of electron, neutron, 
and proton, respectively. $T_n$ is the relatively small ($\sim$ a few keV) 
kinetic energy of the recoiled neutron. Besides the energy of the positron, 
$E_{prompt}$ also contains the mass of electron that annihilates with the positron.
The straight-forward method of reconstructing the neutrino energy is 
a very attractive feature for the study of neutrino oscillations that require 
good knowledge of the neutrino energy.

The basis for the MH determination using the $\bar{\nu}_e$ disappearance 
has been extensively discussed in Refs.
~\cite{Petcov:2001sy,Learned:2006wy,Minakata:2006gq,Zhan:2008id,Parke:2008cz,Zhan:2009rs,
Qian:2012xh,Ciuffoli:2012iz,Ge:2012wj,Li:2013zyd,Capozzi:2013psa}, 
and is 
illustrated in Eq.~\eqref{eq:mphi} in Sec.~\ref{sec:introduction}. In order
to further elaborate on this point, we define the effective mass-squared 
splitting as:
\begin{equation}
\Delta m^2_{eff} = \frac{4 E_{\nu}}{L} \cdot (2\Delta_{32} \pm \phi_{ee}) 
= \left\{ \begin{array}{lr}
2\Delta m^2_{32} + \Delta m^2_{\phi} & ({\rm Normal~Hierarchy}) \\
2\Delta m^2_{32} - \Delta m^2_{\phi} & ({\rm Inverted~Hierarchy})
\end{array}\right.
\end{equation}
which corresponds to the fourth term in Eq.~\eqref{eq:mphi}. Here, we define
$\Delta m^2_{\phi}:=4\cdot {\phi}_{ee} \cdot E_{\nu} / L$. 

\begin{figure}[H]
\begin{center}
\includegraphics[width=0.49\textwidth]{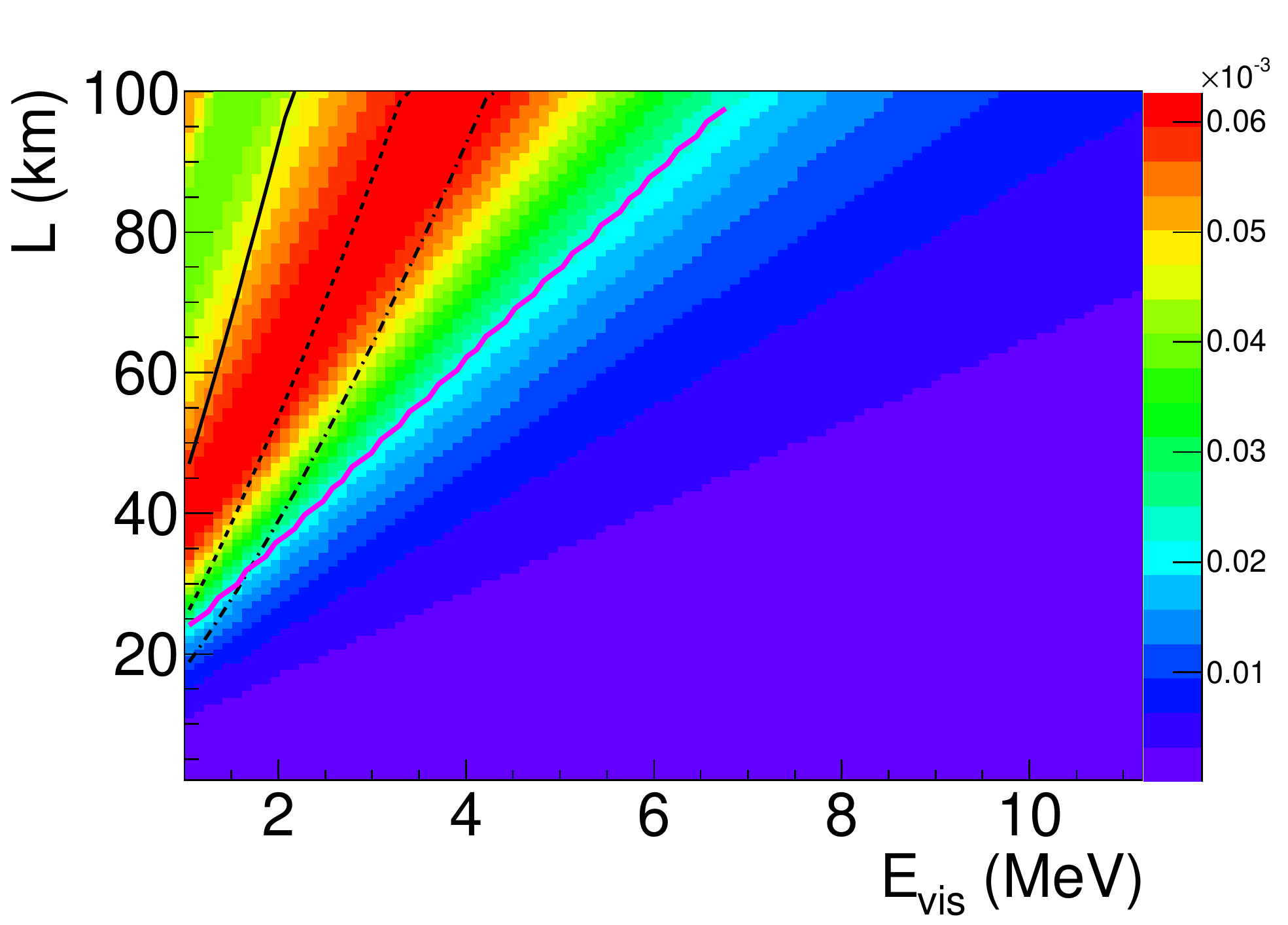}
\includegraphics[width=0.49\textwidth]{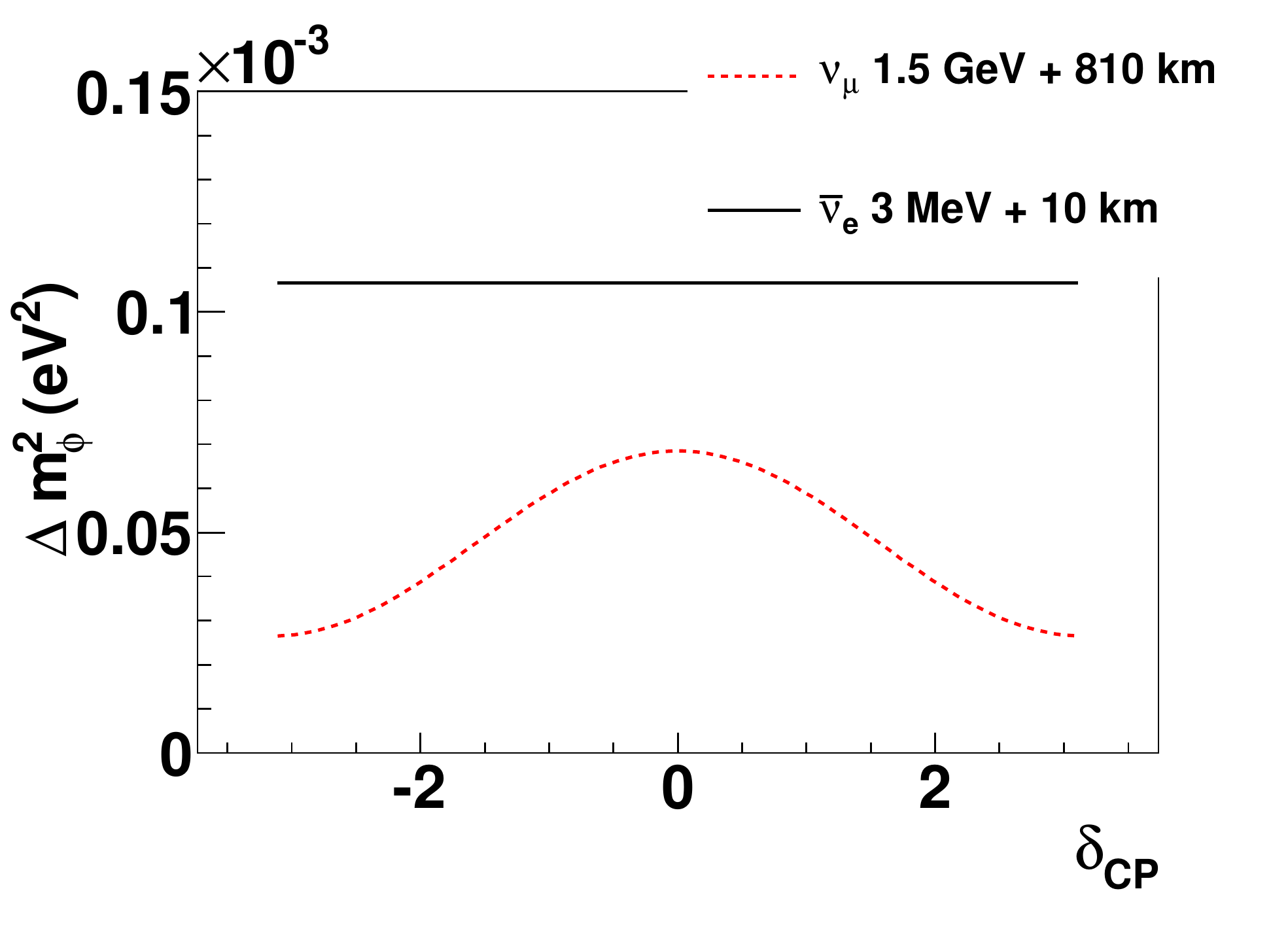}
\end{center}
\caption{\label{fig:mh:contour} 
(Left) The effective mass-squared difference $\Delta m^2_{\phi}$, as a
function of baseline (y-axis) and visible energy $E_{\rm vis} \simeq
E_{\nu}-0.8\,{\rm MeV}$ (x-axis). The color 
represents the size of $\Delta m^2_{\phi}$
in eV$^2$. The solid, dashed, and dotted lines 
are three choices of detector energy resolution with 
2.8\%, 5.0\%, and 7.0\% at 1 MeV, respectively. The left side of these lines
(low values of $E_{\rm vis}$) will yield negligible 
contributions to the differentiation of the MH due to large values of $L/E_{\nu}$ 
The purple solid line represents the approximate boundary of the degenerate mass-squared
difference. The right side of the purple line alone will also yield negligible 
contributions to the differentiation of the MH.
(Right) $\Delta m^2_{\phi}$ for both electron
antineutrino ($\Delta m^2_{\phi~ee}$) and muon disappearance 
($\Delta m^2_{\phi~\mu\mu}$) are shown as a function of the unknown CP phase 
$\delta_{CP}$. Reproduced with permission from Ref.~\protect\cite{Qian:2012xh}.}
\end{figure}

In the left panel of Fig.~\ref{fig:mh:contour}, the $\Delta m^2_{\phi}$
is plotted as a function of the visible energy $E_{vis} \equiv E_{prompt}$ as well as the baseline $L$.
At a baseline $L\sim60$ km, $\Delta m^2_{\phi}$ has a clear energy dependence. 
In particular, the $\Delta m^2_{\phi}$ at low energy (2-4 MeV) is larger than that 
at high energy (4-8 MeV), providing an opportunity to determine MH. 
For the normal (inverted) MH, the $\Delta m^2_{eff}$ measured at low energy (2-4 MeV) 
would be higher (lower) than that measured at high energy (4-8 MeV). 
However, in order to perform such a measurement successfully, the energy resolution 
of the detector must be very good (better than 1.9\% for $\delta E/E$ at 2.5 MeV), 
since the oscillations corresponding to $\Delta m^2_{eff}$ are very fast at low energy
and a cruder energy resolution would make the oscillation pattern to disappear.
On the other hand, at short baseline 
($L<20$ km), $\Delta m^2_{\phi}$ is essentially a constant, thus it would be
impossible to determine MH through comparing 
$\Delta m^2_{eff}$ values at low and high energy regions. The purple line
in the left panel of Fig.~\ref{fig:mh:contour}  represents the approximate boundary 
of degenerate mass-squared difference. The right side of the purple line alone
will yield negligible contributions to MH. 

Alternatively, if one could measure $\Delta m^2_{eff}$ very precisely in both the electron 
antineutrino disappearance channel and in the muon neutrino and antineutrino disappearance 
channels, one can be sensitive to MH~\cite{Nunokawa:2005nx,Minakata:2006gq}. This is because
the $\Delta m^2_{\phi}$ values in these two channels are different. 
More specifically, as shown in the right panel of Fig.~\ref{fig:mh:contour},
$\Delta m^2_{\phi~ee}$ is larger than $\Delta m^2_{\phi~\mu\mu}$. Therefore, 
we have
\begin{equation}
\left\{
\begin{array}{rl}
\Delta m^2_{eff~ee} = 2 \Delta m^2_{32} + \Delta m^2_{\phi~ee}> \Delta m^2_{eff~\mu\mu} = 2\Delta m^2_{32} + \Delta m^2_{\phi~\mu\mu} & ({\rm Normal~Hierarchy}) \\
\Delta m^2_{eff~ee} = 2\Delta m^2_{32} - \Delta m^2_{\phi~ee} < \Delta m^2_{eff~\mu\mu} = 2\Delta m^2_{32} - \Delta m^2_{\phi~\mu\mu} & ({\rm Inverted~Hierarchy})
\end{array}\right.
\end{equation}
Since the difference between the $\Delta m^2_{\phi}$ in these two channels 
is quite small ($\sim5\times10^{-5}$ eV$^2$), the required precision of 
$\Delta m^2_{eff}$ in both channels would be about $\sim1\times10^{-5}$ eV$^2$,
which is very challenging. Therefore, this method is likely to serve as a 
validation once the MH is already determined by other means. 

%% file: stat.tex
\subsection{Interpretation of Mass Hierarchy Sensitivity}\label{sec:stat}

The determination of the neutrino MH is equivalent to finding the 
sign of $\Delta m^2_{32}=m^2_3-m^2_2$.  Thus, the determination of MH 
represents the test of two discrete hypotheses (normal hierarchy vs. 
inverted hierarchy or NH vs. IH in short).  In this context, there are 
two different but related questions: i) given the data from an experiment, 
how to quantify the finding regarding MH? and ii) when designing an experiment, 
how to evaluate its sensitivity to MH?

The usual approach to the first question is as follows. Given the result of 
a measurement, we compare the data (represented by ``x") with 
expectations from both the NH and IH hypotheses.  In practice, one commonly 
defines a test statistic $T_{NH}:=-2\log{L_{NH}}$, where $L_{NH}$ is the likelihood of NH for 
data ``x". $T_{NH}$ is further minimized over all nuisance parameters, 
including unknown parameters and systematic uncertainties, to obtain $T_{NH}^{min}$.  
Similarly, another test statistic $T_{IH}^{min}$ is calculated from the likelihood of 
IH for data "x".  In order to evaluate whether the experimental data favors NH or IH, a test 
statistics $\Delta T = T_{IH}^{min} - T_{NH}^{min}$ is defined.  It is easy to see 
that a positive $\Delta T$ would favor the NH hypothesis, and a negative $\Delta T$ 
would favor the IH hypothesis.  In addition, the absolute size of $\Delta T$ 
contains the information of how much data favor one hypothesis relative to the other one.

Regarding the second question, when designing an experiment, the community sometimes 
report the average expectation $\overline{\Delta T}$~\footnote{The notation of 
$\overline{\Delta \chi^2}$  or sometimes $\Delta \chi^2$ is commonly used to 
represent $\overline{\Delta T}$.}, which is calculated using the following procedure: 
\begin{itemize}
\item Assuming that the true MH is NH, 
a data set can be generated based on the best known experimental parameters.  
Such a data set does not include any statistical fluctuations and is 
usually called the typical data set or the Asimov data set~\cite{asimov}.  
\item When this typical data set is compared with the NH hypothesis, we naturally have 
$T_{NH}^{min}=0$.  Therefore, $\overline{\Delta T_{MH=NH}}=T_{IH}^{min}$.  
The subscript labels represent the assumption that the true MH is NH.  
Similarly, one can calculate $\overline{\Delta T_{MH=IH}}$ assuming that the true MH is IH.
Here, $\overline{\Delta T}$ can be interpreted as either the most probable value of 
$\Delta T$ or the average value of $\Delta T$.
\end{itemize}

\begin{figure}
\begin{center}
\begin{tabular}{cc}
\includegraphics*[width=0.49\textwidth]{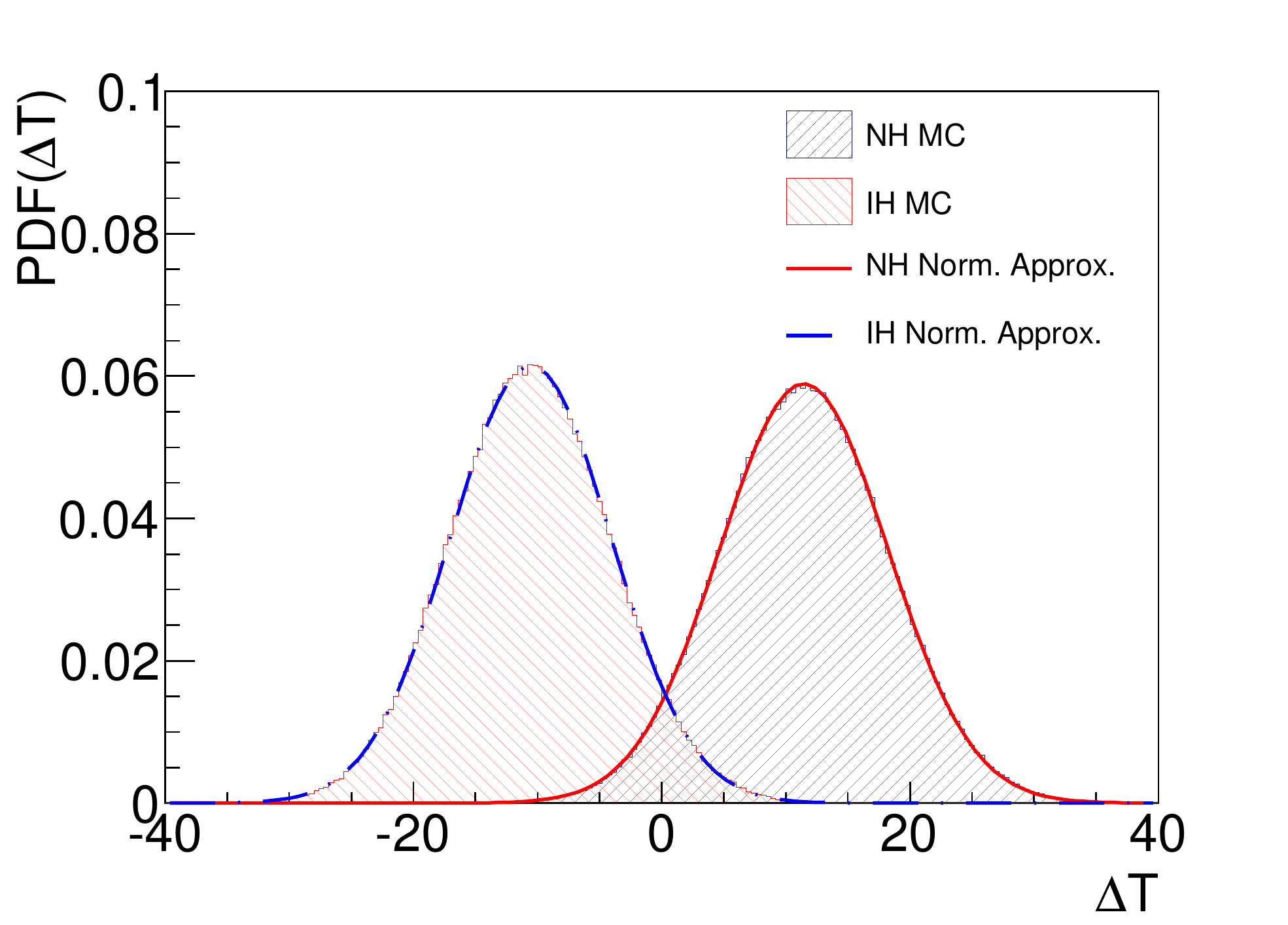}
&
\includegraphics*[width=0.49\textwidth]{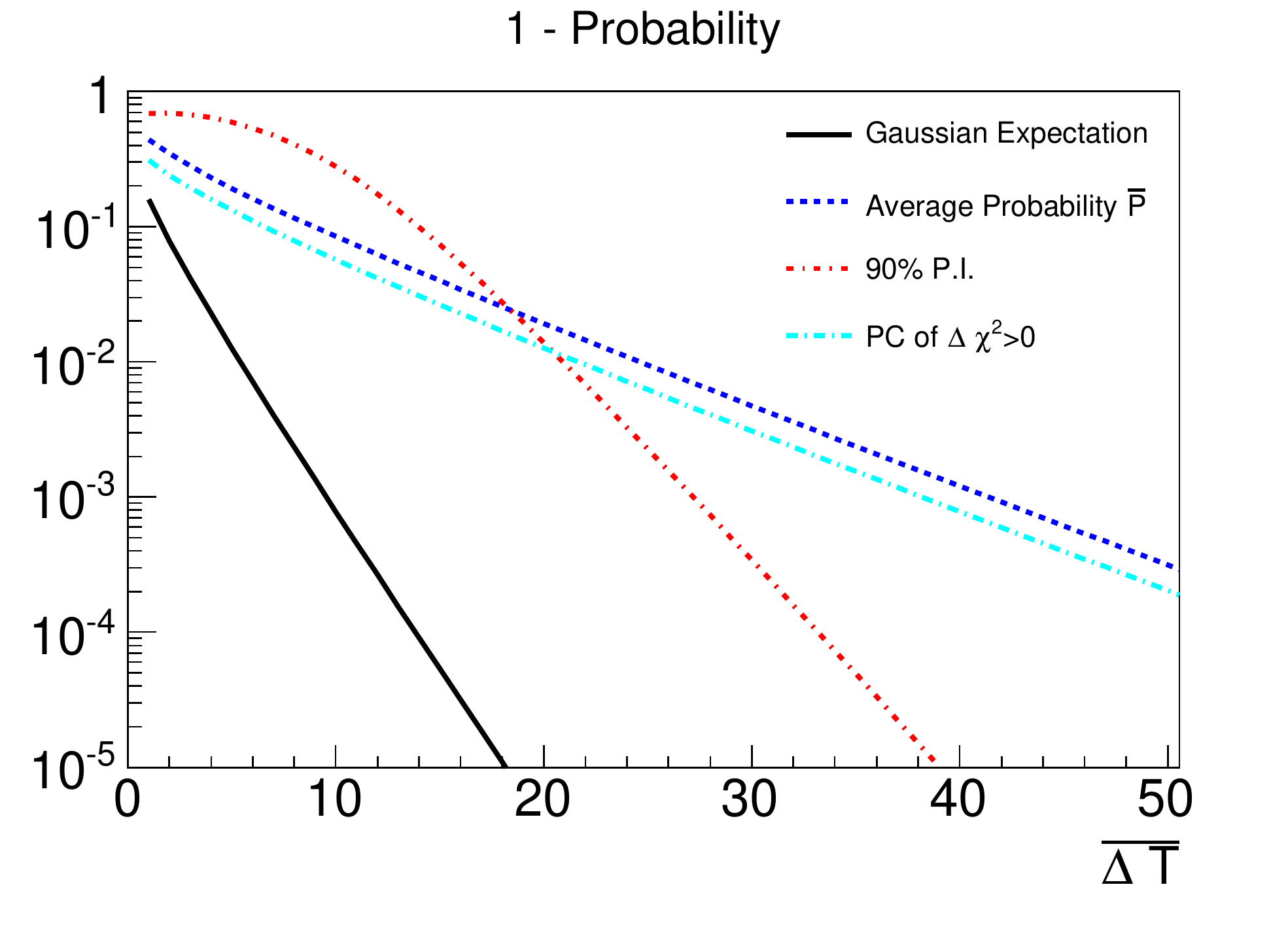}
\end{tabular}
\end{center}
\caption{\label{fig:mh:stat}  (Left panel) The  probability
distribution function of $\Delta T$ with \protect{$\overline{\Delta T}
\approx 11$} is shown. The overlap is due to natural spread from 
statistical and systematic uncertainties. 
(Right panel) Several sensitivity metrics in the
Bayesian statistics with respect to \protect{$\overline{\Delta T}\approx
\Delta\chi^2_{\rm MH}$} are shown. See the text for details. Reproduced with 
the permission from Ref.~\protect\cite{Qian:2012zn}.}
\end{figure}

In reality, the data from any experiments always contain statistical fluctuations 
as well as variations due to systematics.  Therefore, if the experiment would be repeated many 
times, a distribution of $\Delta T$ appears.  Due to the nature of two discrete 
hypotheses of testing MH, when the observed counts in the experiment are large enough, 
one can show that the distribution of $\Delta T$  approximately follows a Gaussian 
distribution with the mean and the standard deviation as $\overline{\Delta T}$, and 
2$\sqrt{|\overline{\Delta T}|}$, respectively~\cite{Qian:2012zn,Blennow:2013oma,Qian:2014nha}.  
For example, the left panel of Fig.~\ref{fig:mh:stat} shows the expected distribution of 
$\Delta T$ for experiments with $|\overline{\Delta T_{MH=NH}}| \approx 
|\overline{\Delta T_{MH=IH}}| \approx 11$.  There will be about 4.8\% of experiments which 
favor a wrong MH preference.   Therefore even if 
$\overline{\Delta T}$ is large, there remains a probability that a small $\Delta T$
is observed in an experiment.  For example, in a situation where NH and IH are such that 
$|\overline{\Delta T}| = 25$ , the chance is about 1\% that the experiment generates observed 
data favoring the wrong mass hierarchy.  It should be noted that such a result naturally 
has a small $\Delta T$  value.  

\subsubsection{Bayesian Approach to the Mass Hierarchy Determination}~\label{sec:bayes}

As illustrated above given data from a measurement one can calculate 
$T_{NH}:-2\log(L_{NH})$  and $T_{IH}:-2\log(L_{IH})$. $T_{NH}$ and $T_{IH}$ are further 
marginalized over all nuisance parameters including unknown parameters and systematic 
uncertainties of the experiment to obtain $T_{NH}^{marginalized}$ and $T_{IH}^{marginalized}$, 
respectively.~\footnote{In the marginalization process, one integrates the likelihood 
function over the entire parameter space of the nuisance parameters. }  
Assuming that the prior information of NH vs. IH is 50\% vs. 50\%,
i.e. that both possibilities are equally expected, the probability ratio 
of the IH vs. NH, which is P(IH;x) vs. P(NH;x), can be calculated as $e^{-\Delta \tau/2}$ vs. 1,
with $\Delta \tau = T_{IH}^{marginalized}-T_{NH}^{marginalized}$~\cite{Qian:2012zn}.  Here, $e^{-\Delta \tau/2}$ 
is commonly referred to as the Bayes factor.  As illustrated in Ref.~\cite{Qian:2012zn}, 
an approximation $\Delta \tau \approx \Delta T$ can be made in practice.  
More generally, $\Delta \tau$ can be explicitly calculated through MC simulations or other 
advanced integration techniques.  Table~\ref{table:sens} lists a few values of $\Delta \tau$  
and their corresponding probability ratios.  In the Bayesian approach, the final results 
are presented in terms of a probability ratio which is a natural and simple way to present 
results for the MH determination~\cite{Qian:2012zn,Blennow:2013kga}. 

\begin{table}
\centering
    \begin{tabular}{| l | l | l | l | l| l |}
    \hline
$\Delta \tau \approx \Delta T$ & 1 & 4 & 9 & 16 & 25 \\\hline
P(IH;x) vs. P(NH;x) & 38\% vs. & 12\% vs. & 1.1\% vs.  &
0.034\% vs.  & 3.7 $\times$ 10$^{-6}$ vs.  \\
& 62\% & 88\%  & 98.9\% & 99.966\% & 100\% \\\hline
P(IH;x)/P(NH;x) & 0.61 & 0.136 & 0.011 & 3.4$\times10^{-4}$ & $3.7\times10^{-6}$ \\\hline
     \hline
    \end{tabular}
\caption{\label{table:sens} Probability ratios are shown with respect to $\Delta \tau$
values.}
\end{table}

Ref.~\cite{Qian:2012zn} also provides a few sensitivity metrics based on the Bayesian approach 
that illustrate the capability of an experiment; these are useful in designing experiments.  
One of them is $\overline{P_{MH=NH}^{NH}}$: the average probability to obtain the correct 
mass hierarchy.  The second one is  $P_{MH=NH}^{90\%}$, which is based on probability 
intervals.  The interpretation of $P_{MH=NH}^{90\%}$  is the following: given that NH is 
the truth, 90\% of the potential data would yield a probability of NH, P(NH;data x), higher 
than  $P_{MH=NH}^{90\%}$. Right panel of Fig.~\ref{fig:mh:stat} shows the metrics   
$\overline{P_{MH=NH}^{NH}}$ (dotted blue line) and $P_{MH=NH}^{90\%}$  (dotted red line) 
with respect to $\overline{\Delta T}$, assuming $\overline{\Delta T} \approx 
|\overline{\Delta T_{MH=NH}}|\approx |\overline{\Delta T_{MH=IH}}|$  and 
$\Delta \tau \approx \Delta T$.

\subsubsection{Frequentist Approach to the Mass Hierarchy Determination}~\label{sec:freq}

Ref.~\cite{Agashe:2014kda} (Particle Data Group) contains a very nice review 
regarding hypothesis tests.  Given a null hypothesis H$_0$ and the alternative 
hypothesis H$_1$, {\it type-I error rate} $\alpha$ is defined as the probability of 
rejecting the null hypothesis H$_0$, if H$_0$ is true.  On the other hand, {\it type-II error 
rate} $\beta$ is defined as the probability of accepting the null hypothesis H$_0$, 
if H$_1$ is true.  Furthermore, according to {\it Neyman-Pearson lemma}, the likelihood ratio 
test is shown to be the most powerful test given that H$_0$ and H$_1$ are both simple 
hypotheses.~\footnote{A simple hypothesis should not contain any undetermined parameters. }   
This provides the basis to use $\Delta T$ defined above to perform the MH hypothesis test.   
Before data are obtained, we choose the type-I error rate $\alpha$, 
and this will decide the rule of the hypothesis test.  The $\alpha$ value will determine 
the rejection region, and it will in turn determine the corresponding type-II error rate 
$\beta$.  In our MH problem, one can choose the null hypothesis to be NH, and the 
alternative hypothesis would be IH.  The result of this test will tell us whether the NH 
would be rejected or not, given the pre-defined rule.  Similarly, one should also perform 
a second hypothesis test by choosing null hypothesis to be IH with the alternative hypothesis 
being NH.  The result of the second test will tell us whether the IH would be rejected or not.  
Unlike the Bayesian approach, the Frequentist approach does not directly address the question 
how much one MH hypothesis is favored over the other MH hypothesis given the experimental 
data.  

Furthermore, given experimental data and test statistics, one can obtain the p-value.  
The p-value represents the probability of obtaining potential data that are less compatible 
with the null hypothesis compared to the current experimental data.  We emphasize 
that one should not confuse the p-value in Frequentist approach with the probability ratios 
in the Bayesian approach.  For the MH determination, a nice feature exists for the p-value 
of the null hypothesis H$_0$ being the disfavored MH hypothesis.  In order to illustrate 
this point, we use the simple model shown in the left panel of Fig.~\ref{fig:mh:stat}.  
Given $\Delta T$ from data and $\overline{\Delta T}$  from the design of the experiment, 
the number of standard deviations which corresponds to the 1-sided p-value of the 
null hypothesis H$_0$ being the disfavored MH hypothesis can be written as:
\begin{equation}
(number~of~\sigma) = \frac{(|\Delta T| + \sqrt{|\overline{\Delta T}|})}{2\sqrt{|\overline{\Delta T}|}} \ge \sqrt{|\Delta T|}
\end{equation}  
The equal sign is reached when $|\Delta T|=|\overline{\Delta T}|$.  
In other words, given the observed data $\Delta T$, there is a maximum p-value of the 
null hypothesis H$_0$ being the disfavored MH hypothesis, which corresponds to 
$\sqrt{|\Delta T|}$  in terms of the number of standard deviations.~\footnote{When transformed 
into number of standard deviations, a smaller p-value would give a bigger number.}   
We should further point out that this maximum p-value does not depend on $\overline{\Delta T}$.

The sensitivity of an experimental design can be represented using the type-I error rate 
$\alpha$ at certain type-II error rate $\beta$.   Given a design of an experiment, 
the type-I error rate $\alpha$ and type-II error rate $\beta$ are directly connected. In 
other word, $\alpha$ can be expressed as a function of $\beta$.  
A smaller type-I error rate $\alpha$ would lead to a larger type-II error rate $\beta$ and 
vice versa.  
The community commonly uses $\sqrt{\overline{\Delta \chi^2}}$ to represent the MH 
sensitivity of an experimental design. A recent discussion can be found in Ref.~\cite{Vitells:2013uza}. 
In a simplified model shown in the left panel of 
Fig.~\ref{fig:mh:stat}, the choice of $\sqrt{\overline{\Delta \chi^2}}$ corresponds 
to choosing the value of type-I error rate $\alpha$ as the sensitivity metric, 
while type-II error rate $\beta$ is fixed at 0.5 ($\beta=0.5$).

Through investigating the statistical problem regarding Spin discrimination of the 
Higgs boson, whose nature is very similar to the MH discrimination, authors of 
Ref.~\cite{Cousins:2005pq} suggested to consider the value of
$\alpha(\beta=\alpha)$, which is the value of $\alpha$ when $\alpha=\beta$,
as the sensitivity metric. This can be compared to the conventional sensitivity
metric $\alpha(\beta=0.5)$, which is the value of $\alpha$ at $\beta = 0.5$. 
In the simplified model as shown in the left panel of 
Fig.~\ref{fig:mh:stat}, the latter sensitivity coincides with the sensitivity metric 
$F_{MH=NH}$ in Ref.~\cite{Qian:2012zn}, which is the probability of obtaining a $\Delta T$ 
favoring the correct MH.  Right panel of Fig.~\ref{fig:mh:stat} shows $F_{MH=NH}$   
with respect to $\overline{\Delta T}$ (cyan dash-dotted line). 

In general, the high energy physics community is more familiar with the Frequentist 
approach than the Bayesian approach.  In particular, there exist traditional criteria 
regarding the evidence and discovery of new physics: the simple 3$\sigma$ (evidence) and 
5$\sigma$ (discovery) rule.  If the 1-sided p-value of data when compared to the null 
hypothesis is smaller than 0.13\%, one can claim that the evidence of new physics is observed.  
Similarly, if the 1-sided p-value is smaller than 3$\times10^{-7}$ , one can claim the 
discovery of new physics.  For example, in the case of the non-zero $\theta_{13}$ 
discovery~\cite{An:2012eh}, the null hypothesis is $\theta_{13}=0$.  In the case of the 
discovery of Higgs boson~\cite{Aad:2012tfa,Chatrchyan:2012ufa}, the null hypothesis is the
non-existence of Higgs boson.  

Recently, there have been some inspiring discussions~\cite{Lyons:2013yja,RCousins:2013yja} 
on whether the 5$\sigma$ rule should be universally applied as 
the criteria for discoveries.   The 5$\sigma$ rule was invented to take into account 
i) look-elsewhere effect and ii) hidden unknown systematic uncertainties.  
For the MH determination, there is clearly no look-elsewhere effect.  
In this case, the discovery criteria should be adjusted accordingly.  
We further note that in practice it is strongly preferable that the designed experiment
is not dominated by systematic uncertainties.

%% file: experiments.tex
\section{Experiments}\label{sec:experiments}

In this section, we briefly describe the existing and planned experiments that have sensitivities
to the neutrino mass hierarchy. In the first part, the current status is discussed. While 
the analysis of the atmospheric neutrino data by the Super-Kamiokande experiment slightly 
favors the normal hierarchy, the comparison of the experimentally determined $\Delta m^2_{32}$ 
mass splitting using the muon and electron neutrino disappearance favors the inverted hierarchy. 
These two hints, that so far lack statistical significance, are obviously incompatible 
with each other. The running NO$\nu$A experiment will be able to shed some light on 
this issue, but needs to collect more data. Also, there is a significant degeneracy between 
MH and the value of the CP phase $\delta_{CP}$ due to the baseline of NO$\nu$A.

In the second part of the section we describe the proposed and planned experiments that 
promise to resolve MH. They are based on the study 
of the accelerator neutrinos (DUNE/LBNF), atmospheric neutrinos (PINGU, ORCA, INO, and 
Hyper-Kamiokande), and reactor neutrinos (JUNO, RENO-50).  These next-generation experiments are
in various stages of planning and funding. 

\begin{figure}[H]
\begin{centering}
\includegraphics[width=0.4\textwidth]{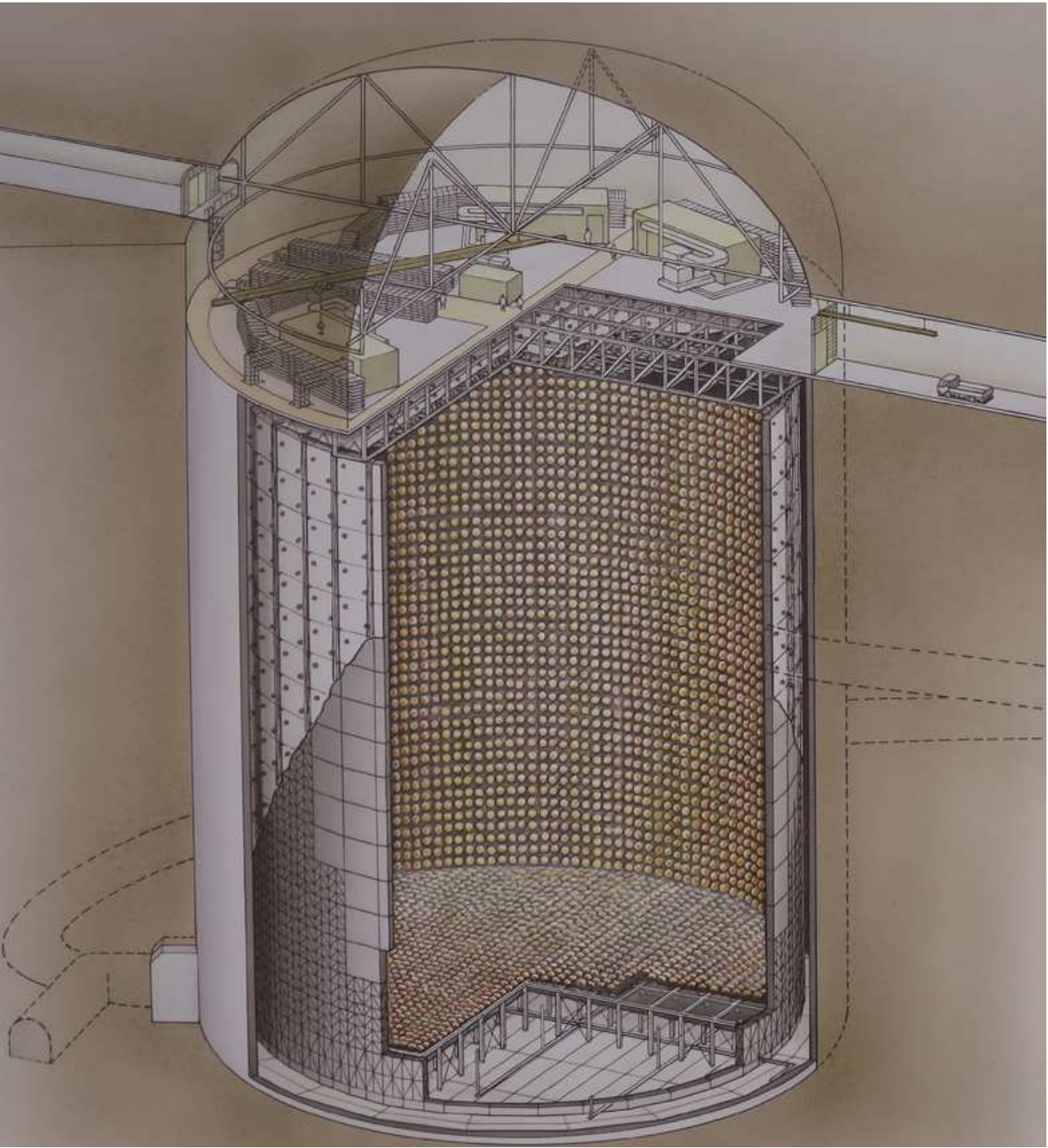}
\includegraphics[width=0.4\textwidth]{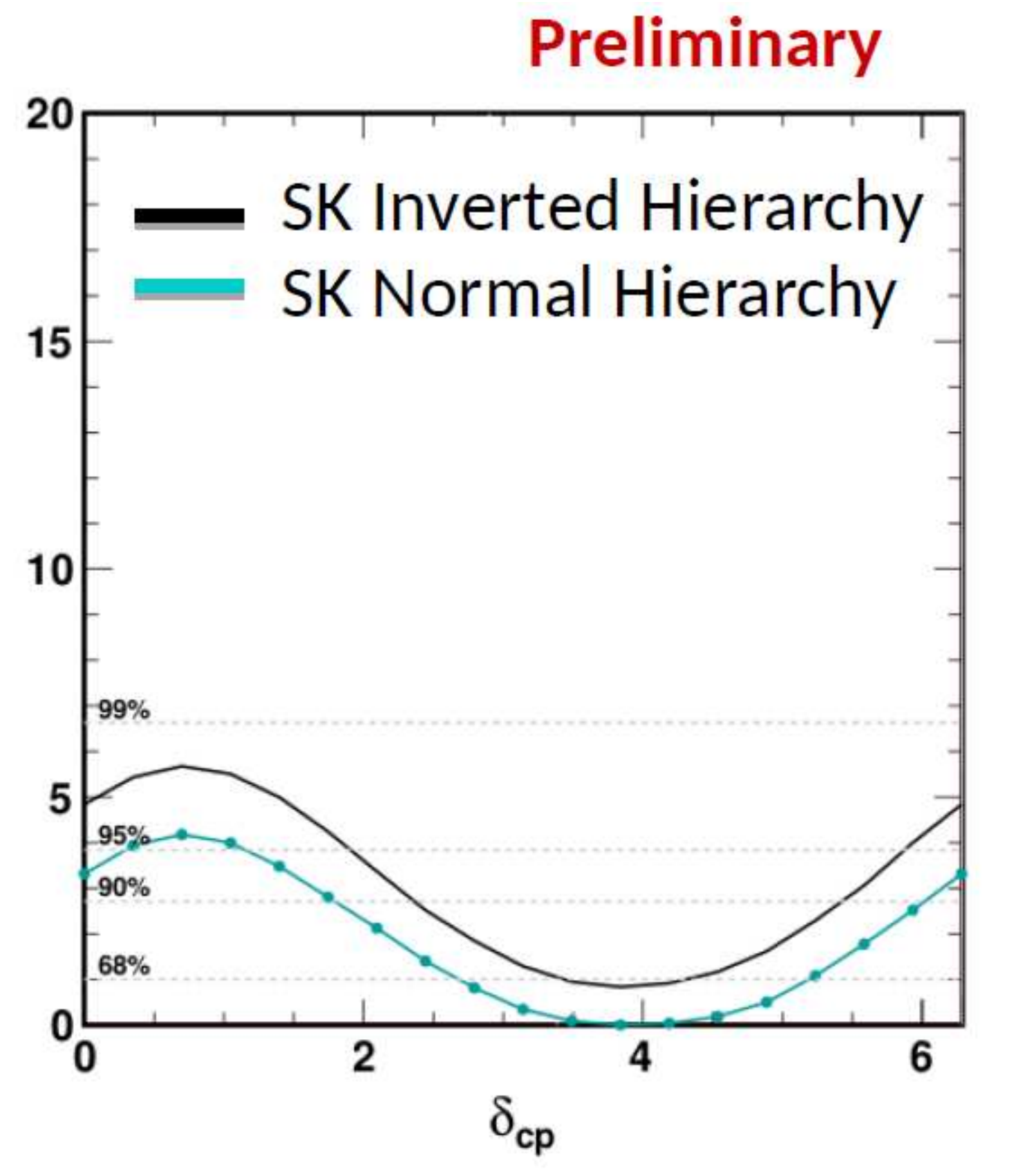}
\par\end{centering}
\caption{\label{fig:sk_det} (Left) A schematic view of the Super-Kamiokande
detector. Reproduced with the permission of Kamioka Observatory, ICRR (Institute 
for Cosmic Ray Research), the university of Tokyo.
(Right) Preliminary result of Super-Kamiokande mass hierarchy measurement in terms of the 
negative-two-log-likelihood ratio (or $\Delta \chi^2 = \chi^2 - \chi^2_{min}$)
are plotted as a function of the unknown CP phase $\delta_{CP}$.
The normal mass hierarchy is slightly favored by $\Delta \chi^2 \sim 0.9$,
corresponding to about 60\% vs. 40\% probability with the Bayesian 
interpretation. This plot is reproduced with the permission from Ref.~\protect\cite{wendell_neutrino2014}.}
\end{figure}

\subsection{Current Status}
\input{sk.tex}

\input{dyb.tex}

\input{nova.tex}

\subsection{Proposed and Planned Experiments}
\input{LBNF.tex}

\input{PINGU.tex}

\input{juno.tex}

%% file: sk.tex
\subsubsection{Atmospheric neutrino results from Super-Kamiokande}

The analysis of the atmospheric neutrino data from the Super-Kamiokande experiment 
currently provides a hint of the mass hierarchy. The Super-Kamiokande detector 
(left of Fig.~\ref{fig:sk_det}) is a cylindrical 50-kton (22.5-kton fiducial volume) 
water Cherenkov detector located 1000 m underground in western Japan.  The signal readout uses an 
array of regularly spaced large photo-multiplier tubes sensitive to 
single photo-electrons~\cite{Fukuda:2002uc}. From the pattern of the 
detected Cherenkov light, one can reconstruct the energy, the direction, 
as well as the particle type of the charged leptons produced by the 
charged-current  neutrino-nuclear interaction. Using the principle illustrated
in Sec.~\ref{sec:atm}, Super-Kamiokande's atmospheric neutrino data
are sensitive to MH. Preliminary result
was presented at Neutrino 2014~\cite{wendell_neutrino2014,Wendell:2014dka} and shown in
the right panel of Fig.~\ref{fig:sk_det}. The normal mass hierarchy
is slightly favored by $\Delta \chi^2 \sim 0.9$. When interpreted in
the Bayesian framework, this preliminary result corresponds to 
60\% (normal) vs. 40\% (inverted) probability in favor of the normal hierarchy.

%% file: dyb.tex
\subsubsection{Global Analysis of the $|\Delta m^2_{32}|$ Mass-squared Splitting}

\begin{figure}[H]
\begin{centering}
\includegraphics[width=0.49\textwidth]{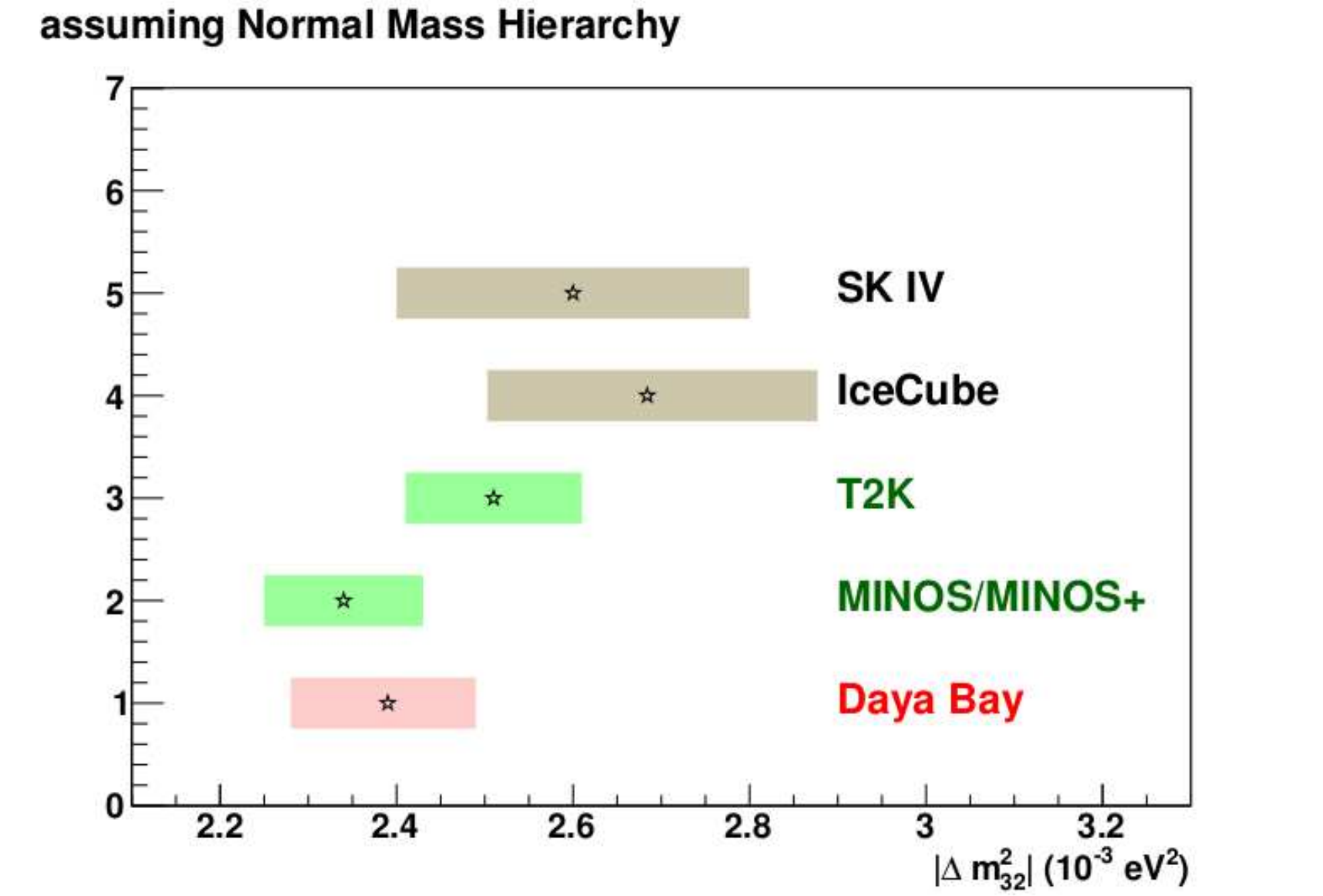}
\includegraphics[width=0.49\textwidth]{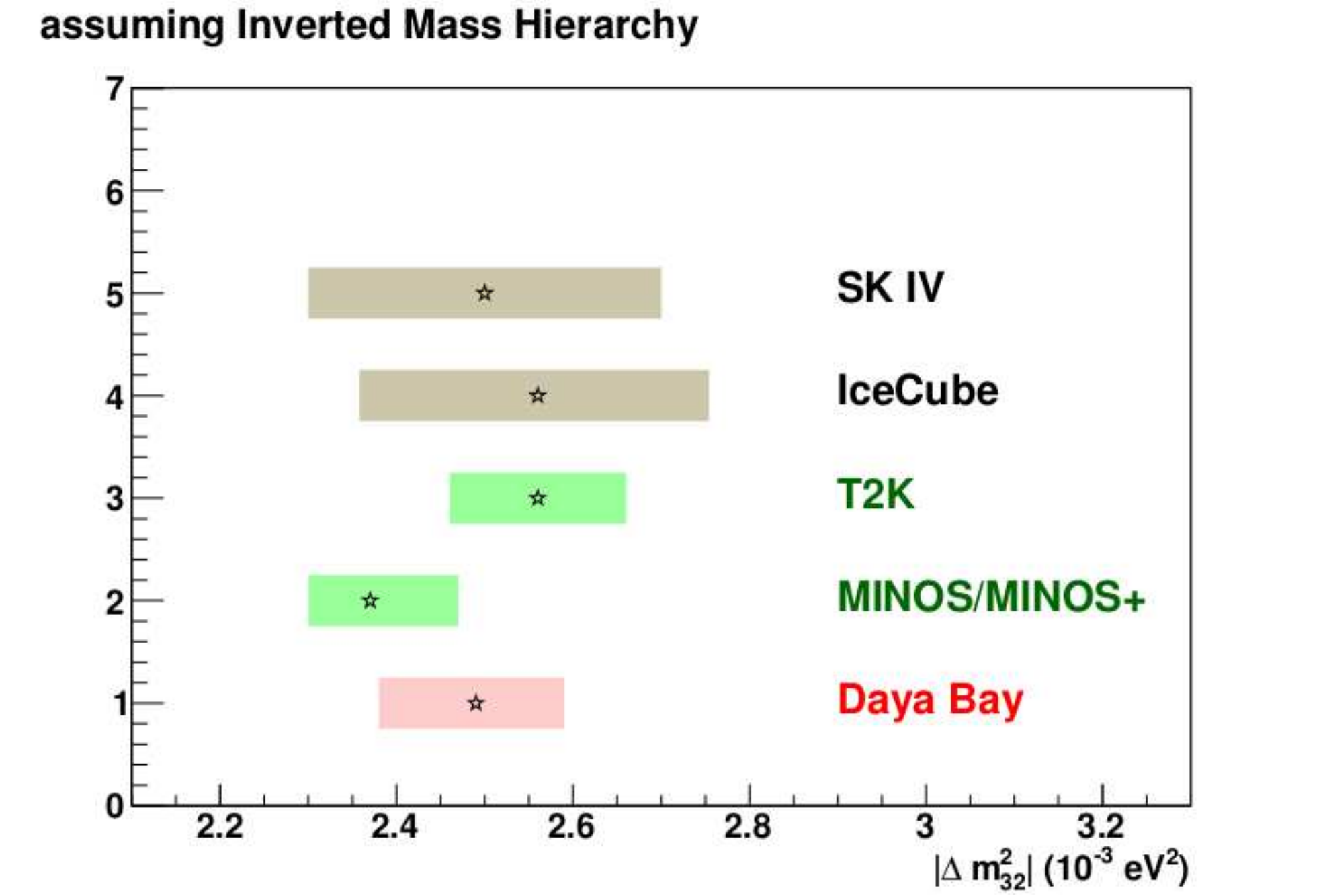}
\par\end{centering}
\caption{\label{fig:global_dm2} $|\Delta m^2_{32}|$ reported at the Neutrino 2014 
conference are summarized assuming normal (left) and inverted (right) mass hierarchy scenarios. While 
Daya Bay measures the $\bar{\nu}_e$ disappearance using the reactors, 
the rest of experiments measure the $\nu_{\mu}$ and/or $\bar{\nu}_{\mu}$ disappearance using accelerator 
neutrinos (T2K and MINOS/MINOS+) and  atmospheric neutrinos (Super-Kamiokande and 
IceCube).}
\end{figure}

As illustrated in Sec.~\ref{sec:reactor}, MH information
can be also extracted by comparing the measurement of the effective atmospheric
mass-squared splitting, i.e. the value of $\Delta m^2_{32}$, between the electron 
and muon neutrino disappearance channels. For the normal (inverted) mass hierarchy, 
the effective mass-squared splitting in electron neutrino disappearance channel is 
larger (smaller) than that in the muon neutrino disappearance channel. 
In other word, the $\Delta m^2_{32}$ value extracted from these two channels 
would be more consistent with each other when testing the true MH hypothesis than the incorrect MH hypothesis.
Figure~\ref{fig:global_dm2} shows the extracted $|\Delta m^2_{32}|$ from Super 
Kamiokande~\cite{wendell_neutrino2014} in Japan, 
IceCube~\cite{Hill_neutrino2014} at the South Pole, 
T2K~\cite{walter_neutrino2014} in Japan, MINOS/MINOS+~\cite{Sousa_neutrino2014} in the U.S., 
and Daya Bay~\cite{zhang_neutrino2014,Zhang:2015fya} in China
at the Neutrino 2014 conference under both normal and inverted MH scenarios. While 
Daya Bay measures the $\bar{\nu}_e$ disappearance from reactors, the rest of experiments
measure the $\nu_{\mu}$ and/or $\bar{\nu}_{\mu}$ disappearance. 
Figure~\ref{fig:global_dm2} shows that IH 
are slightly favored compared to NH. 
Assuming all results are uncorrelated, the fit of NH results yields a $\chi^2/NDF$ value of 4.4/4. On the other hand, the 
fit of IH results yield a $\chi^2/NDF$ value of 1.85/4. 
Therefore, the IH is favored by $\Delta \chi^2 \sim 2.55$, 
which corresponds to a probability ratio of 78\% vs. 22\% in the Bayesian framework.

%% file: nova.tex
\subsubsection{NO$\nu$A}

\begin{figure}[H]
\begin{centering}
\includegraphics[width=0.4\textwidth]{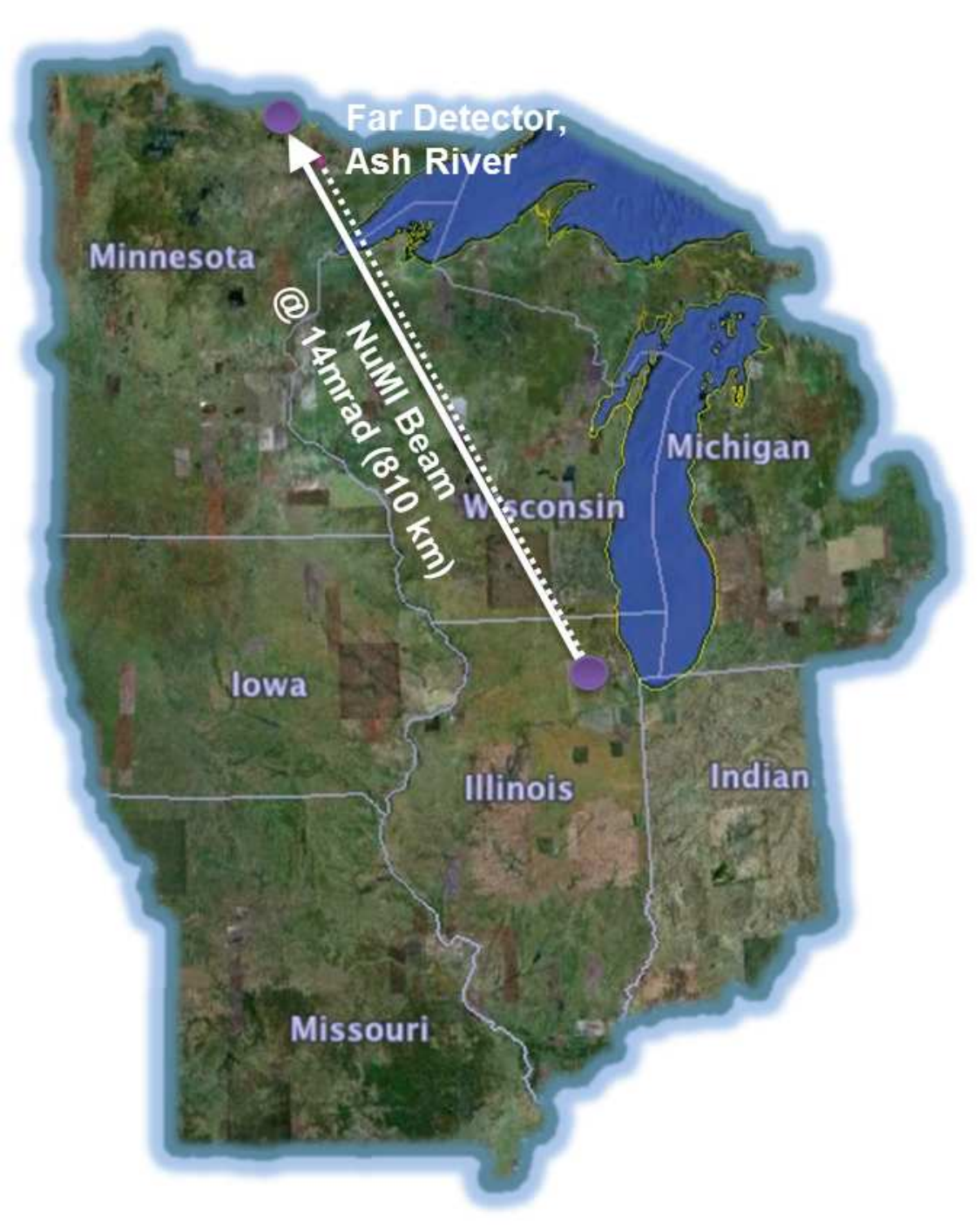}
\includegraphics[width=0.49\textwidth]{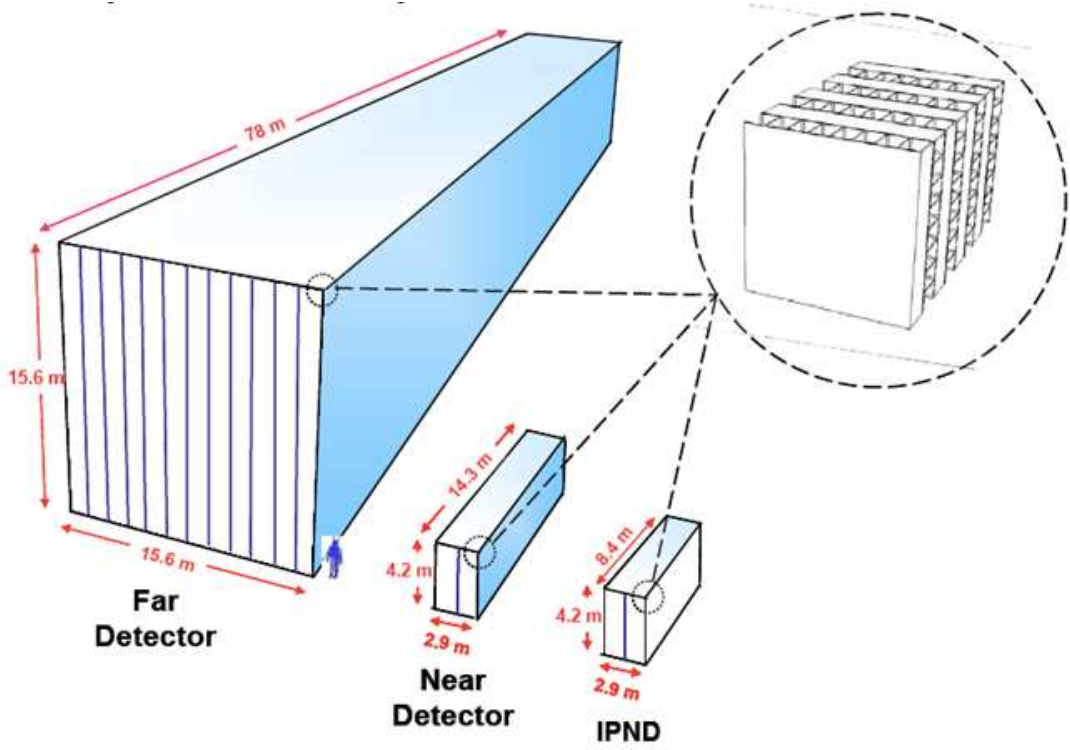}
\par\end{centering}
\caption{\label{fig:nova_conf} (Left) A map illustrating the NO$\nu$A experiment. 
(Right) The detector configuration of NO$\nu$A from Ref.~\protect\cite{nova_report}.
Reproduced with the permission from Ref.~\protect\cite{nova_neutrino2014}.}
\end{figure}

The NO$\nu$A (NuMI Off-axis $\nu_e$ Appearance) experiment~\cite{Ayres:2002ws} 
is an accelerator neutrino experiment in the United States. It is currently 
taking data. NO$\nu$A measures the $\nu_e$ ($\bar{\nu}_e$) appearance as well as the $\nu_{\mu}$ ($\bar{\nu}_{\mu}$)
disappearance in a $\nu_{\mu}$ ($\bar{\nu}_{\mu}$) beam. The neutrino beam is generated 
by the 120 GeV proton beam from the Fermilab's Main Injector. 
The system consists of two detectors, with the detector near the production point 
characterizing the neutrino beam and the far detector at the baseline of 
810 km studying neutrino oscillations. Left panel of Fig.~\ref{fig:nova_conf}
shows the overall configuration of the NO$\nu$A experiment. The far detector 
is located at 14 mrad off-axis of the incident proton beam to obtain a narrow band 
neutrino beam around 2 GeV energy.
The 810 km baseline corresponds to the maximal oscillation at 2 GeV.
In addition, the long baseline offers the experiment a chance to determine the 
mass hierarchy through the matter effect as discussed in Sec.~\ref{sec:acc}.

\begin{figure}[H]
\begin{centering}
\includegraphics[width=0.6\textwidth]{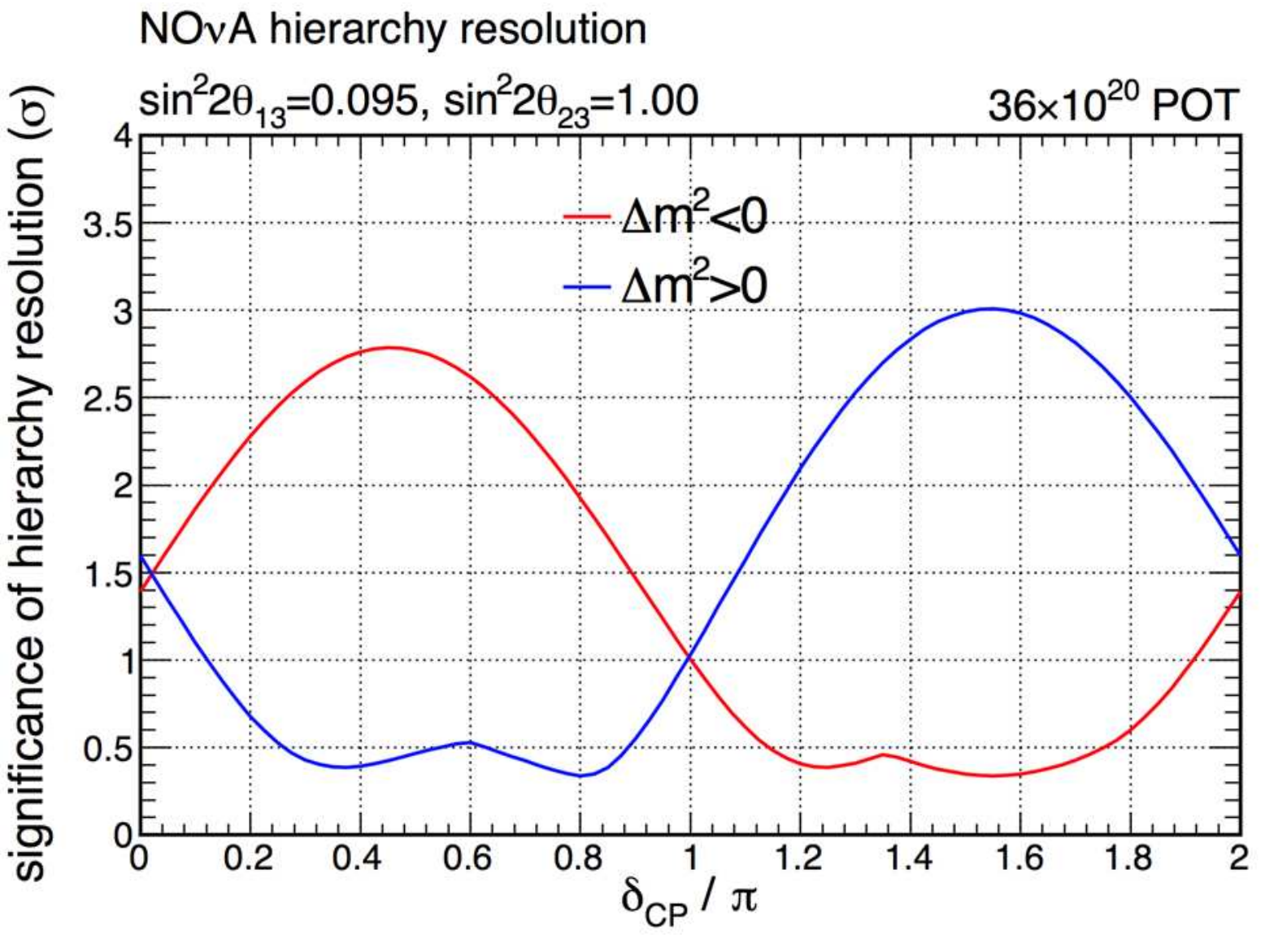}
\par\end{centering}
\caption{\label{fig:nova_sen} Mass hierarchy sensitivity from NO$\nu$A (taken from 
Ref.~\protect\cite{nova_neutrino2014}) is shown. The ``significance of hierarchy resolution''
is defined as $\sqrt{\Delta \chi^2}:=\sqrt{\overline{T}}$ as introduced in Sec.~\protect\ref{sec:stat}. }
\end{figure}

Right panel of Fig.~\ref{fig:nova_conf} shows the detector configuration of 
NO$\nu$A~\cite{nova_report}. The near (0.3 kt) and far (14 kt) detectors are 
structurally and functionally identical, despite the difference in their 
absolute sizes. Such a configuration allows to minimize the systematic uncertainties in 
the neutrino beam, the neutrino interaction cross section, and the detecting efficiency.
The near detector consists of 18 thousand PVC cells filled with liquid scintillator and arranged in 
alternating X/Y layers with an overall dimension of $14.3 \times 4.2 \times 4.2 ~\mbox{m}^3$.
The light is transported through a wavelength shifting fiber and then read out with avalanche 
photodiode. The far detector is approximately 60 times larger with 
the overall dimension of 
$60 \times 15.6 \times 15.6 ~\mbox{m}^3$. Since it is located on the surface, the far detector
is exposed to a large flux of cosmic muons. Nevertheless, a factor of 4$\times10^{7}$ suppression has 
been achieved to reject the cosmic ray background, taking into account the
timing and the direction of the $\nu_{\mu}$ and $\nu_e$ beam. 
Figure~\ref{fig:nova_sen} shows the expected MH sensitivity from NO$\nu$A 
after collecting a total of 36$\times10^{20}$  protons on target (POT), 
which corresponds to about six years (three years neutrino and three years antineutrino running) 
of data taking at 750 kW beam power. 
The mass hierarchy resolution has a large dependence on the value of the unknown 
CP phase $\delta_{CP}$. This is so because both the matter effect and the CP phase 
$\delta_{CP}$ can induce differences between neutrino and antineutrino appearance oscillations, which 
leads to a degeneracy between the MH determination and the determination of CP
phase $\delta_{CP}$ for certain values of $\delta_{CP}$. At the most favorable CP values, the sensitivity 
of MH determination can reach about 3$\sigma$. At this sensitivity, there is a 50\% chance that 
NO$\nu$A will be able to reject the unfavored MH hypothesis by 3$\sigma$. Meanwhile, 6.7\% of measurements will yield a result 
favoring the 
wrong MH hypothesis.

%% file: LBNF.tex
\subsubsection{DUNE/LBNF}
In this subsection the long baseline experiments that aim at the determination of the
neutrino mass hierarchy, among other things, are described. There were two 
proposals, the Long-Baseline Neutrino Experiment (LBNE)~\cite{Adams:2013qkq} in the US 
and the Large Apparatus for Grand Unification and Neutrino Astrophysics and Long 
Baseline Neutrino Oscillations (Laguna-LBNO or LBNO)~\cite{::2013kaa} in Europe, designed to 
separate the signal of the two hierarchies with matter effect. While the fundamental goal 
of the two proposals is similar, they differ in important details, in particular in the 
length of the baseline. Due to the limited available resource and expected costs of these 
projects, LBNE and LBNO collaborations combined efforts and merged into a new experiment 
located in the US: the Deep Underground Neutrino Experiment at the Long-Baseline Neutrino Facility (DUNE/LBNF). 
We concentrate on the DUNE experiment and will describe it in details. There is no firm 
schedule for DUNE at the present time. Tentatively, the construction of DUNE/LBNF might 
begin in 2017 and the neutrino beam might be ready by $\sim$2023.

\begin{figure}[H]
\begin{centering}
\includegraphics[width=0.9\textwidth]{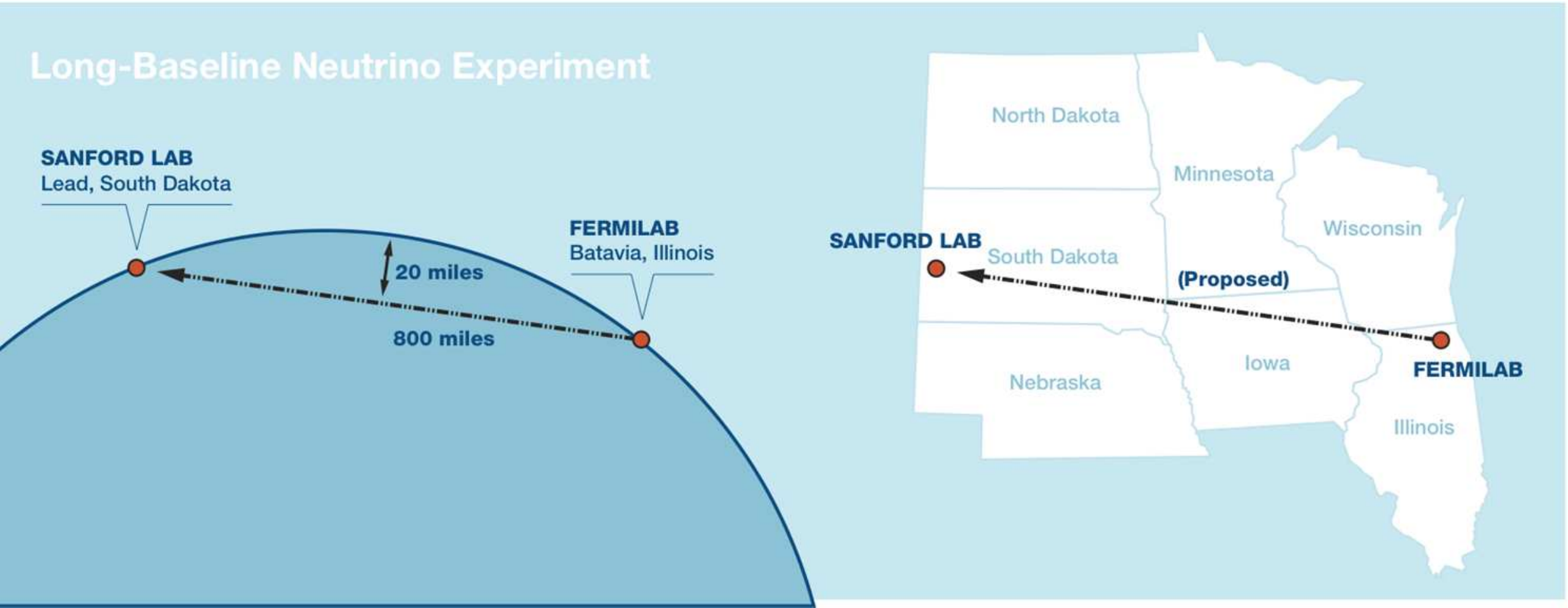}
\par\end{centering}
\caption{\label{fig:lbne_site} Illustration of the DUNE/LBNF (formerly known as the LBNE) 
experiment~\cite{lbne_setup}.}
\end{figure}

DUNE is an accelerator neutrino experiment under development in the United States. It consists 
of a horn-produced broad band beam with 60-120 GeV protons at $\sim$1 MW from Fermilab, 
$>$35 kton fiducial volume Liquid 
Argon Time Projection Chambers (LArTPC) 4850-feet underground at Sanford Laboratory
in Lead, South Dakota, as the far detector, and high-resolution near detector(s) at Fermilab. 
Figure~\ref{fig:lbne_site} shows the overall configuration of DUNE. In order to optimize the 
physics sensitivity to the CP phase $\delta_{CP}$, the baseline of DUNE is chosen to be 
1300 km~\cite{Bass:2013vcg}, which is longer than any existing accelerator neutrino experiments. 
The large underground LAr detector also represents an excellent opportunity for the underground science
including proton decay searches, detection of atmospheric neutrinos and supernova neutrino burst.
In particular, the detection of atmospheric neutrinos will provide additional information 
regarding MH. 

\begin{figure}[H]
\begin{centering}
\includegraphics[width=1.0\textwidth]{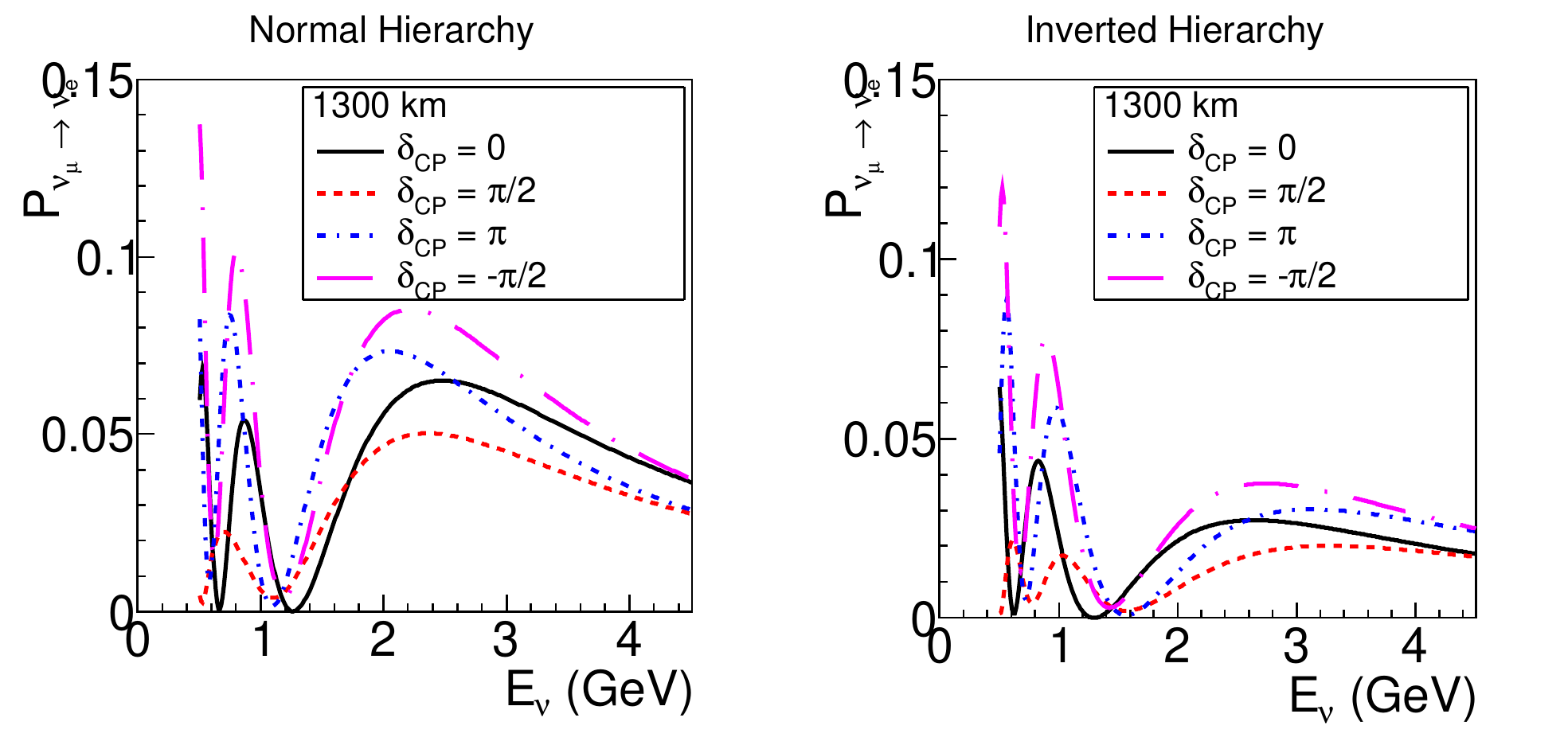}
\par\end{centering}
\caption{\label{fig:lbne_spec} The $\nu_{\mu}\rightarrow\nu_e$ oscillation probabilities
are shown for the normal (left) and the inverted (right) mass hierarchies. Different curves 
represent different values of currently unknown CP phase $\delta_{CP}$. The first and second 
oscillation maxima locate at about 2.5 and 0.8 GeV, respectively.
The matter density $\rho$ and the electron fraction $Y_e$ are assumed to be 
2.7 g/cm$^3$ and 0.5, respectively.}
\end{figure}

DUNE is designed to precisely measure parameters that govern $\nu_{\mu}\rightarrow \nu_e$, $\bar{\nu}_{\mu}\rightarrow \bar{\nu}_e$, 
$\nu_{\mu}\rightarrow \nu_{\mu}$, and $\bar{\nu}_{\mu}\rightarrow \bar{\nu}_{\mu}$
oscillations, including the second mixing angle $\theta_{23}$, the 
third mixing angle $\theta_{13}$, the CP phase $\delta_{CP}$, 
and MH. Figure~\ref{fig:lbne_spec} shows the $\nu_{\mu}\rightarrow\nu_e$ 
oscillation probabilities for different values of $\delta_{CP}$ for both normal and inverted
mass hierarchies. The principle of the MH determination in DUNE relies on the matter effect, 
introduced in Sec.~\ref{sec:acc}. For NH, the $\nu_e$ 
($\bar{\nu}_e$) appearance probability at first oscillation maximum ($\sim$2.5 GeV) is enhanced (suppressed),
while for IH, the opposite is true, i.e., 
the $\nu_e$ ($\bar{\nu}_e$) appearance probability at first 
oscillation maximum is suppressed (enhanced). At 1300 km, the matter effect in DUNE is large 
enough so that there is no fundamental degeneracy between the matter effect and the 
$\delta_{CP}$-induced CP effect. In other word, given enough statistics, these two effects 
can be totally disentangled.
Furthermore, the precision measurements of the oscillation 
probability will also enable a measurement of the currently unknown CP phase $\delta_{CP}$
that may hold the key to explain the large matter over anti-matter asymmetry in our 
universe~\cite{Fukugita:1986hr}.

\begin{figure}[H]
\begin{centering}
\includegraphics[width=1.0\textwidth]{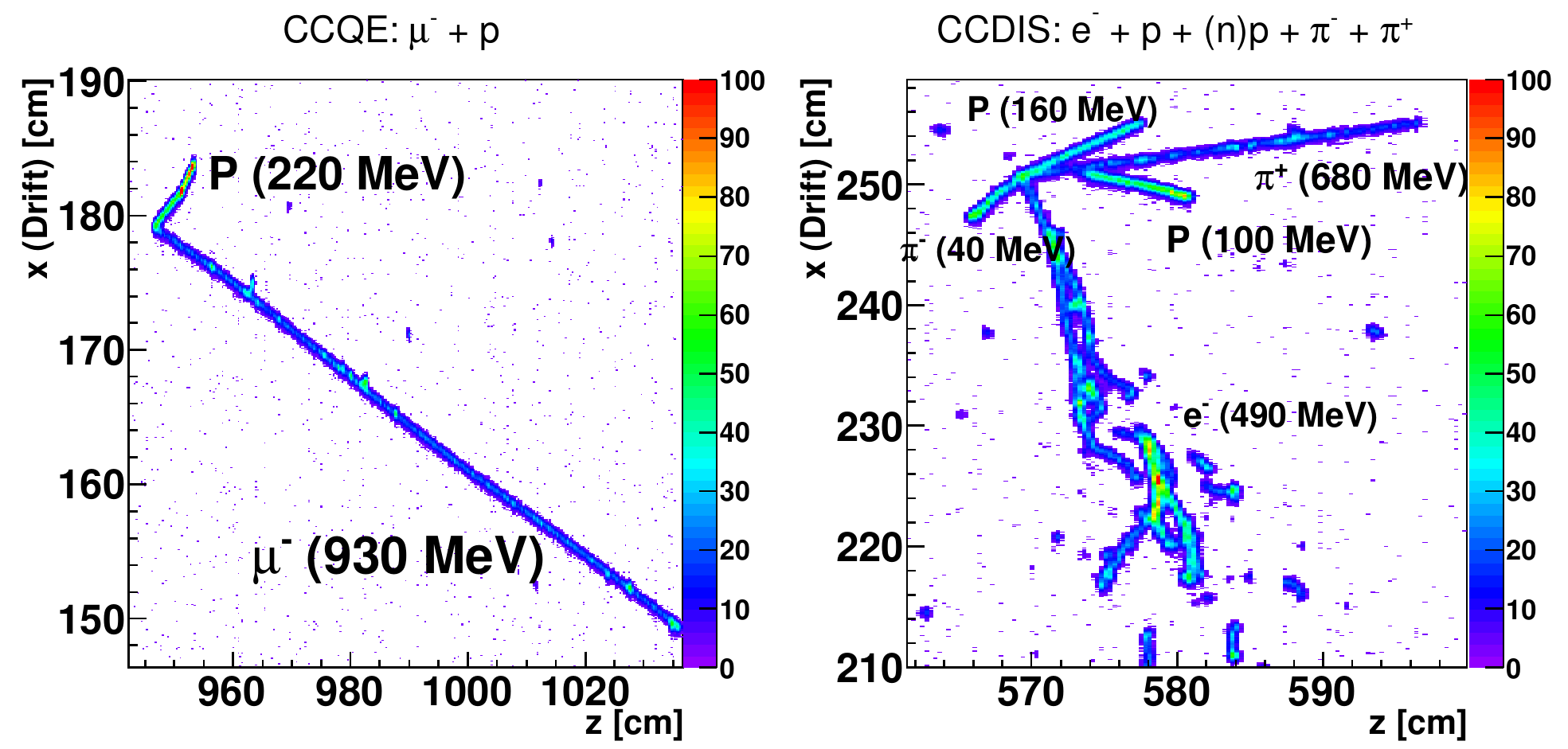}
\par\end{centering}
\caption{\label{fig:lar_ed} Event display of two neutrino events from Monte-Carlo. 
(Left) Charged-Current Quasi Elastic (CCQE) 
interaction with final states containing a muon and a proton. 
(Right) Charged-Current Deep-Inelastic-Scattering 
(CCDIS) with final states containing an electron shower, two protons 
(one being generated by neutron final state
interaction), one positive pion, and one negative pion. }
\end{figure}

The detector technology in DUNE is chosen to be the liquid argon time projection chamber 
(LArTPC)~\cite{Willis:1974gi}. This technology was firstly proposed by C. Rubbia for 
the neutrino physics~\cite{lartpc} in 1977. 
As the most abundant noble gas  
in the atmosphere ($\sim$1.3\% by weight), argon can be extracted economically, and is an ideal material 
for the TPC.  Electrons have very high mobility in LAr~\cite{Shibamura:1975zz}.
At 500 V/cm electric field and 87 K temperature, the electron drift velocity is about 
1.6 km/s~\cite{Kalinin,Walkowiak:2000wf,Amoruso:2004ti,Aprile:1985xz},  
which is equivalent to about 0.64 ms for electrons to travel 1 m.  
Electrons have low diffusions in LAr~\cite{Derenzo:1974ji}; 
the longitudinal and transverse diffusion coefficients are about 5 and 13 
cm$^2$/s~\cite{shibamura,derenzo,Cennini:1994ha,atrazhev}, respectively at 500 V/cm electric field.  
For a 1.6 ms drift time, these numbers correspond to a position resolution (one standard deviation) 
of about 1.3 and 2.0 mm  for longitudinal and transverse diffusions, 
respectively.  LAr also has relatively high density ($\sim$1.4 g/mL at 87K), advantageous 
for neutrino experiments.  

LAr is also a good 
scintillation material.  The scintillation light (emission peaks at $\sim$128 nm~\cite{Boulay}) 
can provide the absolute event time to determine the drift distance along the electric field 
direction.  The coordinate information perpendicular to the electron drift direction can be 
determined with a high resolution detector (e.g. wire planes).  The spatial resolution 
could reach a few mm.  Finally, the ionization energy required 
to produce an electron-ion pair is about 23.6 eV~\cite{Shibamura:1975zz,Miyajima}.  
Some of the pairs will recombine in LAr, and the remaining electrons will be collected at 
wire planes.  The charge collected per wire will be closely related to the energy deposition 
per unit distance (dE/dx), which can be used for the particle identification (PID).  In addition, 
the event topology can be used for PID as well.  LArTPCs are excellent detectors of neutrino 
interactions.   Fig.~\ref{fig:lar_ed} shows event displays for 
two neutrino events from Monte Carlo as an example.  The high-resolution nature of LArTPC is self-evident.

\begin{figure}[H]
\begin{centering}
\includegraphics[width=0.49\textwidth]{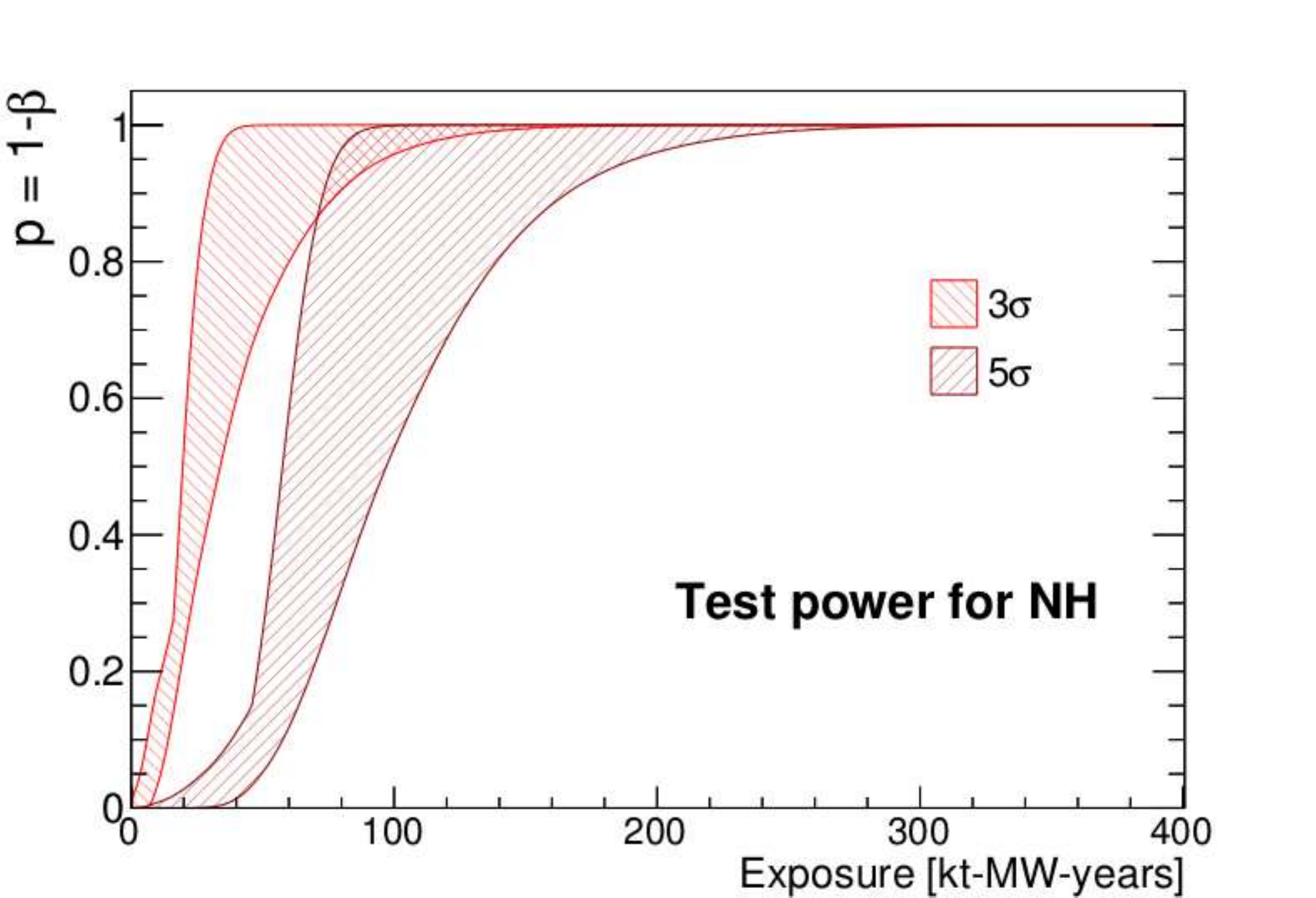}
\includegraphics[width=0.49\textwidth]{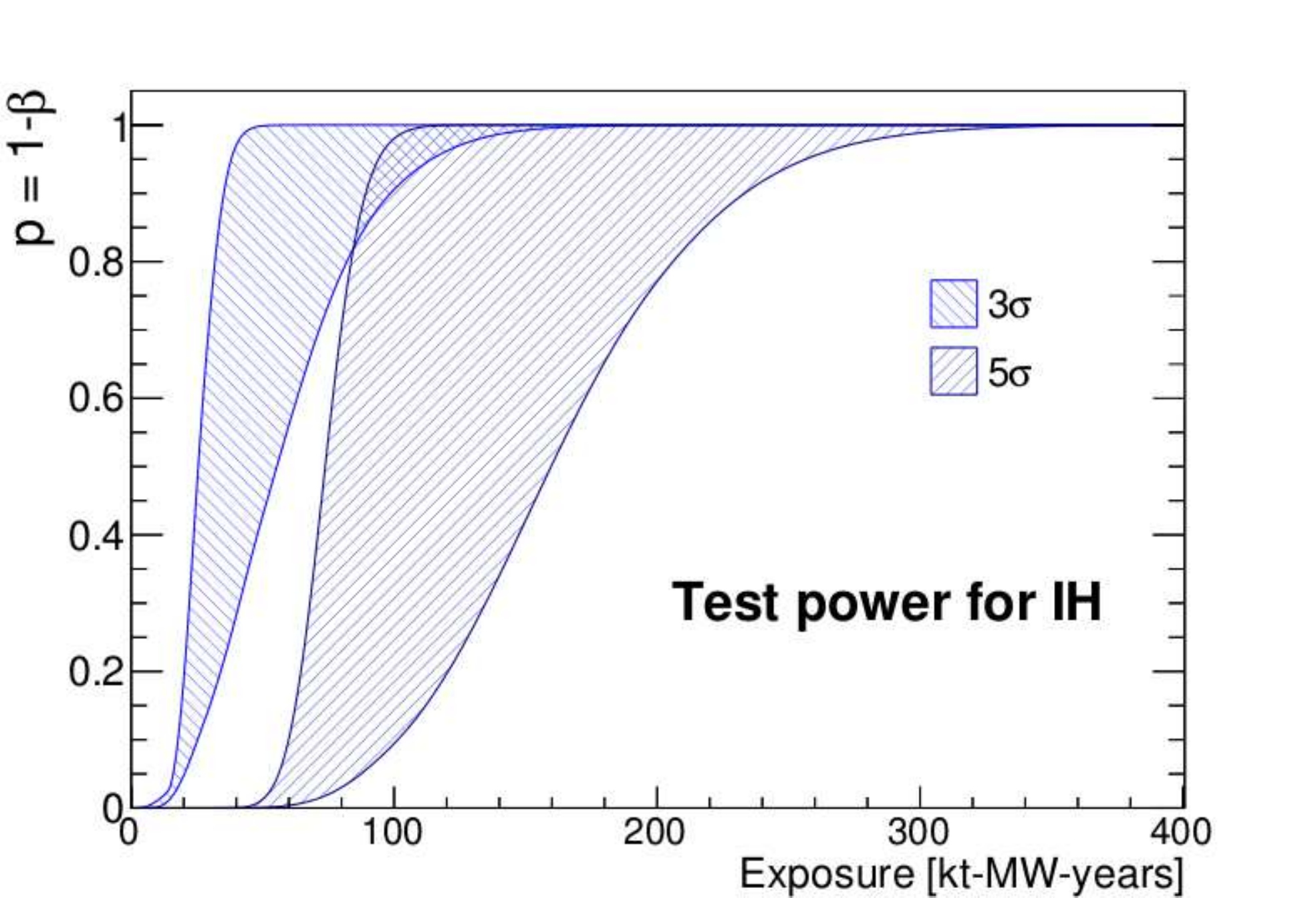}
\par\end{centering}
\caption{\label{fig:lbne_mh} The test power
($p=1-\beta$ as introduced in Sec.~\protect\ref{sec:stat}), which represents the probability of accepting 
the correct (NH) hypothesis while 
excluding the incorrect (IH) hypothesis at the 3 and 5$\sigma$ level, shown as a function of exposure in
kt-MW-years~\cite{lbnf_loi}. The width of the bands represent the range of CP values. 
Sensitivities are for true NH (left) and true IH (right). }
\end{figure}

\begin{figure}[H]
\begin{centering}
\includegraphics[width=0.49\textwidth]{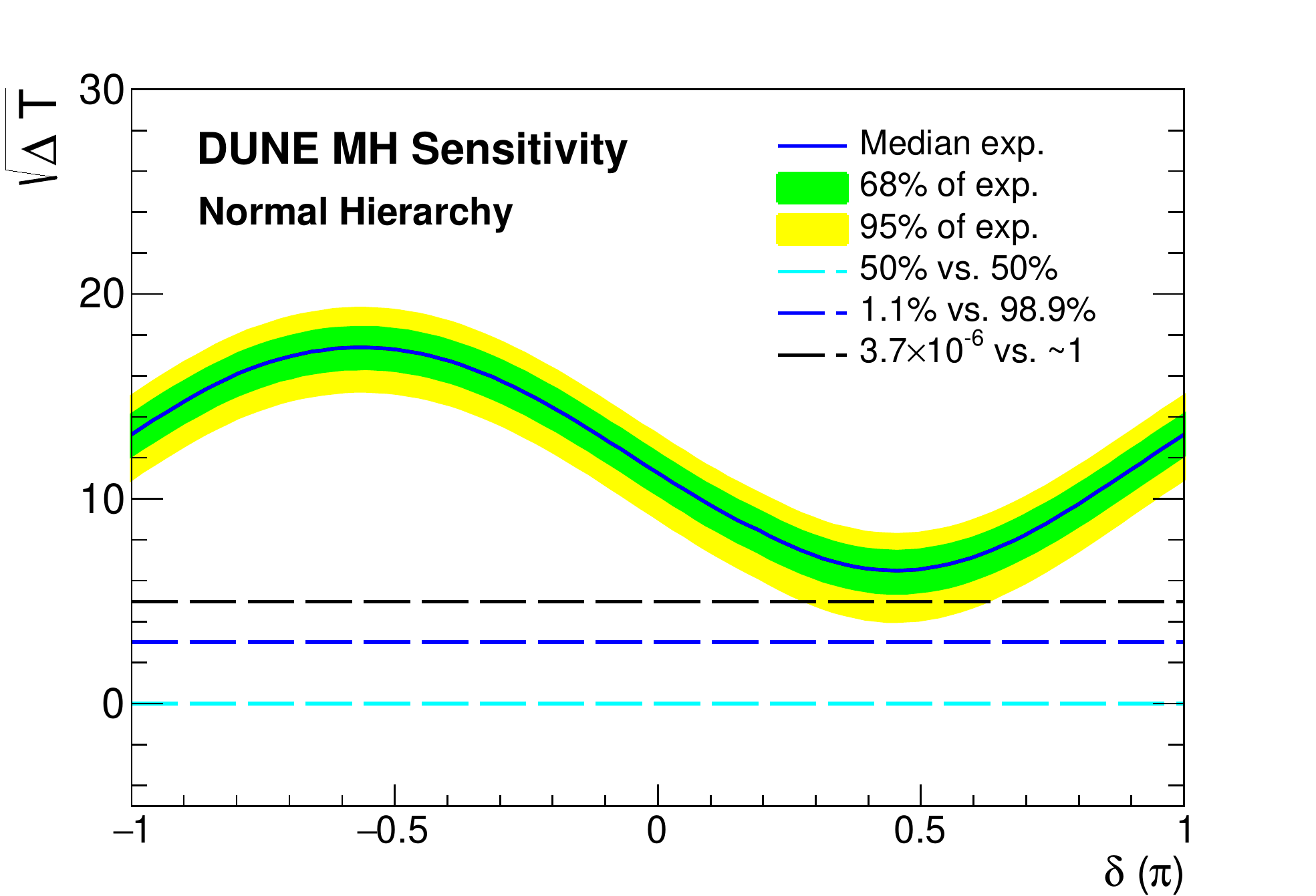}
\includegraphics[width=0.49\textwidth]{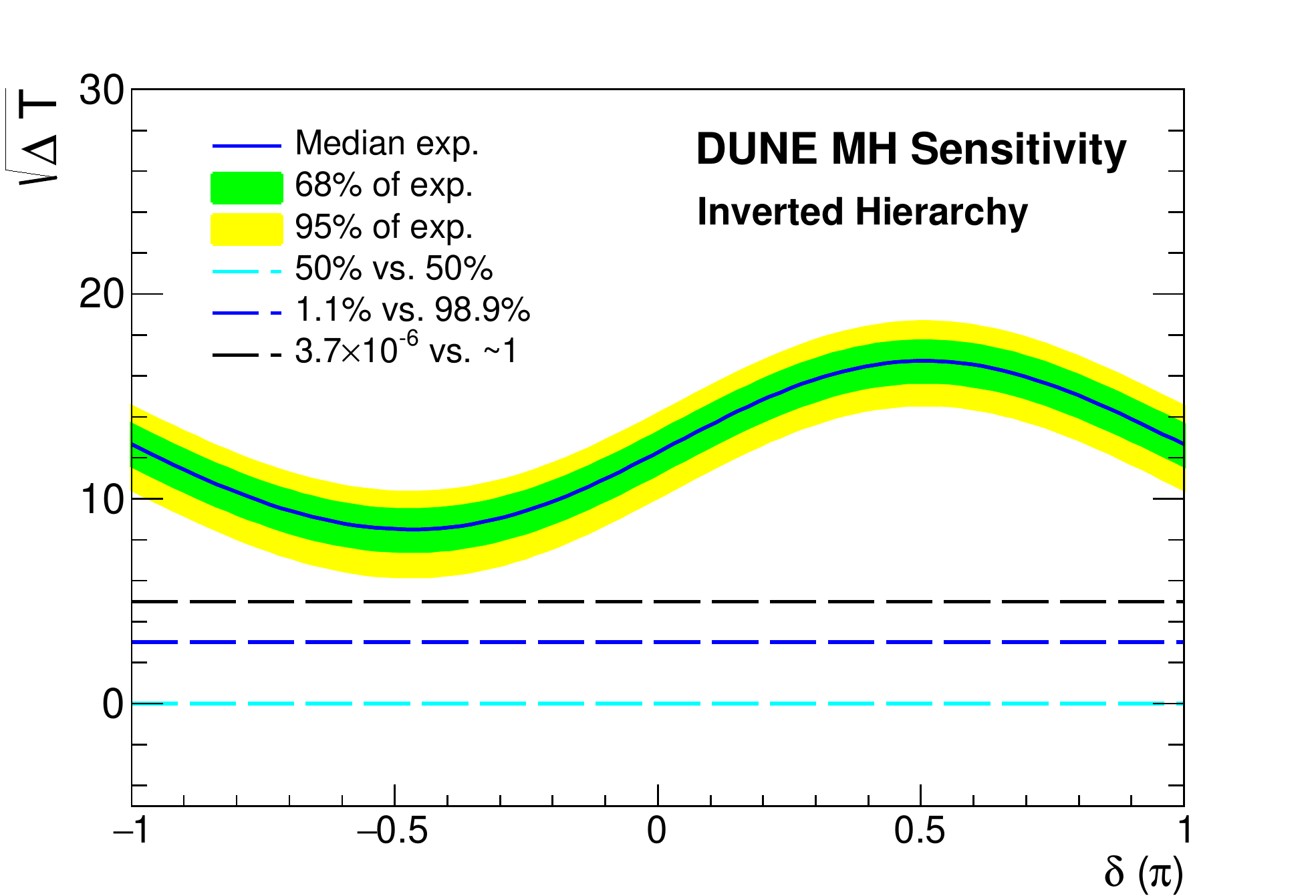}
\par\end{centering}
\caption{\label{fig:lbne_mhcp} The sensitivity for a typical experiment (blue line) is
compared to the bands within which 68\% (green) and 95\% (yellow) of experiments are expected
to fall~\cite{lbnf_loi}. The dashed lines represent the value of the $\sqrt{\Delta T}$ metric 
for which the probability of determining the correct MH is 50\% (cyan), 98.9\% (blue), 
or $3.7\times10^{-6}$ (black) in the Bayesian framework. The normal and inverted hierarchy are assumed 
to be true in the left and right plots, respectively.}
\end{figure}

In order to suppress the uncertainties in neutrino interaction cross sections and the 
neutrino flux, a fine-grained neutrino detector will be installed at FNAL near the beam source. 
The near detector will measure the absolute neutrino flux
via i) neutral-current $\nu_{\mu}+e^{-}\rightarrow \nu_{\mu} + e^-$ elastic scattering, ii) inverse-muon
decay $\nu_{\mu} + e^{-} \rightarrow \mu^- + \nu_{e}$ for the high energy neutrino component, 
and iii) charged-current quasi-elastic scattering $\nu_{\mu} + n \rightarrow \mu^- + p$ and 
$\bar{\nu}_{\mu} + p \rightarrow \mu^+ + n$. In addition, it will also measure the
spectra of all four neutrino species, $\nu_{\mu}$, $\bar{\nu}_{\mu}$, $\nu_e$, 
and $\bar{\nu}_{e}$ 
through the low-$\nu_0$ method~\cite{lownu,Bodek:2012uu} to 
enable an accurate prediction for each species of the far/near flux ratio as a function of energy.
The fine grained detector will be able to precisely measure the four-momenta of secondary hadrons, 
such as the charged and neutral mesons produced, and to constrain neutrino interaction cross sections, which are 
crucial inputs to the oscillation analysis. Furthermore, the precise measurement of these hadrons produced 
in the neutral- and charged-current interactions will significantly improve the background estimation 
in the oscillation signal.


Fig.~\ref{fig:lbne_mh} shows the expected sensitivity of the MH determination as a function of the 
accumulated exposure in units of kt-MW-years. As introduced in Sec.~\ref{sec:freq}, the test
power $1-\beta$ is plotted at values of {\it type-I error} $\alpha$ corresponding to 3 and 5$\sigma$
significance. Fig.~\ref{fig:lbne_mhcp} show the sensitivity in terms of test statistic $\Delta T$ as 
introduced in Sec.~\ref{sec:bayes}. At 300 kt-MW-years exposure, most of experiments will yield a MH
determination better than $3.7\times10^{-6}$ vs. 1 in the Bayesian framework.

\begin{figure}[H]
\begin{centering}
\includegraphics[width=0.9\textwidth]{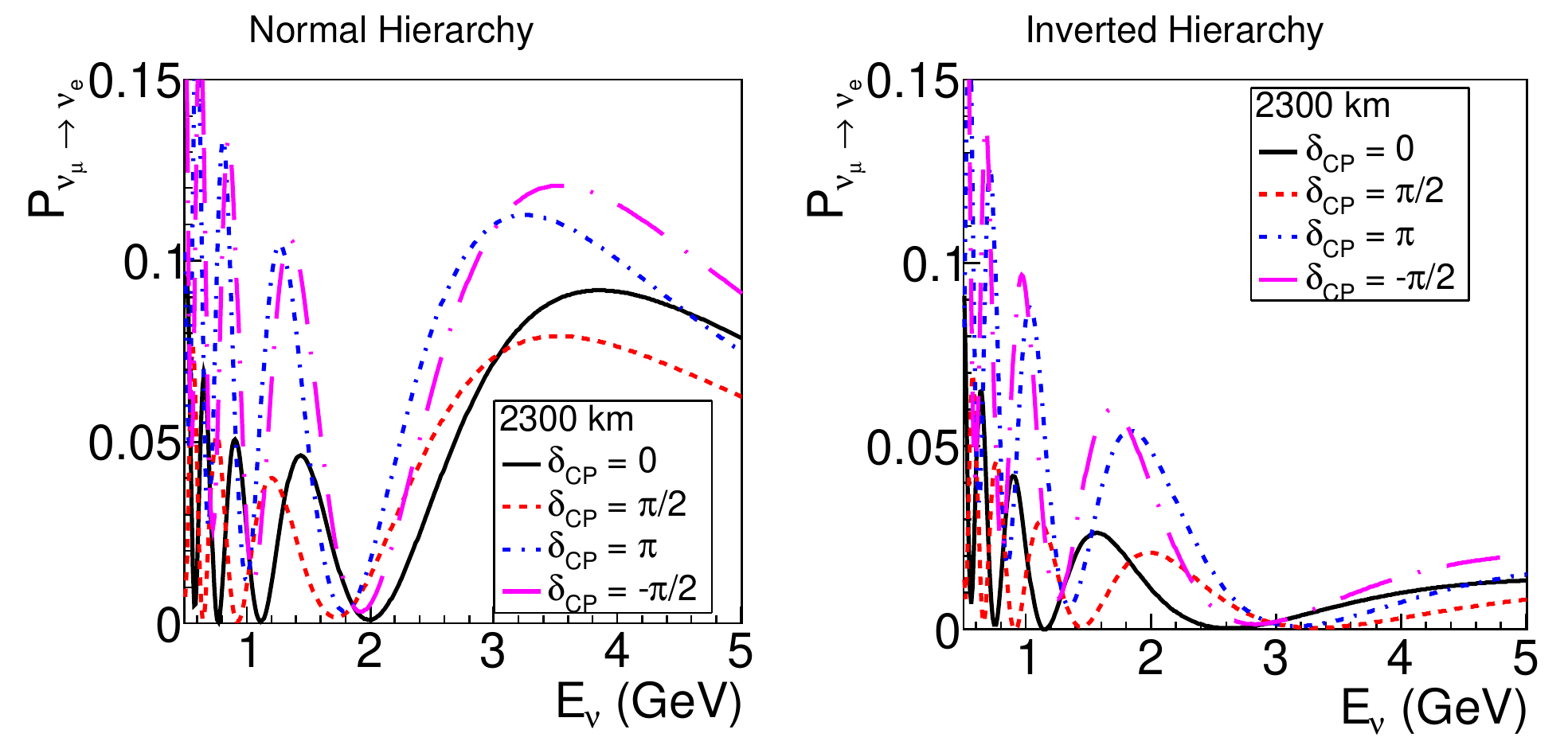}
\par\end{centering}
\caption{\label{fig:lbno_spec} The expected $\nu_{\mu}\rightarrow \nu_e$ oscillation patterns in LBNO
at a baseline of 2300 km. The matter density $\rho$ and the electron fraction $Y_e$ are assumed to be 
3.2 g/cm$^3$ and 0.5, respectively.
}
\end{figure}

LBNO, before merging into DUNE/LBNF with LBNE, was a proposed experiment in Europe.  
The initial neutrino beam is expected to be produced from 
CERN with a 400 GeV proton beam with a power of 750 kW. The primary choice of the far detector location is the
Pyh\"{a}salmi mine in Finland. Various other sites are also under consideration. 
A multi-tens of kilotons double-phase liquid argon time projection chamber is 
expected to reside deep underground. The baseline from CERN to Pyh\"{a}salmi is 2300 km, 
much longer than that of LBNE (1300 km). The $\nu_{\mu}\rightarrow \nu_e$ oscillation patterns are shown in 
Fig.~\ref{fig:lbno_spec}. Comparing to those of 1300 km (Fig.~\ref{fig:lbne_spec}), the matter effect 
(leading to the MH determination) at 2300 km 
is better due to longer baseline and corresponding higher neutrino energies. 
Therefore, the MH sensitivity is 
expected to be much larger than that of 1300 km. Given a similar overall exposure in kt-MW-years, 
it has a factor of $\sim$ 4 increase in the test statistics $\overline{\Delta T}$.


%% file: PINGU.tex
\subsubsection{ PINGU, ORCA, INO and Hyper Kamiokande}

As explained in Sec.~\ref{sec:atm}, the study of atmospheric neutrinos with
very large detectors may be used to determine the neutrino mass hierarchy.  The long
baseline of the upward going  atmospheric neutrinos and the corresponding large matter effect,
make this program possible. We briefly describe three proposed experiments that
are aiming, among other things, to resolve MH.

The Precision IceCube Next Generation Upgrade (PINGU) is a proposed new multi megaton 
array for IceCube upgrade at the South Pole Station. Its primary purpose is the determination 
of neutrino mass hierarchy using the detection of atmospheric neutrinos. The initial 
design will have a dense array of 40 strings (20 m of average string spacing) with 60 
optical modules each (5 m module spacing on the string), located in the Deep Core region 
of the IceCube detector. This will allow reconstruction of the Cherenkov tracks of events, 
i.e. of the energy and the zenith angle, for a relatively low energy region, 5 - 15 GeV, where 
the effect of MH is most evident. Note that the zenith angle determination is equivalent 
to the neutrino path length determination. The initial design of the array, and description 
of the corresponding simulation can be found in Ref.~\cite{Pingu}. The PINGU detector will 
also improve other measurements of neutrino oscillation parameters, in particular of the 
mixing angle $\theta_{23}$. It will be the world's largest detector of the possible supernova 
neutrino burst. It will also be able to search for the indirect signal from the WIMP dark matter
annihilation or decay in the Sun or Galactic Center. In a favorable scenario, PINGU installation
might be completed in 2021 or 2022.

Although the atmospheric flux contains both neutrinos and antineutrinos that are difficult to 
separate experimentally, as well as both $\nu_{\mu}$ and $\nu_e$, the large size of the PINGU 
detector, and the correspondingly large event statistics, makes it possible to overcome these challenges. At the relatively low energies, most of the secondary particles will be 
contained in the detector volume. Simulation indicates that $\sim$ 90\% of the charged-current 
$\nu_e$ and $\nu_{\mu}$ induced processes, and of the neutral current events, can be identified. 

\begin{figure}[H]
\begin{centering}
\includegraphics[width=0.90\textwidth]{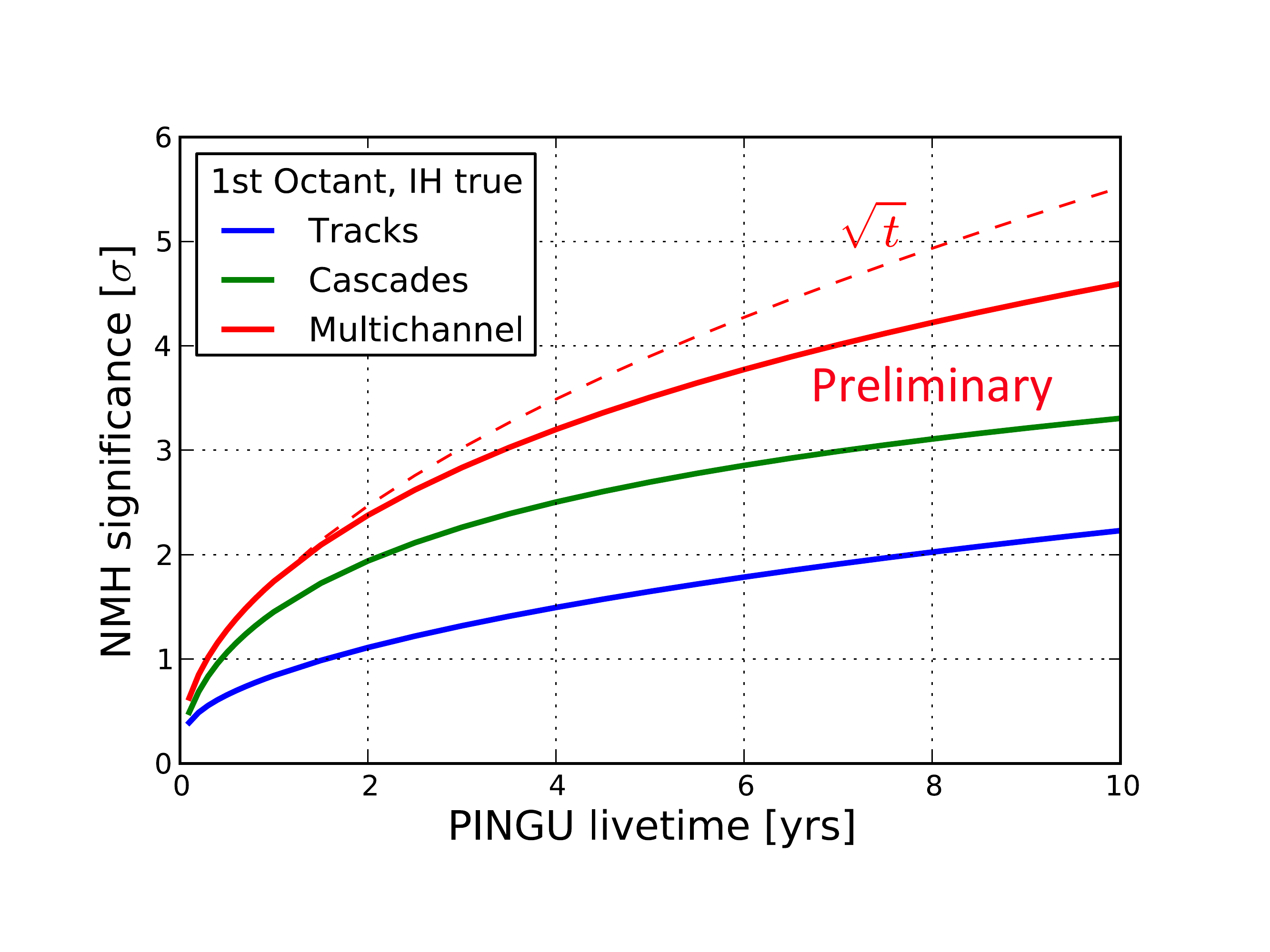}
\par\end{centering}
\caption{\label{fig:pingu} 
MH sensitivity of PINGU vs. the running time is shown.
The blue line (tracks) is based on the identification of the muon track events and the
green line is based on the hadronic and electron events. The red line represents the
total sensitivity. For comparison the dashed red line show the ideal scaling without
systematics. Reproduced with permission from Ref.~\protect\cite{Pingu}.}
\end{figure}

As shown in Sec.~\ref{sec:atm}, it turns out that the survival probabilities for a given $E_{\nu}$ 
and $\cos \theta_{zenith}$ for the $\nu_{\mu}$ in the case of NH is essentially the 
same as the probability 
for $\bar{\nu}_{\mu}$ with IH, and vice versa. This appears to be an insurmountable 
problem.  However, the differences in the cross section and kinematics of the $\nu$ and $\bar{\nu}$ 
interaction with nuclei, as well as the differences in the atmospheric $\nu$ and $\bar{\nu}$ fluxes,
makes it possible to separate the two hierarchies. As described in Ref.~\cite{Pingu2}, it is possible 
to identify the ``distinguishability" region in the space of  $E_{\nu}$ and $\cos \theta_{zenith}$ where 
the effect of MH is most pronounced. It is these regions where the most useful information of the MH can 
be extracted.

Simulation indicates that the PINGU sensitivity to the MH will reach a 3$\sigma$ significance with 
about 3.5 years of data collection. In Fig.~\ref{fig:pingu},
the preliminary plot of the MH significance versus the PINGU detector running time is shown. Note that 
only when all available information is combined a significant determination of MH is possible.

ORCA (Oscillation Research with Cosmics in the Abyss)~\cite{Katz:2014tta} is a proposed 
option aiming to determine the MH using the deep-sea neutrino telescope technology developed for 
the KM3NeT project (A multi-km$^3$ sized Neutrino Telescope)~\cite{KM3NeT}. The experimental principle
of ORCA is similar to that of PINGU. Instead of deep ice in the south pole, ORCA is expected to deploy 
large 3-dimensional arrays of photo-sensors to detect Cherenkov lights in the deep Mediterranean Sea.
A 3-5$\sigma$ MH sensitivity is targeted for a $\sim$20 Mton$\cdot$year overall exposure given
the neutrino detecting threshold around 5 GeV.

\begin{figure}[H]
\begin{centering}
\includegraphics[width=0.49\textwidth]{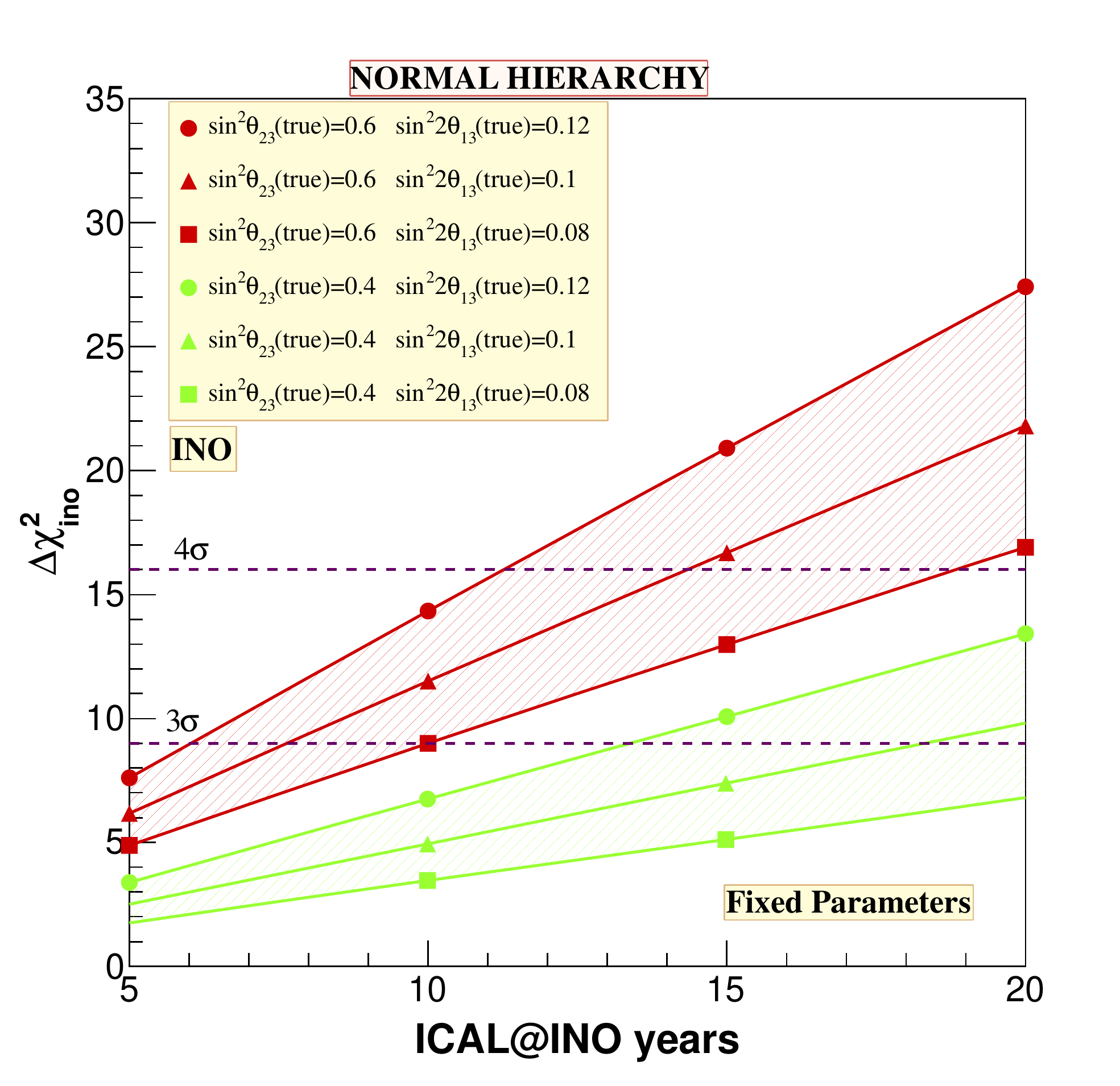}
\includegraphics[width=0.49\textwidth]{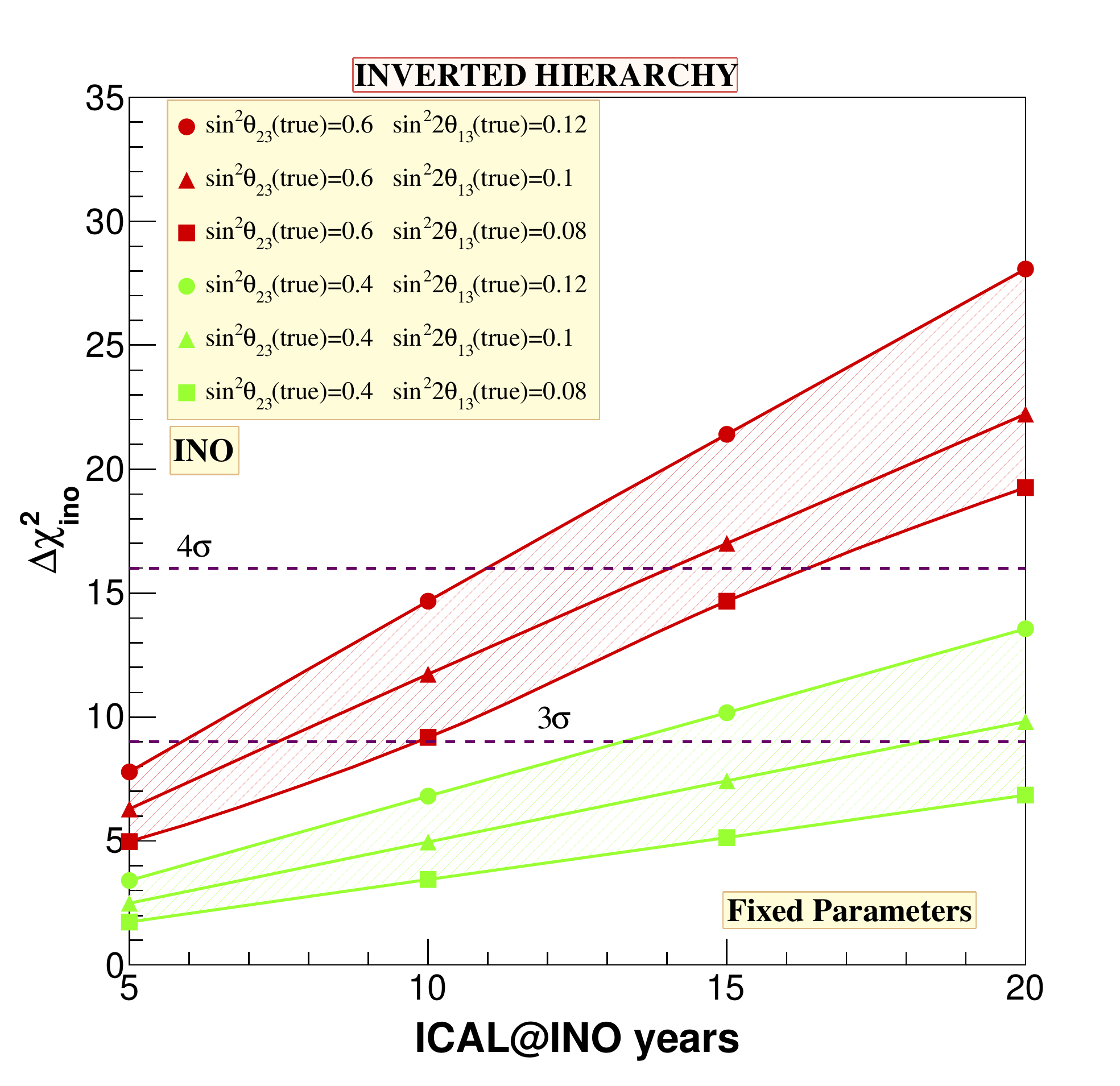}
\par\end{centering}
\caption{\label{fig:ino} 
MH sensitivities of INO with respect to the running time are shown as 
$\Delta \chi^2_{ino}:=\overline{\Delta T}$. Left and right panels assume 
the true MH is normal and inverted, respectively. Different lines represent 
different values of $\sin^2\theta_{23}$ and $\sin^22\theta_{13}$. During the 
analysis, all the oscillation parameters are fixed at their input true values. 
Reproduced with permission from Ref.~\protect\cite{Ghosh:2012px}. }
\end{figure}

The India-based Neutrino Observatory (INO)~\cite{ino,Ahmed:2015jtv} is a planned underground 
experiment at Pottipuram, Theni (India) to study the atmospheric neutrinos. The 
project has been approved by the Indian government with the tentative
start of data taking by 2020. The 
proposed detector is a magnetized iron calorimeter with 52 kiloton mass and
1.5 Tesla magnetic field, divided into three detector modules. Each module contains
151 layers of 5.6 cm iron plates with the resistive plate chambers (RPC) as active 
detector elements in the gaps between the plates. INO will have excellent muon energy 
and direction resolutions as well as a charge reconstruction capability. 
Besides MH, INO will be also sensitive to $\theta_{23}$ and $\Delta m^2_{32}$. 
Fig.~\ref{fig:ino} shows the projected MH sensitivity of INO vs. data taking time~\cite{Ghosh:2012px}. 
An improved analysis taking into account the hadron energy reconstruction together with
all oscillation parameters marginalized, show an enhanced sensitivity~\cite{Devi:2014yaa}.
The $\overline{\Delta T}$ is expected to reach 9.5 with 10 years of data taking for 
$\sin^22\theta_{13}=0.1$ and $\sin^2\theta_{23}=0.5$~\cite{Devi:2014yaa}.
Note that the MH sensitivity strongly depends on the true values of $\theta_{23}$ and 
$\theta_{13}$ which, however, might be precisely known by the time the detector begins 
data taking. The MH sensitivity is significantly larger if the mixing angle $\theta_{23}$ is 
in the second octant, i.e. if $\sin^2 \theta_{23} > 0.5$. The same also applies to
MH sensitivity in PINGU.

\begin{figure}[H]
\begin{centering}
\includegraphics[width=0.6\textwidth]{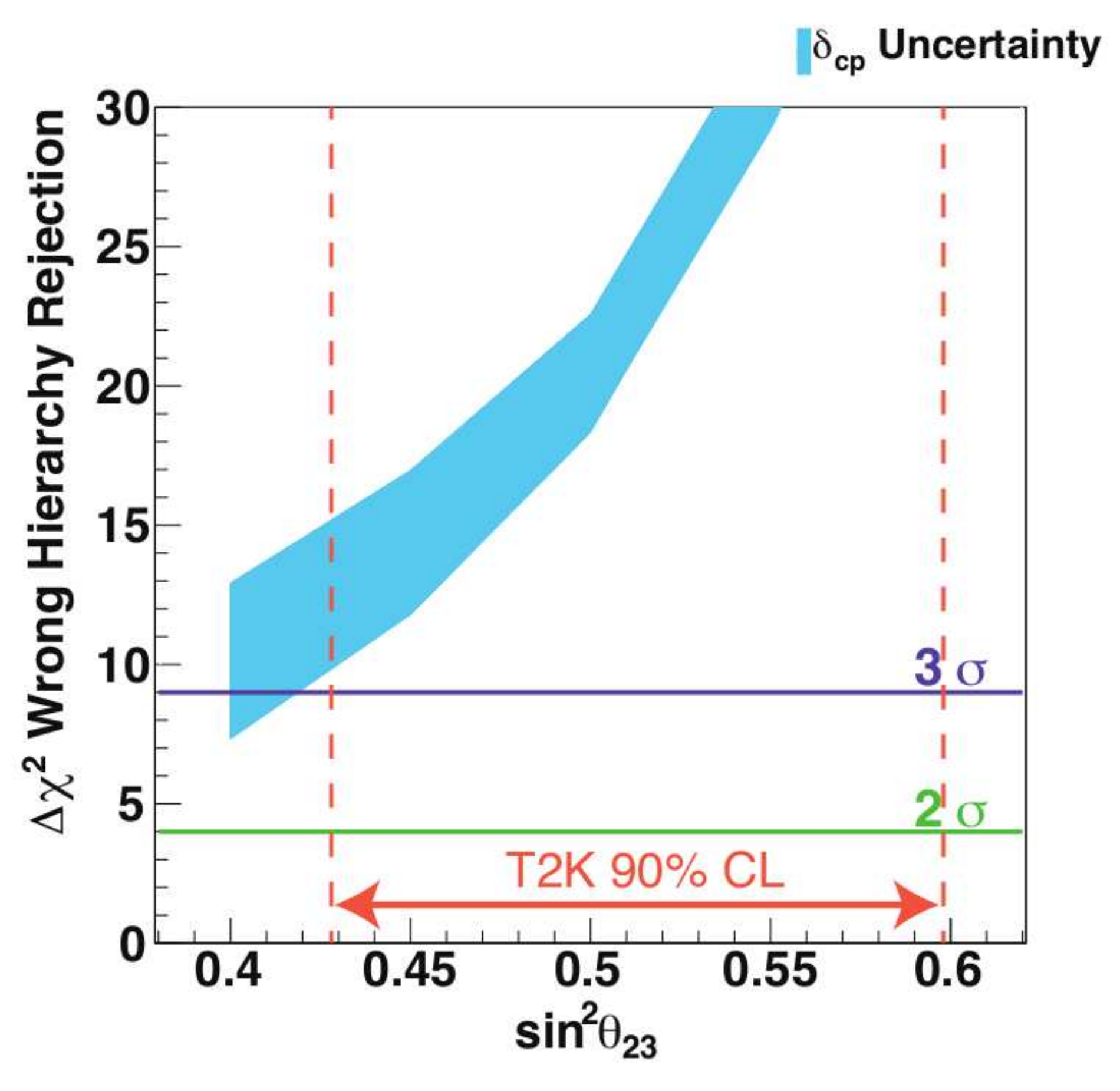}
\par\end{centering}
\caption{\label{fig:hk} 
The MH sensitivity of Hyper-Kamiokande with respect to the true value of $\sin^2\theta_{23}$
is shown as $\Delta \chi^2:=\overline{\Delta T}$. The blue band represents the uncertainty
of the currently unknown CP phase $\delta_{CP}$. A ten-years exposure is assumed. 
Reproduced with permission from Ref.~\protect\cite{Abe:2014oxa}.}
\end{figure}

Hyper-Kamiokande~\cite{Abe:2014oxa} is a proposed experiment  near the site of
the previous Kamiokande and the current Super-Kamiokande detectors in Japan. The 
master plan at present suggests that the construction and the operation 
might begin in 2018 and 2025, respectively. 
The detector technology will be water Cherenkov similar as in Super-Kamiokande, 
only many times larger, with a total mass of 1 Megatons (560 kiloton fiducial mass). 
While the main experimental program at Hyper-Kamiokande is the precision 
measurement of $\nu_{\mu} \rightarrow \nu_{e}$ and $\bar{\nu}_{\mu} \rightarrow \bar{\nu}_{e}$ oscillations with 
accelerator neutrinos at a relative short baseline (295 km) to search for CP violation 
in the lepton sector, Hyper Kamiokande is also sensitive to MH with atmospheric neutrinos. Fig.~\ref{fig:hk}
shows the atmospheric neutrino MH sensitivity with ten years of data taking. 
Again, the MH sensitivity is significantly larger if the mixing angle $\theta_{23}$ 
happens to be in the second octant. 

%% file: juno.tex
\subsubsection{Medium baseline reactor experiments: 
JUNO and RENO-50} 

As described in the Sec.~\ref{sec:reactor}, reactor experiments at baselines $\approx$ 60 km are 
well suited for the MH determination. Two such experiments have been proposed and are currently at 
various stages of preparation and construction. Here we describe them in more details, 
using the JUNO experiment as the main example.

\begin{figure}[H]
\begin{centering}
\includegraphics[width=0.7\textwidth]{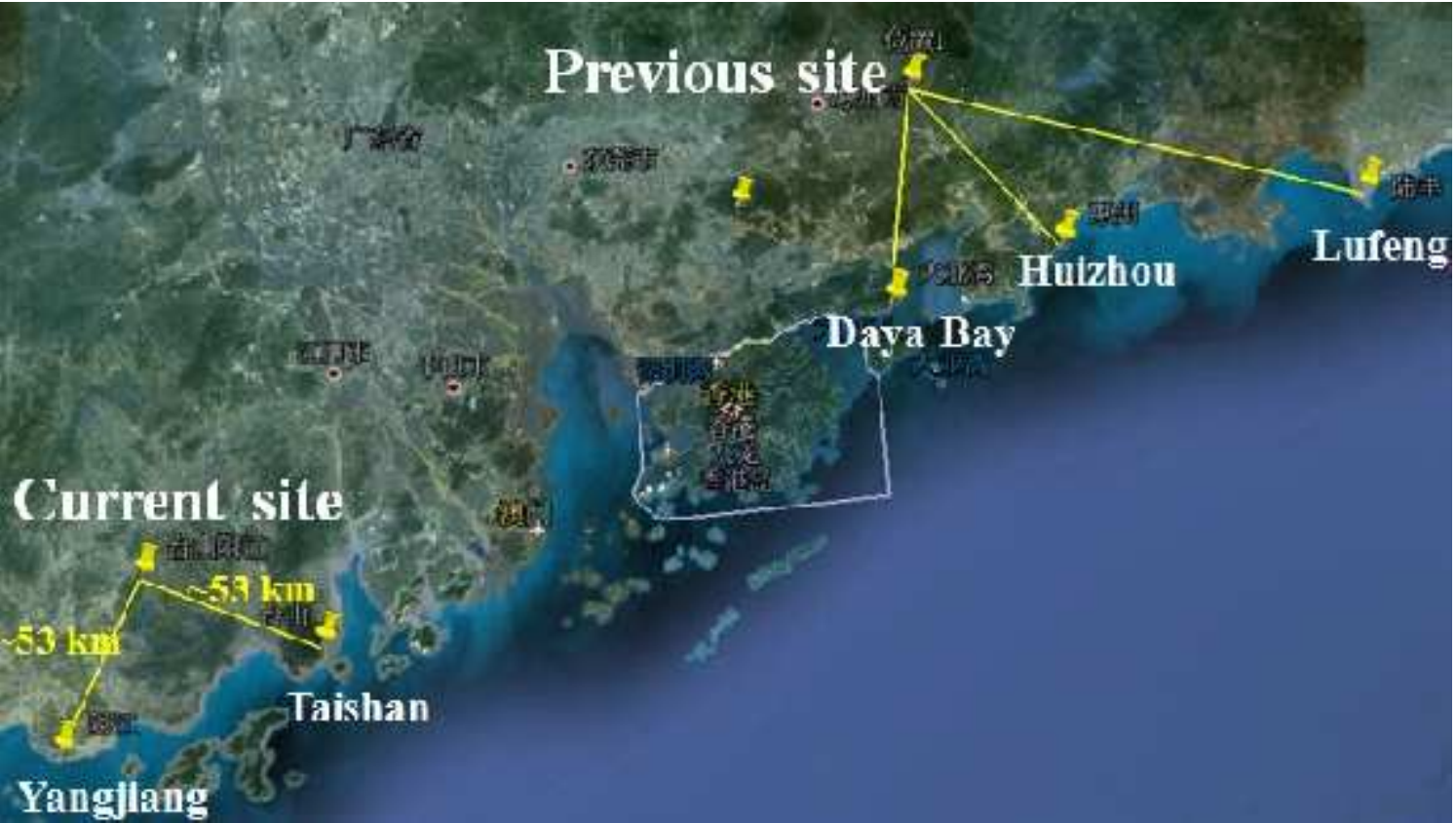}
\par\end{centering}
\caption{\label{fig:juno_site} Experimental site of JUNO. Hongkong and Daya 
Bay reactor complex are labeled. Reproduced with the permission from Ref.~\protect\cite{Yifang:2013july}.}
\end{figure}

Jiangmen Underground Neutrino Observatory  (JUNO) is a reactor neutrino experiment under construction in Jiangmen City, 
Guangdong Province, China~\cite{Yifang:2013july,liangjian_nu2014}. The groundbreaking for 
JUNO started in January 2015. The plan calls for the construction to last until 2017,
assembly and installation in 2018-2019 and start of the data taking in 2020. 
Fig.~\ref{fig:juno_site} shows the JUNO site. The JUNO 
detector will mostly receive $\bar{\nu}_e$ from two reactor complexes at 
Taishan and Yangjiang. The Taishan reactor complex contains six reactor 
cores with a total thermal power of 17.4 GW. The Yangjiang reactor 
complex has two reactor cores with a total thermal power of 9.2 GW.
There are two additional reactor cores (9.2 GW) planned at the Yangjiang
site. The average baseline of JUNO is $\sim$52.5 km with a RMS (root 
mean square) of 0.25 km. 

\begin{figure}[H]
\begin{centering}
\includegraphics[width=0.8\textwidth]{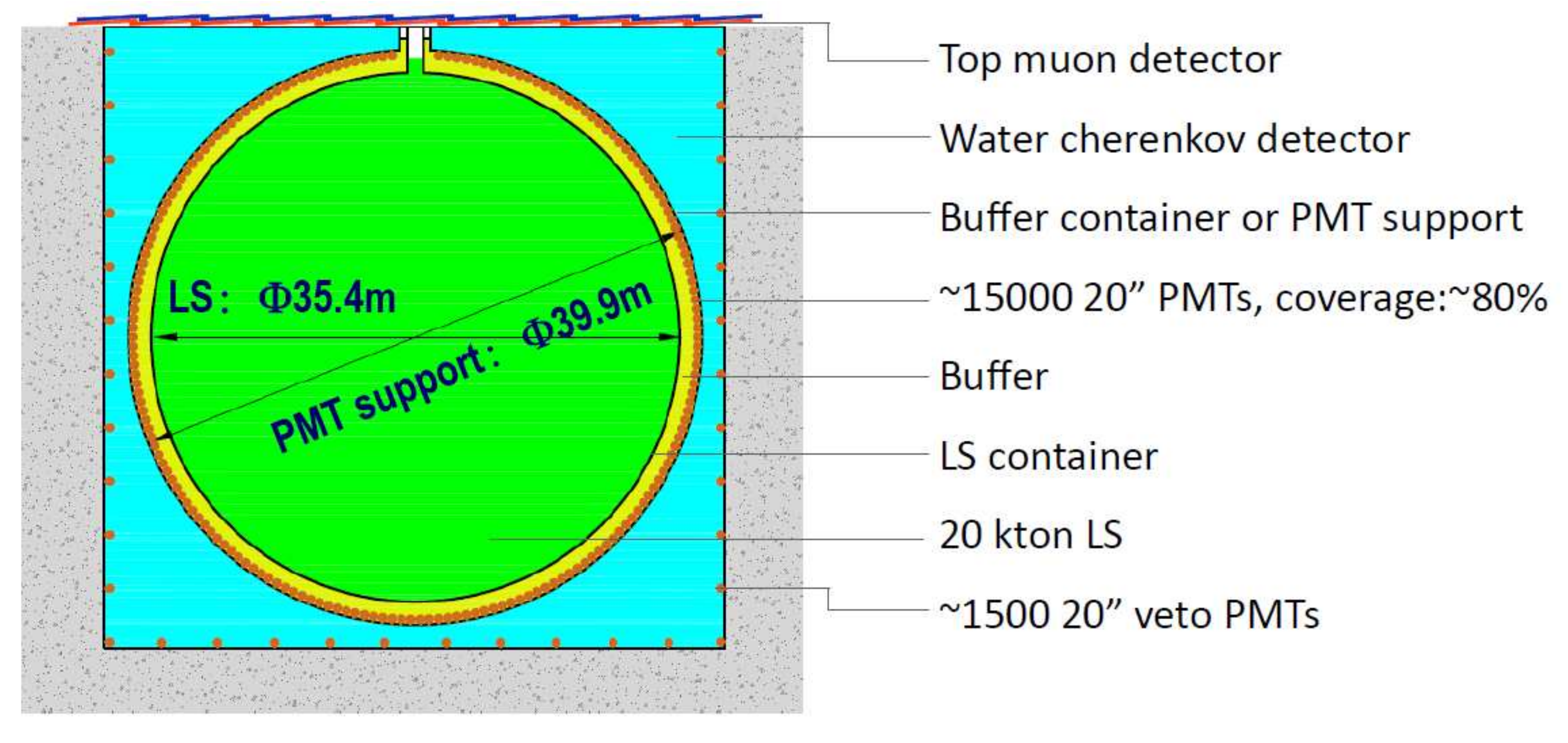}
\par\end{centering}
\caption{\label{fig:juno_det} One conceptual design of JUNO detector. Reproduced with the permission from Ref.~\protect\cite{Yifang:2013july}.}
\end{figure}

The central detector of JUNO is a 20 kton liquid scintillator (LS) detector with 
a total overburden of 1850 meter water equivalent. Figure~\ref{fig:juno_det} shows a conceptual 
design of the spherical LS central detector. The photo-cathode coverage is expected 
to reach ~75\%. Together with high performance LS (high photon yield with $>$14,000 
photons per MeV, optical attenuation length of 30 m or better) and high quantum 
efficiency PMTs (peak quantum efficiency reaching 35\%), JUNO aims to achieve an energy 
resolution of better than 1.9\% at 2.5 MeV, which is essential for the MH determination. 
This level of energy resolution requires a comprehensive calibration system to minimize 
position dependent corrections to the reconstructed energy. The spherical central detector 
will be placed inside a water pool instrumented with PMTs to identify and veto cosmic muons and 
provide shielding from radioactive backgrounds in the underground environment. 
A muon tracker on top of the detector will further enhance the muon identification.

\begin{figure}[H]
\begin{centering}
\includegraphics[width=0.5\textwidth]{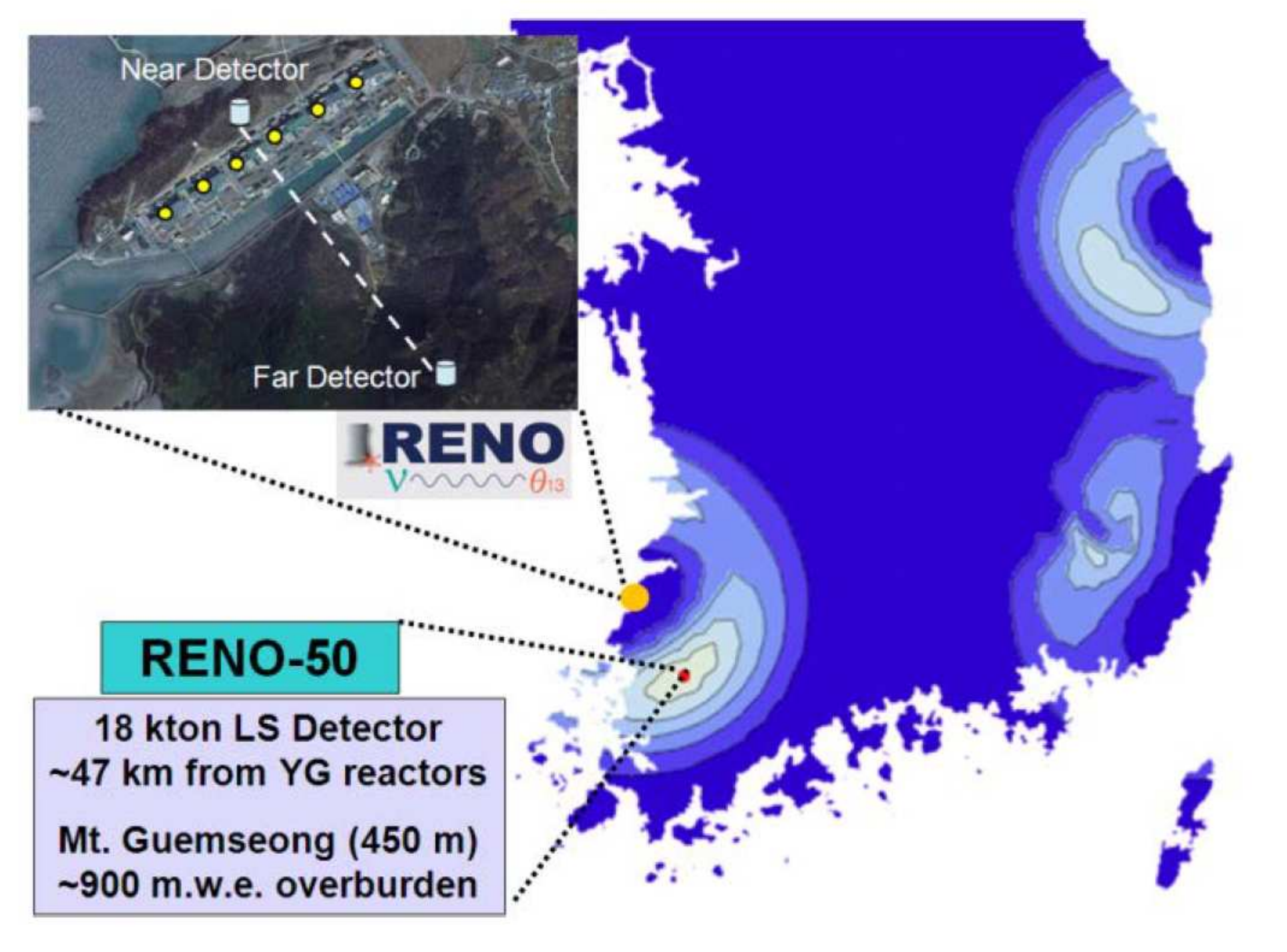}
\par\end{centering}
\caption{\label{fig:reno_site} 
The RENO-50 detector will be located at underground of Mt. Guemseong in a city of Naju, 
47 km from the Hanbit nuclear power plant. Reproduced with the permission from
Ref.~\protect\cite{Kim:2014rfa}.
}
\end{figure}

RENO-50 is a proposed large (18 kton) liquid scintillator detector in 
Korea~\cite{Kim:2014rfa}. At the present time, the schedule for the facility 
and detector construction aims at years 2015-2020 with the start of operation 
in 2021. The central detector is expected to be equipped with 15k
high quantum efficiency 20-inch PMTs. Fig.~\ref{fig:reno_site} shows the proposed
site of RENO-50, roughly 50 km away from the Hanbit nuclear power 
plant. The physics goal of RENO-50 is similar to those of JUNO, including the 
determination of MH and the precision measurements of neutrino mixings. 
In order to reach the desired energy resolution for MH determination, 
RENO-50 is carrying out R\&D to i) enhance the light yield of liquid scintillator 
from $\sim$ 500 to $>$ 1000 photoelectrons/MeV, ii) increase the attenuation length of the liquid 
scinillator from $\sim$ 15 to $\sim$25 m, and iii) enhance the quantum efficiency of PMT 
from 20\% to 35\%. Additional R\&D include the reduction of liquid scintillator 
radioactivity to 10$^{-16}$g/g for U and Th.

\begin{figure}[H]
\begin{centering}
\includegraphics[width=0.5\textwidth]{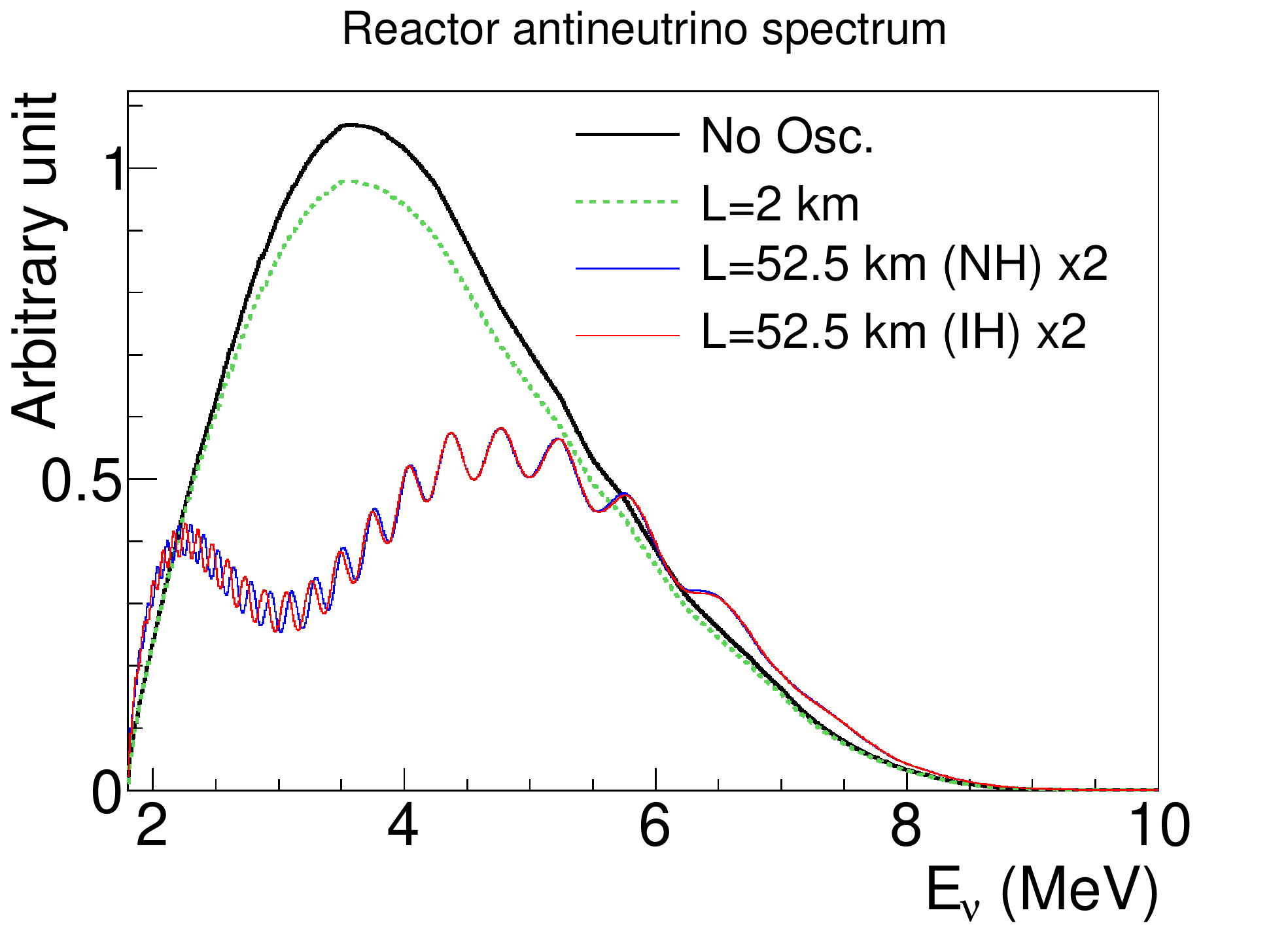}
\par\end{centering}
\caption{\label{fig:juno_spec1} 
The spectral distortion at JUNO (arbitrary scale in 
the vertical axis) for both normal and inverted hierarchies 
with a perfect energy resolution. $\Delta m^2_{32}$ is assumed to be
2.41$\times10^{-3}$ eV$^{2}$ and $-2.53\times10^{-3}$ eV$^2$ for NH and IH, respectively. 
The reactor spectrum without 
oscillation and the oscillated spectrum at 2 km are shown for 
comparison. }
\end{figure}

The principle of the MH determination in a medium baseline reactor 
neutrino experiment has been illustrated in Sec.~\ref{sec:introduction}. 
Figure~\ref{fig:juno_spec1} shows the distortions of the ideal spectrum 
at JUNO for both normal and inverted MH with the perfect 
energy resolution. The effective mass-squared splitting $\Delta m^2_{ee}$ 
(fast oscillation frequency) at low energy (2-4 MeV) is higher than that 
at high energy (4-8 MeV) for NH, and vice versa for 
IH. As the oscillation at the low energy region 
(2-4 MeV) is very rapid, a good energy resolution is required to extract
the oscillation signal. In general, a better than $3\%$/$\sqrt{E({\rm MeV})}$ 
energy resolution has been discussed in the literature as the requirement. 
In reality, the energy resolution is generally parameterized as:
\begin{equation}
\frac{\Delta E}{E} = \sqrt{a + \frac{b}{E} + \frac{c}{E^2}}.
\end{equation}
Here, the $1/E$ term represents the contribution from photon statistics; 
the $1/E^2$ term represents the contribution from PMT dark noise; the 
constant term represents the contribution from residual non-uniformity after
calibration. As the MH signal is around 2.5 MeV, the requirement on the 
energy resolution is naturally converted to better than 1.9\% at 2.5 MeV
prompt energy. Beside the energy resolution, a $\ll$1\% absolute energy scale
uncertainty~\cite{Qian:2012xh,Kettell:2013eos} is also required in order
to unambiguously extract the MH signal. Such a requirement can be addressed
with a dedicated calibration system with a small positron accelerator (such as a 
Pelletron).

\begin{figure}[H]
\begin{centering}
\includegraphics[width=0.49\textwidth]{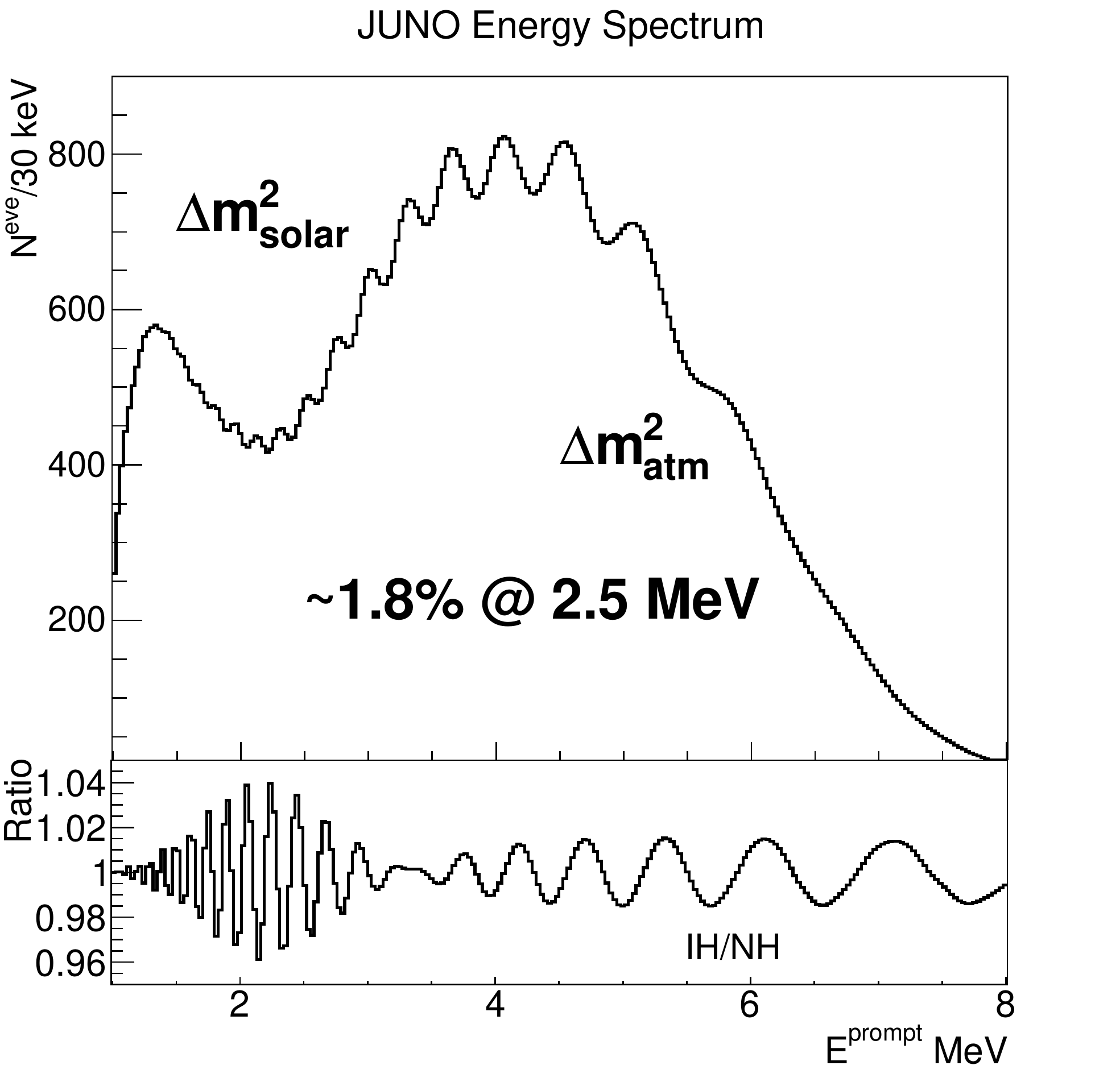}
\includegraphics[width=0.49\textwidth]{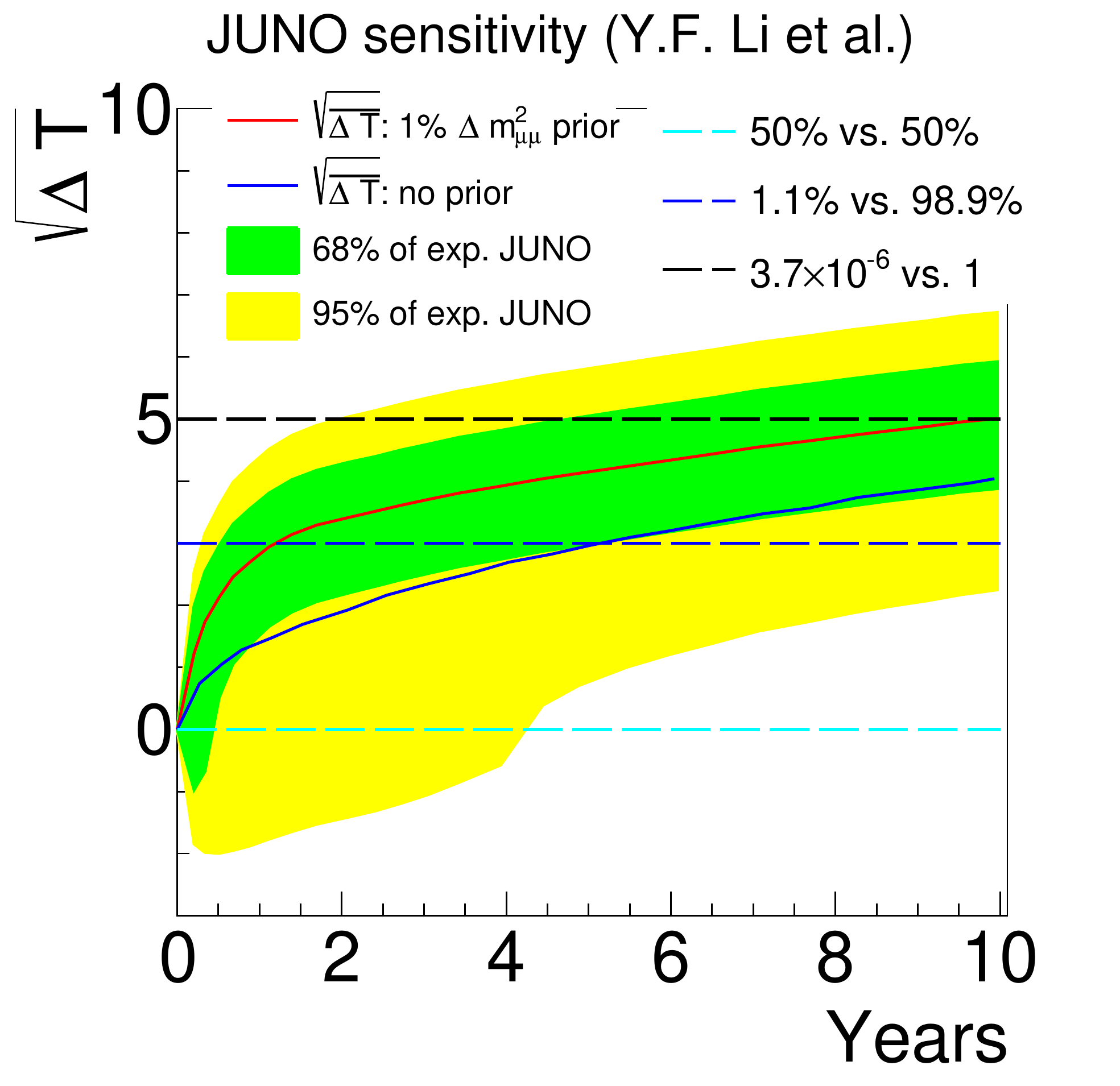}
\par\end{centering}
\caption{\label{fig:juno_spec} (Left) The expected prompt energy
spectrum of JUNO for six years of data
taking with a 20 kt detector and 36 GW$_{th}$ reactor power (a total
of 100k IBD events). The energy resolution is assumed to be about
1.8\% at 2.5 MeV. The bottom panel shows the MH signal as the ratio
of spectrum of inverted hierarchy over the normal hierarchy. 
(Right) JUNO's sensitivity evolution with respect to calendar years.~\protect\cite{Li:2013zyd}. 
 The $\Delta T$ is a test statistics consisting likelihoods
of normal and inverted MH for data $x$. The green and yellow bands
represent the 68\% and 95\% expectations, respectively, taking into 
account the fluctuations in statistics and variations in systematic. 
The dashed lines correspond to the probability ratios of the normal 
vs. inverted MH in the Bayesian framework.}
\end{figure}

In addition to MH, JUNO will perform precision 
measurements of the neutrino mixing, which is a powerful tool to test
the standard 3-flavor neutrino framework (or $\nu$SM). The precision
measurement of $\sin^22\theta_{12}$ will i) lay the foundation for 
a future sub-1\% direct unitarity test of the PMNS 
matrix~\cite{Antusch:2006vwa,Qian:2013ora},  
ii) constrain the allowed region of the effective neutrino mass 
associated with  the decay rate of the neutrinoless double beta decay,
and iii) test various phenomenological models of neutrino masses and mixing~\cite{King:2014nza}, 
such as 
$\theta_{12} = 35^\circ + \theta_{13}\cos\delta_{CP}$, 
$\theta_{12} = 32^\circ + \theta_{13}\cos\delta_{CP}$, and
$\theta_{12} = 45+\theta_{13}\cos\delta_{CP}$, with $\delta_{CP}$ being the CP phase in 
PMNS matrix. The precision measurement 
of $\Delta m^2_{ee}$ (or $\Delta m^2_{32}$) will
i) test an important sum rule  $\Delta m^2_{13} + \Delta m^2_{21} + \Delta m^2_{32} = 0$ and 
ii) reveal additional information regarding MH,
when combined with the precision $\Delta m^2_{\mu\mu}$ measurements 
from muon neutrino and antineutrino disappearance in accelerator experiments. 
JUNO's expected precision of $\Delta m^2_{21}$, $\Delta m^2_{32}$, and $\sin^22\theta_{21}$ 
will be about $\sim$0.6\%~\cite{Kettell:2013eos}. These precision measurements 
are complementary to the atmospheric and long-baseline accelerator neutrino programs. 
Together, the next-generation reactor neutrino and accelerator neutrino 
experiments will access all seven neutrino mixing parameters governing the 
oscillation phenomenon within the three-neutrino framework.

To summarize, the major physics goals of the next generation
medium baseline reactor experiments such as JUNO and RENO-50 are:
\begin{itemize}
\item to be the first experiment to simultaneously observe 
neutrino oscillations from both the atmospheric and the solar neutrino 
mass-squared splittings (see Fig.~\ref{fig:juno_spec}), 
\item to be the first experiment to observe more than two oscillation 
cycles of the atmospheric mass-squared splitting (see 
Fig.~\ref{fig:juno_spec}), 
\item determination of the neutrino mass hierarchy, i.e. whether $\Delta m^2_{32}$
is positive or negative, through the precision measurement of the spectral distortion.
Right panel of Fig.~\ref{fig:juno_spec} shows the expected sensitivity of 
JUNO~\cite{Li:2013zyd} vs. the running time.
\item  precision measurements of $\sin^22\theta_{21}$, $\Delta m^2_{32}$, 
and $\Delta m^2_{21}$ to better than 1\%. We should note that 
the precision measurement of $\Delta m^2_{32}$ requires the 
knowledge of the neutrino mass hierarchy. 
\end{itemize}
Besides these, the 20 kt underground liquid scintillator detector offers a rich physics 
program including the search for proton decay, geoneutrinos, supernova neutrinos, 
and many exotic neutrino physics topics.

%% file: summary.tex
\section{Summary and Outlook}

In this work we review the theoretical and experimental issues one has to resolve in 
order to determine the neutrino mass hierarchy. The motivations and the theoretical 
framework are depicted in Sec.~\ref{sec:introduction}. Then, in Sec.~\ref{sec:methods}
the practical methods that can be used in the MH determination are described. Finally, 
in Sec.~\ref{sec:experiments} the various  existing and proposed or planned experiments 
that have the capacity to shed light on the MH are briefly reviewed.

Determining the neutrino mass hierarchy is a challenging problem. Even though one could 
naively expect that the difference between the normal and inverted hierarchies, the two 
possible mass ordering shown in Fig.~\ref{fig:pattern}, is going to be significant,
in fact the difference is rather subtle in the real data. That is so because the majority 
of experimental findings, at least for the most common case of neutrinos propagating in 
vacuum, can be described in a reasonable approximation as stemming from mixing of just two 
neutrino states; in that case the oscillation pattern depends only on the absolute values 
of the mass-squared differences, not on their sign. Thus, in order to determine the MH, i.e. 
the sign of $\Delta m^2_{3x}$, it is necessary to go beyond that approximation and consider 
the effects associated with mixing of the three neutrino states. The corresponding effects are, 
however, suppressed by the two small parameters present in the oscillation formalism, the mixing 
angle $\theta_{13} \sim 8.5^o$ and the mass square difference ratio $\Delta m^2_{21}/\Delta m^2_{32}
\sim 1/30$. The fact that it is possible to approach the problem in this way at all is the 
result of successful experiments that have recently shown that the mixing angle $\theta_{13}$ is 
not as small as many people believed. That finding opens the way to the MH determination.  
At the same time, that finding also shows that the seemingly simple and attractive possibility of 
the so-called tribimaximal mixing matrix, originally suggested in Ref.~\cite{tribi} and discussed 
in many papers since, is not realized in nature. Even though the PMNS  matrix, as determined from 
the oscillation experiments, looks similar to the simple form suggested, with $|U_{ij}|^2$ expressed in
rational numbers 2/3,1/2, 1/3,1/6 and 0, the reality is more complicated. 

The other way to approach the MH is to use the neutrino propagation in matter, with the additional 
interaction that depends on the mass ordering. The mass ordering of neutrinos $\nu_1$ and $\nu_2$ 
($m_{\nu_1} < m_{\nu_2}$) is determined in this way thanks to the results of the solar neutrino 
experiments. Solar neutrinos propagate from the central region of the Sun, over large distance with 
high electron density, so the matter effects are very pronounced. 
However, the corresponding oscillation length $L_{matter}$ to determine the mass ordering of neutrinos
$\nu_1$ and $\nu_3$ (see Eq.~\eqref{eq:matter}) is very long O($10^3$) km for the typical terrestrial 
electron densities, so the practical application requires very intense neutrino sources and large and 
complicated detectors.

Instead of neutrinos from accelerators, one could use the atmospheric neutrinos that come from all directions, 
including the upward going ones that traverse the Earth diameter. In that case the matter effect is large, 
but the MH determination is hindered by the fact that the atmospheric neutrino flux contains both $\nu_{\mu}$ 
and $\nu_e$ and also both $\nu$ and $\bar{\nu}$ and thus one needs a way to recognize the flavor and polarity 
of the interaction products, or employ the fact that the different components of the atmospheric neutrino flux 
have different intensities and different interaction cross sections.

Having enumerated the difficulties we can understand why the MH remains unknown. But that situation 
can change. As discussed in Sec.~\ref{sec:methods} one can approach the MH problem with neutrinos from 
accelerators and from the atmosphere through the matter effect, or from nuclear reactors through exploring
difference between $\Delta m^2_{32}$ and $\Delta m^2_{31}$. Since the effects are subtle, it is necessary to 
use adequate devices and reliable statistical analysis methods. Also, it would be desirable if all, or 
at least several, of the possible methods were used in practice. The results would be complementary and 
increase the confidence that the conclusions are indeed correct.

 While some of the existing experiments, described in Sec.~\ref{sec:experiments}, have a chance to 
resolve the MH problem soon, it is likely that eventually the more complex and costly devices described 
there will be needed. Again, complementarity will be desirable and very useful. Realistically, some of the 
listed experiments might begin operation within this decade, with
 results after few years of operation. So stay tuned. Rewards are coming only to those who are patient.